\documentclass[10pt, prd, twocolumn]{revtex4}

\usepackage{graphicx,amssymb,amsmath,amsthm,amsfonts}
\usepackage[usenames]{color}
\usepackage{mathtools}

\usepackage{dcolumn}
\usepackage{bm}
\usepackage{hyperref}
\usepackage{physics}
\usepackage{mathtools}
\usepackage{subcaption}

\allowdisplaybreaks

\usepackage{graphicx,amssymb,amsmath,amsthm,amsfonts}
\usepackage[usenames]{color}
\usepackage{xcolor}
\usepackage{mathtools}

\begin{document}

\title{Hot spaces with positive cosmological constant in
the canonical ensemble:  \\  de Sitter solution, 
Schwarzschild-de Sitter black hole, and Nariai universe}

%





\author{Jos\'{e} P. S. Lemos}
\affiliation{Center for Astrophysics and Gravitation -
CENTRA, Departamento de F\'{\i}sica,
Instituto Superior T\'{e}cnico - IST, Universidade de Lisboa - UL,
Avenida Rovisco Pais 1, 1049-001, Portugal, email: joselemos@ist.utl.pt}

\author{Oleg B. Zaslavskii}
\affiliation{Department of Physics and Technology,
Kharkov V.~N.~Karazin National
University, 4 Svoboda Square, Kharkov 61022, Ukraine,
email: zaslav@ukr.net}

\begin{abstract}

\vskip 0.4cm

In a space with fixed positive cosmological constant $\Lambda$, we
consider a system with a black hole surrounded by a heat reservoir at
radius $R$ and fixed temperature $T$, i.e., we analyze the
Schwarzschild-de Sitter black hole space in a cavity. We use results
from the Euclidean path integral approach to quantum gravity to study,
in a semiclassical approximation, the corresponding canonical ensemble
and its thermodynamics. We give the action for the Schwarzschild-de
Sitter black hole space and calculate expressions for the
thermodynamic energy, entropy, temperature, and heat capacity. The
reservoir radius $R$ gauges the other scales. Thus, the temperature
$T$, the cosmological constant $\Lambda$, the black hole horizon
radius $r_+$, and the cosmological horizon radius $r_{\rm c}$, are
gauged to $RT$, $\Lambda R^2$, $\frac{r_+}{R}$, and $\frac{r_{\rm
c}}{R}$. The whole extension of $\Lambda R^2$, $0\leq \Lambda R^2\leq
3$, can be split into three ranges. The first range, $0\leq \Lambda
R^2<1$, includes York's pure Schwarzschild black holes. The other
values of $\Lambda R^2$ within this range also have black holes. The
second range, $\Lambda R^2=1$, opens up a folder containing Nariai
universes, rather than black holes. The third range, $1< \Lambda
R^2\leq 3$, is unusual. One striking feature here is that it
interchanges the cosmological horizon with the black hole horizon. The
end of this range, $\Lambda R^2=3$, only existing for infinite
temperature, represents a cavity filled with de Sitter space inside,
except for a black hole with zero radius, i.e., a singularity, and
with the cosmological horizon coinciding with the reservoir
radius. For the three ranges, for sufficiently low temperatures, which
for quantum systems involving gravitational fields can be very high
when compared to normal scales, there are no black hole solutions and
no Nariai universes, and the space inside the reservoir is hot de
Sitter. The limiting value $RT$ that divides the nonexistence from
existence of black holes or Nariai universes, depends on the value of
$\Lambda R^2$. For each $\Lambda R^2$ different from one, for
sufficiently high temperatures there are two black holes, one small
and thermodynamically unstable, and one large and stable. For $\Lambda
R^2=1$, for any sufficiently high temperature there is the small
unstable black hole, and the neutrally stable hot Nariai
universe. Phase transitions can be analyzed, the dominant phase has
the least action. The transitions are between Schwarzschild de Sitter
black hole and hot de Sitter phases and between Nariai and hot de
Sitter. For small cosmological constant, the action for the stable
black hole equals the pure de Sitter action at a certain black hole
radius and temperature, and so the phases coexist equally. For
$0<\Lambda R^2<1$ the equal action black hole radius is smaller than
the Buchdahl radius, the radius for total collapse, and the
corresponding Buchdahl temperature is greater than the equal action
temperature. So above the Buchdahl temperature, the system collapses
and the phase is constituted by a black hole. For $\Lambda R^2\geq1$ a
phase analysis is also made.


\end{abstract}

\maketitle

\centerline{}
\newpage

\centerline{}
\newpage

\section{Introduction}

\centerline{}
\vskip -0.20cm
One of the most fascinating aspects of an event horizon consists in
the fact that it possesses entropy as well as other quantum and
thermodynamic properties.  These understandings emerged through the
initial works of Bekenstein \cite{bek} and Hawking \cite{h}, and were
endorsed by Gibbons and Hawking \cite{ghpath} within a Euclidean path
integral approach that led to the statistical mechanics canonical
ensemble formalism for black holes.  The Euclidean path integral
approach was extended by Gibbons and Hawking \cite{gh} and Ginsparg
and Perry \cite{gp} to include cosmological horizons, in particular
the de Sitter one, that further permitted the analysis of
semiclassical effects in such spaces.
The path integral and the ensemble theory were put on a firm basis by
York \cite{y86} who realized that proper boundary conditions on the
walls of a heat reservoir that encloses a cavity with a black hole
inside is a well posed problem. This idea was implemented by Whiting
and York \cite{wy} by advancing the correct manner to constraining the
problem.
Hayward \cite{hwds} examined the approach in spaces with
cosmological de Sitter horizons,
Braden, Brown, Whiting,
and York \cite{bbwy}  included electrically charged black holes
in the formalism, 
Zaslavskii enlarged it
further to ensembles with arbitrary configurations of self-gravitating
systems \cite{can},
Lemos applied the approach to 
the two-dimensional black hole of
the Teitelboim-Jackiw theory
\cite{jpsl}, Zaslavskii
studied
the extreme state of a charged black hole in a
grand canonical ensemble \cite{prl},
and
also analyzed the geometry of nonextreme
black holes near the extreme state \cite{prd97}. Pe\c ca and Lemos
implemented the formalism to the grand canonical ensemble of
electrically charged
black holes in anti-de Sitter spaces \cite{pecalemos},
Andr\'e and Lemos extended the results
to $d$ dimensions \cite{Andre:2020,Andre:2021},
Fernandes and
Lemos constructed the grand canonical ensemble
of the electric charged $d$-dimensional
case \cite{fernandeslemos}, and 
Lemos and Zaslavskii studied the interaction
between black holes and matter in the canonical
ensemble \cite{lz2023}

\vskip 0.25cm
The de Sitter space with its cosmological horizon is in itself
fascinating, and its study is now of central importance since it is
realized as the asymptotic solution for our real expanding universe,
as well as forming the basis for the inflationary models of the early
universe. It is thus of significance to learn not only the classical
aspects related to it, but also its quantum and thermodynamic
properties. A key feature of this space is that its cosmological
horizon radiates through quantum processes, but since this radiation
is due to a fixed cosmological constant, it radiates on and on, and
so, unlike a black hole horizon, the de Sitter horizon does not
evanesce. In this sense, the de Sitter cosmological horizon is quantum
stable.
One might want to add a central black hole to the de Sitter space,
obtaining thus the Schwarzschild-de Sitter space, which is a solution
of general relativity.  The Schwarzschild-de Sitter solution has an
appeal of its own since it has two horizons, namely, the black hole
horizon and the cosmological horizon.
These two horizons generically have different temperatures, and so
there is no possible thermodynamic equilibrium solution. This means
that in this setting the problem
should be treated as a nonequilibrium case and no
thermodynamics can be properly devised. However, some insights to
bypass this obstacle to a thermodynamic formulation have been given.
One way to have a proper thermodynamics is to apply
a reservoir kept at some temperature and with boundary at some
radius,  and use York's Euclidean path integral formalism.
Developments in this direction can be mentioned.
Wang and Huang \cite{wanghuang1,wanghuang2}
made a study
of the thermodynamics of the Schwarzschild-de Sitter space in York's
formalism and also extended to the thermodynamics of
Reissner-Nordstr\"om-de Sitter. Ghezelbash and Mann \cite{ghezelbashmann}
analyzed
the action and entropy of Schwarzschild-de Sitter black holes. Saida
\cite{saida}
treated
some aspects of the Schwarzschild-de Sitter thermodynamics in the
canonical ensemble, and  Draper and Farkas \cite{draperfarkas}
discussed 
de Sitter black holes in the Euclidean path integral approach.
Banihashem and Jacobson \cite{banihashem}
considered
thermodynamic ensembles with cosmological horizons, Banihashemi,
Jacobson, Svesko, and Visser \cite{bjsv}
explored
further the  minus sign that enters the thermodynamic energy in
the first law of thermodynamics for de Sitter horizons, Jacobson and
Visser \cite{jacviss2,jacviss}
built
the partition function for a volume of space as well as the partition
function and the entropy of causal diamond ensembles, and Morvan, van
der Schaa, and Visser,
examined
the Euclidean action of de Sitter black holes \cite{morvan}.

\vskip 0.25cm
In this work, we want to further understand the thermodynamics of the
Schwarzschild-de Sitter black hole space in the canonical ensemble
within York's formalism. In using it one has to choose whether the
heat reservoir, put in-between the two horizons, is a reservoir for
the inside, i.e., is a reservoir for the black hole horizon region, or
is a reservoir for the outer universe, i.e., for the region that
includes the cosmological horizon.  Here we are interested in the
first situation, and will study the thermodynamics of the black hole
region in a cavity with a heat reservoir outside.  The black hole
horizon and its temperature together with the reservoir and its
temperature play a principal role in the thermodynamic analysis,
indeed the equilibrium situations are established by them.  The
cosmological horizon has no major role in this setting, actually its
place is a function of the black hole horizon location, and it is
directly determined once the location off this latter has been found
through thermodynamic computation.
In this setup there are three main
scales, the scale set by the size of the
reservoir, the scale set by the temperature of the
reservoir, and the scale set by the cosmological constant,
which in turn yield 
the scale set by the size of the black hole horizon and
the scale set by the size of the cosmological horizon.  It is thus
expected that the existence of these various scales yields new,
interesting, and important properties of the system.
One that we can advance now, is that the set of ensembles is comprised
not only of the Schwarzschild-de Sitter black hole but also of the
Nariai universe, which arises naturally when the cosmological length
scale and the reservoir length scale are equal.  This intermediate
case divides the ensembles into a set of ensembles with low
cosmological constant that has familiar properties, and another
set with high cosmological constant that is new and in which the black
hole and cosmological horizons exchange roles. Other thermodynamic
properties become quite unusual as compared with the no
cosmological constant black hole case in the canonical ensemble.

\centerline{}
We would like to mention several other different attempts devised to
understand the quantum and thermodynamic nature of black holes in de
Sitter space that are interesting on themselves but that do not have a
direct bearing with our work here.  Davies \cite{pcwdavies} studied
the black hole mechanics and thermodynamics of the Kerr-Newman-de
Sitter family of solutions, and Romans \cite{romans} performed an
important analysis of the de Sitter black holes, classifying them as
temperature goes in cold, lukewarm, and warm. Bousso and Hawking
\cite{boussohawking2} put forward the interesting possibility of having
evaporation and anti-evaporation of Schwarzschild–de Sitter black
holes, Maeda, Koike, Narita, and Ishibashi \cite{maeda} found an upper
bound for the entropy of an asymptotically de Sitter spacetime, Wu
\cite{wu} worked out the entropy of black holes with different surface
gravities with applications to Schwarzschild-de Sitter black holes,
Yueqin, Lichun, and Ren \cite{yueqinlichunren} used the brick wall
method of ´t Hooft to calculate the entropy of Schwarzschild-de Sitter
black holes, Bousso \cite{bousso} has defined causal diamonds in de
Sitter black hole spaces and inspected their entropy, Cai \cite{cai1}
considered the Cardy-Verlinde formula in connection to thermodynamics of
de Sitter black holes, Shankaranarayanan \cite{shanka} attempted to set a
scheme where the different temperatures of the two Schwarzschild-de
Sitter horizons could be made consistent, and Teitelboim and Gomberoff
\cite{teitel1,teitel2} examined the de Sitter black holes with
either one of
the two horizons working as a thermodynamic boundary.  Dias and Lemos
\cite{diaslemospaircreationdesitter} analyzed pair creation of de
Sitter black holes on a cosmic string background and the associated
entropy, Cardoso, Dias, and Lemos
\cite{diaslemoscardosonariaiinhigherd} displayed the Schwarzschild-de
Sitter, Nariai, Bertotti-Robinson, and anti-Nariai solutions in higher
dimensions, in particular their temperatures including the lukewarm
cases, Sekiwa \cite{sekiwa} investigated
Schwarzschild-de Sitter spaces with
the cosmological constant as a thermodynamic variable, Choudhury and
Padmanabhan \cite{roychoudhurypadmanabhan} invoked a new concept of
temperature in spaces with several horizons, Myung \cite{myung}
inspected the thermodynamics of the Schwarzschild-de Sitter black hole
and the Nariai solution in five dimensions, Pappas and Kanti
\cite{pappas} treated Schwarzschild–de Sitter spaces and the role of
temperature in the emission of Hawking radiation, Simovic and Mann
\cite{fil} exhibited critical phenomena of certain types of de Sitter
black holes in cavities, Qiu and Traschen \cite{qiutraschen} obtained
new results related to thermodynamics
of black pair production in Schwarzschild-de Sitter spaces,
Singha \cite{singha} developed further the thermodynamics of spaces with
several horizons, Volovik \cite{volovik2} suggested a double Hawking
temperature ansatz to explain the thermodynamics of
a black hole in de Sitter space, and
Akhmedov and Bazarov \cite{akhmedov} made an analysis of the
backreaction issue for a black hole in de Sitter space.

\vskip 0.25cm
An important concept for finite self-gravitating systems,
as the ones we want to consider,
is the Buchdahl bound
that sets a maximum mass or a maximum gravitational radius for the
energy that can be enclosed in a cavity before the system turns
singular and presumably suffers total
gravitational collapse.  Usually, the Buchdahl radius
concerns the mechanical structure of balls or stars in general
relativity, but it should also appear
somehow in connection with
thermodynamics and thermodynamic phases,
since when, in a cavity, there is energy in the form
of matter or radiation with gravitational radius larger than the
gravitational radius permitted by the Buchdahl bound for a given
cavity size, that energy should collapse. Thus, a thermodynamic system
that has too much thermodynamic energy for a given cavity size must
collapse.
Since here, we are interested in self-gravitating systems
in a positive cosmological constant background
in a general relativistic context, the
Buchdahl bound of interest is the
one found by Andr\'easson and
B\"ohmer \cite{andreassonbohmer}.

\vskip 0.5cm

The paper is organized as follows.
In Sec.~\ref{generalresults} we state the main general thermodynamic
results, derived from the canonical ensemble set by the Euclidean path
integral approach, for a cavity containing the black hole horizon
region of the Schwarzschild-de Sitter space inside a heat reservoir.
In Sec.~\ref{Rlamda2<1} we give specific results for the
thermodynamics of Schwarzschild-de Sitter black holes with small values
of the cosmological constant, $\Lambda R^2<1$,
also studying thermodynamic phases and phase transitions.
In Sec.~\ref{Rlamda2=1} we give specific results for the
thermodynamics of Schwarzschild-de Sitter black holes wth the
intermediate value of the cosmological constant, $\Lambda R^2=1$,
which is found to be the Nariai universe, and 
also studying thermodynamic phases and phase transitions.
In Sec.~\ref{Rlamda2>1} we give specific results for the
thermodynamics of Schwarzschild-de Sitter black holes wth large values
of the cosmological constant, $\Lambda R^2>1$, and 
also studying thermodynamic phases and phase transitions.
In Sec.~\ref{diagramsandanalysis} we present important plots and make
a thorough analytic study of all the cases.  In Sec.~\ref{conc} we
draw our conclusions.
In the Appendix~\ref{appendix:sdsnbasics} we state the basic geometric
elements of the Schwarzschild-de Sitter and Nariai spaces.
In the Appendix~\ref{nariailimitapp} the Nariai limit from the
Schwarzschild-de Sitter space in a cavity in a thermodynamic setting
is presented in all detail.
In the Appendix~\ref{fomulassectionvi} we derive explicitly some
expressions of the main text.

\vfill

\section{
Thermodynamics of the Schwarzschild-de Sitter space in the canonical
ensemble: General results for the black hole horizon region inside a
heat reservoir
}
\label{generalresults}

\subsection{Setup and Euclidean metric}

\begin{figure}[h]
\includegraphics*[scale=0.7]{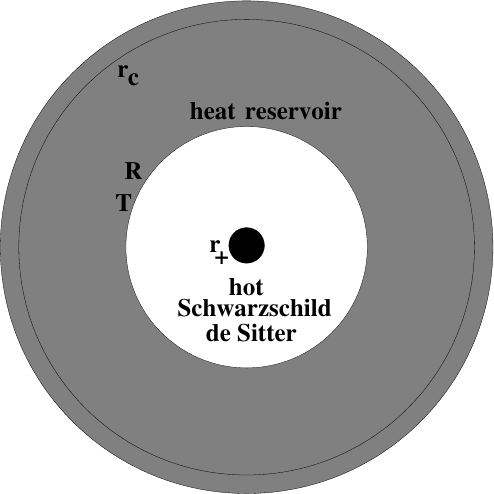}
\caption{
A drawing of a black hole in a cavity
within a heat reservoir at temperature $T$
and radius $R$ in a space with positive cosmological constant.
Outside the black hole radius $r_+$ the geometry is a Schwarzschild-de
Sitter geometry. The cosmological radius $r_{\rm c}$ is beyond the
heat reservoir.  The Euclideanized space and its boundary have $R^2
\times S^2$ and $S^1 \times S^2$ topologies, respectively, where the
$S^1$ subspace with proper length $\beta=\frac1T$ is not displayed.
See text for more details.
}
\label{schwdesitterreservoir1}
\end{figure}

Put,  at some
radius $R$,
the boundary of a spherical cavity with a black hole in a positive
cosmological constant background inside a heat reservoir.
At this boundary, one specifies the data that determine
the ensemble, see Fig.~\ref{schwdesitterreservoir1}.
We fix the temperature at the boundary at $R$, so we are working with
the canonical ensemble. Furthermore, we consider that the heat
reservoir is a heat reservoir for the inner region of the
Schwarzschild-de Sitter space and thermodynamically discard the region
$r>R$. This is the approach first suggested by York for the
Schwarzschild metric \cite{y86,wy,bbwy,hwds}, that was generalized
further to include matter \cite{can}.

In order to implement the Euclidean path integral approach and work
out canonical ensemble results for a black hole in a
positive constant background we use 
the Schwarzschild-de Sitter solution in general relativity. 
Then, for
the Schwarzschild-de Sitter black
hole metric, one Euclideanizes time and fixes its
period $\beta$ at radius $R$ to be the heat reservoir inverse
temperature, $\beta\equiv\frac1T$.
The Schwarzschild-de Sitter space
is characterized by two parameters, namely, the mass $m$ and the
positive cosmological constant $\Lambda$.  Instead of working with $m$
and $\Lambda$, it is sometimes preferable to use the black hole
horizon radius $r_+$ and the cosmological horizon radius $r_{\rm c}$,
the two sets of two parameters are interchangeable by precise
formulas.
In the case we are working, 
the reservoir
is a reservoir for the inner region that contains a black hole,
so that $r_+$ is inside $R$ and
$r_{\rm c}$ is outside $R$.
The Euclidean line element of
the Schwarzschild-de Sitter space in spherical
coordinates $(t,r,\theta,\phi)$ is
obtained by Euclideanizing time, $t\to it$,
to get the  Euclidean Schwarzschild-de Sitter space,
\begin{align}
ds^2=V(r) \,
dt^2+&\frac{dr^2}{V(r)}+r^2(d\theta^2+ \sin^2\theta
\,d\phi^2),\nonumber\\
&0\leq t<\beta^{\rm H}_+\,, \, r_+\leq r\leq R\,,
\label{met1euc+}
\end{align}
where the metric potential $V(r)$
has the form
\begin{equation}
V(r)=1-\frac{2m}{r}-\frac{\Lambda r^2}{3}\,,
\label{meteuc}
\end{equation}
and $0\leq\theta\leq\pi$, 
$0\leq\phi<2\pi$.
The range of coordinates of Euclidean time
$t$ is $0\leq t<\beta^{\rm H}_+$, where
$\beta^{\rm H}_+$ is the period
of the time coordinate such that the line element
given by Eqs.~\eqref{met1euc+} and \eqref{meteuc}
has no conical singularities.
The relation with the Hawking temperature $T^{\rm H}_+$
is
$\beta^{\rm H}_+=\frac1{T^{\rm H}_+}$.
Due to the reservoir at radius $R$ the range
of the radial coordinate is
$r_+\leq r\leq R$.
The line element provided in
Eqs.~\eqref{met1euc+} and \eqref{meteuc}
is the Euclideanized
form of the Schwarzschild-de Sitter
spacetime, see Appendix \ref{appendix:sdsnbasics}, and
its topology is
$R^2\times S^2$.

The black hole horizon radius $r_+$ is one
of the two real zeros 
of $V(r)$ in Eq.~\eqref{meteuc}
and so obeys
\begin{equation}
\frac{r_+}{2}\left(1-\frac{\Lambda r_+^2}{3}\right)=m\,.
\label{V=0euc}
\end{equation}
Then $m$ can be traded for $r_+$ to give $V(r)$
of Eq.~\eqref{meteuc} as
$V(r)=1-\frac{r_+}{r}
-\frac{\Lambda r^2}{3}\left(1-\left(\frac{r_+}{r}\right)^3\right)$,
i.e.,
\begin{equation}
V(r)=\left(1-\frac{r_+}{r}\right)
\left(1-\frac{\Lambda r^2}{3}\left(1+
\frac{r_+}{r}+\left(\frac{r_+}{r}\right)^2\right)\right)\,.
\label{meteucnew}
\end{equation}
The cosmological horizon radius
is the other zero
of $V(r)$ in Eq.~\eqref{meteuc}. 
It
can be found
once $r_+$ is known through the equation
\begin{equation}
r_{\rm c}=
-\frac{r_+}2+\frac{r_+}2\sqrt{\frac{12-3\Lambda
r_+^2}{\Lambda r_+^2}}\,,
\label{rcofr+Lambdaagain}
\end{equation}
or $
r_{\rm c}=-\frac{r_+}2
+\frac12 \sqrt{\frac{3}{\Lambda}}
\sqrt{4-\Lambda r_+^2}\
$.
Since the heat reservoir envelopes the inside,
the primary horizon radius is
the black hole horizon $r_+$, which
has to be found
from thermodynamic considerations. The
cosmological horizon radius $r_{\rm c}$
has a secondary role, being determined
once $r_+$ is known.

Now, the radius of the heat reservoir $R$ sets a scale for our
problem. It is then meaningful to gauge all the length scales involved
in the problem to $R$.  Thus, the heat reservoir temperature $T$, the
cosmological constant $\Lambda$, the black hole horizon radius $r_+$,
and the cosmological horizon radius $r_{\rm c}$, are gauged to
quantities without units as $RT$, $\Lambda R^2$, $\frac{r_+}{R}$, and
$\frac{r_{\rm c}}{R}$.  The extensions of these quantities are
important. They are: $0\leq RT<\infty$, $0\leq \Lambda R^2\leq 3$,
$0\leq\frac{r_+}{R}\leq1$, and $1\leq\frac{r_{\rm c}}{R}<\infty$.  It
is advisable to separate the whole extension of $\Lambda R^2$, $0\leq
\Lambda R^2\leq 3$, into three cases, namely, small values of the
cosmological constant which means $\Lambda R^2<1$, more exactly
$0\leq\Lambda R^2<1$, the intermediate value of the cosmological
constant which means $\Lambda R^2=1$, and large values of the
cosmological constant which means $\Lambda R^2>1$, more exactly
$1<\Lambda R^2\leq3$.
Note that the cosmological constant, which has units of inverse
length square, has associated to
it a natural cosmological length scale
$\ell$ given by $\ell^2=\frac{1}{\Lambda}$.

\subsection{Action, energy, entropy}

We list here the most relevant general formulas
that come out of the path integral approach, one
can look elsewhere to pick up these formulas,
see, e.g., \cite{
y86,
wy,
bbwy,
hwds,
can,
jpsl,
prl,
prd97,
pecalemos,Andre:2020,Andre:2021,fernandeslemos,lz2023}.
The Euclidean action $I$ is 
\begin{equation}
I=\beta R\left(1-\sqrt{V(R)}\right) -\pi r_+^2\,,
\label{actionbh}
\end{equation}
where $\beta=\frac1T$ is the inverse temperature
of the ensemble, i.e., at the
boundary of the heat reservoir, 
and where $V(R)$ is given by Eq.~\eqref{meteuc}
at $r=R$, i.e.,
$V(R)=1-\frac{2m}{R}-\frac{\Lambda R^2}{3}$
or 
\begin{equation}
V(R)=\left(1-\frac{r_+}{R}\right)
\left(1-\frac{\Lambda R^2}{3}\left(1+
\frac{r_+}{R}+\left(\frac{r_+}{R}\right)^2\right)\right)\,.
\label{meteucnewR}
\end{equation}
Note that $I$ in Eq.~\eqref{actionbh}
is $I=I(\beta,R,\Lambda;r_+)$. The statistical
mechanics ensemble
is characterized by $\Lambda$ which is
fixed for each space, by
$T=\frac1\beta$ and $R$
which are fixed for each ensemble,
and by $r_+$ which can vary,
there are particular 
$r_+$ solutions for which $I$ is stationary,
$\frac{dI}{dr_+}=0$,
yielding the thermodynamic solutions of the
problem.

The Euclidean action and the free energy are related by
$
I=\beta F
$,
so that
$
F=R\left(1-\sqrt{V(R)}\right) -T\pi r_+^2
$.
Now, 
$
F=E - TS
$,
so the thermodynamic or quasilocal energy here
also the thermal energy at $R$ at this order
can be found to be 
\begin{equation}
E=R\left(1-\sqrt{V(R)}\right).
\label{Ebh}
\end{equation}
The entropy of the system is
\begin{equation}
S=\pi r_+^2\,, 
\label{entropybh}
\end{equation}
which is the Bekenstein-Hawking entropy.

To find the thermodynamic stability one has to compute
the heat capacity
at constant reservoir area $A$, 
$C_A$. This is given by 
\begin{equation}
C_A=\left(\frac{dE}{dT}\right)_{\hskip -0.15cm A}\,.
\label{heatcapacity}
\end{equation}
If $C_A<0$ the system is thermodynamically unstable,
if $C_A\geq0$ the system is thermodynamically stable,
with the equality giving the marginal case.
Since $A=4\pi R^2$, Eq.~\eqref{heatcapacity}
is equivalent to 
$C_R=\left(\frac{dE}{dT}\right)_{\hskip -0.05cm R}$.

\subsection{Temperature and solutions}

In order that the whole formalism be meaningful,
the line element of 
Eqs.~\eqref{met1euc+} and \eqref{meteuc}
should have no conical singularities
in the Euclidean $r\times t$ plane.
This implies that the time coordinate $t$
has to have period $\beta^{\rm H}_+$ given by
$\beta^{\rm H}_+=\frac{4\pi}
{
\left(\frac{dV(r)}{dr}\right)_{r_{
+}}
}$. It turns out that this period is
related to the Hawking temperature 
$T^{\rm H}_+$ through
$\beta^{\rm H}_+=\frac1{T^{\rm H}_+}$,
so 
$T^{\rm H}_+=\frac
{
\left(\frac{dV(r)}{dr}\right)_{r_{
+}}
}{4\pi}$.
From Eq.~\eqref{meteuc} we obtain
\begin{equation}
T^{\rm H}_+=
\frac{1}{4\pi r_+} \left(1-\Lambda r_+^2\right).
\label{tbh}
\end{equation}
Using
Eq.~\eqref{V=0euc} this can be also put
in the form
$T^{\rm H}_+=\frac{1}{2\pi r_+}\left(\frac{3m}{r_+}-1\right)$.

The stationary points
of the action  Eq.~\eqref{actionbh}
are found through 
$\frac{dI}{dr_+}=0$,
which yields the equation
\begin{equation}
T=\frac{T^{\rm H}_+}{\sqrt{V(R)}}\,,
\label{tn}
\end{equation}
with $T^{\rm H}_+$ being the Hawking temperature given
in Eq.~\eqref{tbh} and $V(R)$ is given in  Eq.~\eqref{meteucnewR}.
This is the Tolman relation for the temperature
at the reservoir and the Hawking temperature.
Since $T$ has the expression given in Eq.~\eqref{tn}
one finds explicitly for this case, using 
Eq.~\eqref{tbh}
and Eq.~\eqref{meteucnewR},
that 
$T=\frac{
\frac{1}{4\pi r_+} \left(1-\Lambda r_+^2\right)}{\sqrt{1-
\frac{2m}{R}-\frac{\Lambda R^2}{3}}}$, i.e., 
$T=\frac{
\frac{1}{4\pi r_+} \left(1-\Lambda r_+^2\right)}{
\sqrt{1-\frac{r_+}{R}
-\frac{\Lambda}{3R}\left(R^3-r_+^3\right)}
}$.
Thus,
$4\pi RT=
\frac{1}{ \frac{r_+}{R}}
\frac{
1-\Lambda R^2\left(\frac{r_+}{R}\right)^2}{
\sqrt{1-\frac{r_+}{R}
-\frac{\Lambda R^2}{3}\left(1-\left(\frac{r_+}{R}\right)^3\right)}}$, 
which can be put in the form
\begin{equation}
4\pi RT=
\frac{1}{ \frac{r_+}{R}}
\frac{
1-\Lambda R^2\left(\frac{r_+}{R}\right)^2}
{
\sqrt{1-\frac{r_+}{R}}\,
\sqrt{\left(1-\frac{\Lambda R^2}{3}\left(1+
\frac{r_+}{R}+\left(\frac{r_+}{R}\right)^2\right)\right)}
}\,.
\label{tbh2}
\end{equation}
We want to find $r_+$ that obey
Eq.~\eqref{tbh2}, and so generically
one has $\frac{r_+}{R}(\Lambda R^2, RT)$.
For fixed $RT$ one has 
$\frac{r_+}{R}(\Lambda R^2)$,
and for fixed 
$\Lambda R^2$ one has 
$\frac{r_+}{R}(RT)$.

Thus, for a given fixed
$T$ at the boundary, for the ensemble, one can look
for solutions $r_+$.
Indeed,
depending on the parameters $(T,R,\Lambda)$,
Eq.~\eqref{tbh2} can have no solution,
one solution, or two solutions $r_+$.
When it has two solutions 
we denote these by 
\begin{equation}
r_{+1}=r_{+1}(R,\Lambda,T),
\label{rh1}
\end{equation}
and 
\begin{equation}
r_{+2}=r_{+2}(R,\Lambda,T),
\label{rh2}
\end{equation}
with $r_{+1}\leq r_{+2}$, say. The functions
$r_{+1}$ and $r_{+2}$ 
have to be worked
out in some way or another. The
fact that for  given $T$ and $R$ there exist two roots $r_{+1}$ and
$r_{+2}$ is similar to that for the Schwarzschild space \cite{y86},
here it is for a given $T$ and $R$, and for a given $\Lambda$.
Moreover,
given the solutions $r_{+1}$ of 
Eq.~\eqref{rh1} and 
 $r_{+2}$ of 
Eq.~\eqref{rh2},
one finds 
from Eq.~\eqref{rcofr+Lambdaagain}
the two corresponding
cosmological horizon radii, namely, 
\begin{equation}
r_{\rm c1}=r_{\rm c1}(r_{+1}(R,\Lambda,T))\,,
\label{rc1}
\end{equation}
and 
\begin{equation}
r_{\rm c2}=r_{\rm c2}(r_{+2}(R,\Lambda,T))\,,
\label{rc2}
\end{equation}
respectively, with 
$r_{\rm c1}\geq r_{\rm c2}$.

In brief, we have already generic results, though not explicit. We
now treat separately the three cases, $0\leq\Lambda R^2<1$, $\Lambda
R^2=1$, and $1<\Lambda R^2\leq3$.
In general there are no analytical solutions. In the first
case, analytical expressions for particular ranges
of $\Lambda R^2$ can be found,
in the second case one finds that one is in the presence of a
Nariai universe which has exact thermodynamic solutions,
and in the third case expressions for particular ranges
of $\Lambda R^2$ can also be found.
The high temperature limit in the three
cases yields analytical solutions.
Thermodynamic phases and phase transitions
can be analyzed in all the three
cases.


\section{Thermodynamics of Schwarzschild-de Sitter black holes in the
canonical ensemble: Small values of the cosmological constant,
$\Lambda R^2<1$}
\label{Rlamda2<1}

\subsection{Solutions}

We treat here the small positive cosmological constant, $\Lambda
R^2<1$, problem. Again, we put the boundary of a spherical cavity with
a black hole in a positive cosmological constant background inside a
heat reservoir, at some radius $R$, where it is also specified a fixed
temperature $T$, see Fig.~\ref{schwdesitterreservoir1} anew.  Small
positive cosmological constant means exactly in this context that
\begin{equation}
0\leq\Lambda R^2<1\,.
\label{range<1}
\end{equation}
Within this range, for fixed $RT$ and generic $\Lambda R^2$, it is
hard to find solutions of Eq.~\eqref{tbh2} for black hole horizon
radii $r_+$ analytically.  However, for very small $\Lambda R^2$ or
for $\Lambda R^2$ very near one, one can make some progress.

For very small $\Lambda R^2$, i.e., for 
$\Lambda R^2\ll 1$, one finds
from Eq.~\eqref{tbh2} that
there are no black hole solutions 
for
\begin{equation}
R T<\frac{\sqrt{27}}{8\pi}
\left( 1-\frac{5}{54}\,\Lambda R^2\right)
\,,\quad\quad\quad\Lambda R^2\ll 1\,,
\label{nosolutionslimit}
\end{equation}
only hot de Sitter space is possible.
Still for very small $\Lambda$, i.e., for 
$\Lambda R^2\ll 1$, there
are two black hole solutions 
for
\begin{equation}
RT
\geq \frac{\sqrt{27}}{8\pi} \left( 1-\frac{5}{54}\,
\Lambda R^2\right)\,,\quad\quad\quad\Lambda R^2\ll 1
\,.
\label{solutionslimit}
\end{equation}
One of the two solutions is the small black
hole $r_{+1}(R,\Lambda,T)$,
and the other solution is the large black hole
$r_{+2}(R,\Lambda,T)$.
For zero cosmological constant,
$\Lambda R^2=0$, one has a pure Schwarzschild black
hole and one recovers York's result of 
$RT
\geq \frac{\sqrt{27}}{8\pi}$ to have black
hole solutions.
The minus sign inside the parenthesis in
Eq.~\eqref{solutionslimit} is what one expects really.
The two solutions merge
into one sole solution when 
the equality sign in
Eq.~\eqref{solutionslimit}
holds.
In this case the coincident
double solution has horizon radius
given by
\begin{equation}
\frac{r_{+1}}{R}=\frac{r_{+2}}{R}=\frac23
\left(1+
+\frac{17}{81}\Lambda R^2\right)\,,
\quad \Lambda R^2\ll1\,.
\label{r+solutionslimit}
\end{equation}
The corresponding
cosmological radius can then be found
from Eq.~\eqref{rcofr+Lambdaagain}
to be given by 
\begin{equation}
\frac{ r_{\rm c1} }{R}=\frac{r_{\rm c2}}{R}=
\sqrt{\frac{ 3}{\Lambda R^2}}
\left(1-\frac{1}{3}\sqrt{\frac{\Lambda R^2}3}\right)\,,
\quad \Lambda R^2\ll1\,.
\label{rcsolutionslimit}
\end{equation}

For $\Lambda R^2$ very near unity, i.e., for 
$(1-\Lambda R^2)\ll 1$, one finds
from Eq.~\eqref{tbh2} that
there are no black hole solutions 
for
\begin{equation}
\hskip -0.2cm
R T<\frac{1}{2\pi}
\left( 1+
\left(\frac{3}{8}\,\left(1-\Lambda R^2\right)\right)^\frac13\right)
\hskip-0.07cm,
\;\; (1
\hskip-0.07cm
-
\hskip-0.07cm
\Lambda R^2)
\hskip-0.07cm
\ll
\hskip-0.07cm
1\,,
\label{solutionsLambdaR2<near1}
\end{equation}
only hot de-Sitter space is possible.
Still for small $\Lambda R^2 -1$, i.e., for 
$(1-\Lambda R^2)\ll 1$, there
are two black hole solutions 
for
\begin{equation}
\hskip-0.2cm
R T\geq\frac{1}{2\pi}
\left( 1+
\left(\frac{3}{8}\,\left(1-\Lambda R^2\right)\right)^\frac13\right)
\hskip-0.07cm,
\;\; (1
\hskip-0.07cm
-
\hskip-0.07cm
\Lambda R^2)
\hskip-0.07cm
\ll
\hskip-0.07cm
1\,.
\label{solutionslimit>LambdaR2<near1}
\end{equation}
When the equality holds
the coincident double solution has horizon radius given by
\begin{equation}
\hskip-0.2cm
\frac{r_{+1}}{R}=\frac{r_{+2}}{R}= 1-\left(\frac38\,
(1-\Lambda R^2)^2\right)^{\frac13}
\hskip-0.1cm,
\;\; (1
\hskip-0.07cm
-
\hskip-0.07cm
\Lambda R^2)
\hskip-0.07cm
\ll
\hskip-0.07cm
1\,.
\label{solutioninitial1}
\end{equation}
The corresponding
cosmological radius can then be found
from Eq.~\eqref{rcofr+Lambdaagain}
to be given by 
\begin{equation}
\hskip-0.2cm
\frac{r_{\rm c1} }{R}=\frac{r_{\rm c2}}{R}=
1+
\left(\frac38\,
(1-\Lambda R^2)^2\right)^{\frac13}
\hskip-0.1cm,
\;\; (1
\hskip-0.07cm
-
\hskip-0.07cm
\Lambda R^2)
\hskip-0.07cm
\ll
\hskip-0.07cm
1\,.
\label{rcsolutionsLambdaR2<near1}
\end{equation}

One could work out in both regimes, i.e., $\Lambda R^2\ll1$ and
$(1-\Lambda R^2)\ll1$, the action $I$, the thermodynamic energy $E$, the
entropy $S$, and the heat capacity $C_A$, given through
Eqs.~\eqref{actionbh}-\eqref{heatcapacity}.  Apart from the entropy
expression $S=4\pi r_+^2$, valid for each of the two black hole
solutions, the calculation of the other quantities is not practical
and not be particularly illuminating.  However, an instance where all
quantities can be worked out, in particular the heat capacity $C_A$,
is the hight temperature limit to which we now turn.

\subsection{High temperature limit: Analytical solutions}

For the range of values of the cosmological constant considered
in this section, $0\leq\Lambda R^2<1$, one can find solutions in
the limit in which $RT$ goes to infinity, see
Eqs.~\eqref{tn} and \eqref{tbh2}. Since $R$ is the quantity
that we consider as the gauge,
$RT$ going to infinity is the same in this context as $T$ going to
infinity.
Let us then find explicit results by taking the limit
of high temperature. In this case the equations can be solved.

For a given $T$ there are two black hole solutions, the small black
hole solution $r_{+1}$ and the large black hole solution $r_{+2}$.
We set
the heat reservoir temperature 
$T$ fixed but very high, in the sense that $T\rightarrow
\infty$. From Eq.~\eqref{tn} there are two possibilities.  Either 
$T^{\rm H}_+\rightarrow \infty$ which corresponds to the small black
hole solution having a very small $r_{+1}$,
or $V(R)\rightarrow 0$ which corresponds to
the large black hole solution $r_{+2}$
approaching the reservoir radius. Let us work one at a time
for $T$ fixed and very high.

\vskip 0.3cm
\noindent  The first solution 
for a very high
heat reservoir temperature,
$T\to\infty$, is $r_+=r_{+1}\rightarrow 0$
with $T^{\rm H}_+\rightarrow \infty$.
It is clear from Eqs.~(\ref{tbh}) together with
Eq.~\eqref{tn}, or directly from Eq.~\eqref{tbh2},
that this requires
that the black hole
solution is of the form
$r_+=r_{+1}\rightarrow 0$. So, in this limit
one has
\begin{equation}
T^{\rm H}_{+1}= \frac{1}{4\pi r_{+1}}\,,
\label{TH+1}
\end{equation}
where the equality sign is valid
within the approximation taken.
From Eq.~\eqref{tbh2} one
finds the small black hole
solution $r_{+1}$ to be of the form 
\begin{equation}
\frac{r_{+1}}{R}= \frac{1}{4\pi RT\sqrt{1-\frac{ \Lambda R^2}{3}}}\,,
\label{r+1Tinfinite}
\end{equation}
where the equality sign is valid within the approximation
taken. 
The expression inside the square root of Eqs.~\eqref{r+1Tinfinite} is
clearly positive.
As a by-product, we also find
from Eq.~\eqref{V=0euc} that in this limit one has
$m_1=\frac{r_{+1}}{2}$.
The corresponding cosmological radius $r_{\rm c1}$ can then be found
directly from Eq.~\eqref{rcofr+Lambdaagain},
yielding a correspondingly far away 
value for $r_{\rm c1}$, see Eq.~\eqref{rcofr+Lambdaagain}.
One could work out in this order, i.e., $T\to\infty$,
the action $I$, the energy $E$, the
entropy $S$, and the heat capacity $C_A$, given
through Eqs.~\eqref{actionbh}-\eqref{heatcapacity}.
The most interesting quantity is
the heat capacity $C_A$,
which yields the criterion for thermodynamic stability,
indeed when $C_A<0$ the solution is thermodynamically
unstable, when $C_A\geq0$ the solution is thermodynamically
stable.
We thus find an explicit expression for $C_A$.
From Eq.~\eqref{heatcapacity}, i.e.,
$C_A=\left(\frac{dE}{dT}\right)_A$ or 
equivalently, 
$C_A=\left(\frac{dE}{dT}\right)_R$, 
we find from Eq.~\eqref{Ebh} that
$
C_A=\frac{1}{2\sqrt{V(R)}}\left(\frac{dr_{+1}}{dT}\right)_R$,
which upon using Eq.~\eqref{r+1Tinfinite} yields
\begin{equation}
C_{A_{+1}}=-\frac{1}{8\pi T^2\left(1-\frac{\Lambda R^2}{3}\right)}<0\,,
\end{equation}
so that $C_A$ for the small black hole
$r_{+1}$
is negative.
The small black hole $r_{+1}$
solution is thus
unstable. Note that actually, the black hole is surrounded by quantum
fields. We neglect their backreaction on the metric. However, if
$T^{\rm H}\rightarrow \infty$, the corresponding energy density
and other components of the stress-energy tensor diverge. To avoid
this, we restrict $r_{+1}$ in the the sense that is has
to be larger than the Planck length scale
$l_{\rm pl}$, i.e., $r_{+1}> l_{\rm pl}$.

\vskip 0.3cm
\noindent The second solution for a very high
heat reservoir temperature,
$T\to\infty$
has 
 $V(R)\rightarrow 0$.
It is clear from Eqs.~\eqref{tn} or \eqref{tbh2}
that the condition $V(R)\rightarrow 0$, implies,
in the case $0\leq\Lambda R^2<1$, that
$r_{+2}= R$ minus a small quantity.
Now, from Eq.~(\ref{tbh}) one has in
this limit 
\begin{equation}
T^{\rm H}_{+2}= \frac{1-\Lambda R^2}{4\pi R}\,,
\end{equation}
where the equality sign is valid within the approximation
taken. In first order, we can
perform a Taylor expansion, and write
$V(R)=\left(\frac{dV}{dr}\right)_{r_{+2}}
\left(R-r_{+2}\right)$ plus higher order terms.
Since $\left(\frac{dV}{dr}\right)_{r_{+2}}=
4\pi T^{\rm H}_{+2}$, one can write
$V(R)= 4\pi
T^{\rm H}_{+2}\left(R-r_{+2}\right)$.
Using Eq.~\eqref{tn}, or Eq.~\eqref{tbh2},
we have
\begin{equation}
\frac{r_{+2}}{R}= 1-\frac{1-\Lambda R^2}{(4\pi RT)^2}\,,
\label{r+2highT}
\end{equation}
where the equality is valid within the approximation taken.
As a by-product, we also
find from Eqs.~\eqref{V=0euc} and \eqref{r+2highT} that in this limit
one has
$m_2=\frac{R}{2}\left[1-\frac{\Lambda R^{2}}{3}-\frac{(1-\Lambda
R^2)^2}{(4\pi RT)^2}\right]$.
The corresponding cosmological radius $r_{\rm c2}$ can then be found
directly from Eq.~\eqref{rcofr+Lambdaagain}, we refrain from showing
the explicit formula here,
noting nevertheless
that $\Lambda R^2$ can be small of order of
zero in which case the cosmological horizon is very far away, or of
order one in which case the cosmological horizon is very near the
reservoir and the black hole horizon.
We could work out in this order, i.e., $T\to\infty$,
the action $I$, the energy $E$, the
entropy $S$, and the heat capacity $C_A$, given
through Eqs.~\eqref{actionbh}-\eqref{heatcapacity}.
Again, the most interesting one is 
the heat capacity $C_A$.
For the
heat capacity $C_A$, given in Eq.~\eqref{heatcapacity}, i.e.,
$C_A=\left(\frac{dE}{dT}\right)_A$, equivalently,
$C_A=\left(\frac{dE}{dT}\right)_R$, we find from Eq.~\eqref{Ebh} that
$C_A=\frac{1}{2\sqrt{V(R)}} \left(\frac{dm_2}{dT}\right)_R$, where it
was used the expression $V(R)=1-\frac{2m_2}{R} -\frac{\Lambda R^2}{3}$
given in Eq.~\eqref{meteuc}.  Thus, using the expression for $m_2$
just found above we have $C_A=\frac{1}{2\sqrt{V(R)}} \frac12\frac{(1-\Lambda
R^2)^2}
{16\pi^2 T^3 R}$ and since $\sqrt{V(R)}=\frac{1-\Lambda
R^2}{4\pi RT}$ it gives
\begin{equation}
C_{A_{+2}}=\frac{1-\Lambda R^2}{4\pi T^2} >0\,,
\label{CAr+2}
\end{equation}
so that $C_{A+2}$ is small and positive.  The large black hole $r_{+2}$
solution is thus stable.

\subsection{Thermodynamic phases and phase transitions between hot
Schwarzschild-de Sitter and hot de Sitter in the $\Lambda R^2<1$ case}

We now work out the thermodynamic phases and phase transitions
for the $\Lambda R^2<1$ case.  The
discussion is valid for the thermodynamically stable
black hole, the black hole $r_{+2}$, since the
unstable one $r_{+1}$ has at most a fleeting existence
and could not count for a phase.
From Eq.~\eqref{actionbh}
we get that the action $I$
for a hot Schwarzschild-de Sitter $r_{+2}$ phase
is 
$I_{\rm SdS}=\beta R\left(1-\sqrt{V(R)}\right) -\pi r_{+2}^2$,
where 
from Eq.~\eqref{meteucnewR} we have $V(R)=\left(1-\frac{r_{+2}}{R}\right)
\left(1-\frac{\Lambda R^2}{3}\left(1+
\frac{r_{+2}}{R}+\left(\frac{r_{+2}}{R}\right)^2\right)\right)$, with
$0\leq\Lambda R^2<1$ here.
The free energy
$F=\frac{I}\beta=IT$ for hot Schwarzschild-de Sitter is
then 
\begin{equation}
F_{\rm SdS}= R\left(1-\sqrt{V(R)}\right) -\pi T r_{+2}^2\,,
\quad\quad\quad 0\leq\Lambda R^2<1\,.
\label{FHSdSLR2<1}
\end{equation}
Another phase that might exist is hot de Sitter,
in which case $r_+=0$,
$V(R)=1-\frac{\Lambda R^2}{3}$
and the action is 
$I_{\rm HdS}=\beta R\left(1-\sqrt{1-\frac{\Lambda R^2}{3}}\right)$.
The free
energy is then
\begin{equation}
F_{\rm HdS}=\left(1-\sqrt{1-\frac{\Lambda R^2}{3}}\right)\,R\,,
\quad\quad\quad 0\leq\Lambda R^2<1\,.
\label{FHdSLR2<1}
\end{equation}
In the canonical ensemble,
for systems characterized by the size
and the temperature of the heat reservoir,
the phase that has lowest $F$
is the phase that dominates.
So, the hot Schwarzschild-de Sitter black hole
phase dominates
over hot de Sitter, or the two phases coexist equally,
when 
\begin{equation}
F_{\rm SdS}\leq F_{\rm HdS}\,,
\quad\quad\quad 0\leq\Lambda R^2<1\,,
\label{FHSdSdominatesSLR2<1}
\end{equation}
i.e., 
$\left(1-\sqrt{V(R)}\right)\,R -\pi T r_{+2}^2\leq
\left(1-\sqrt{1-\frac{\Lambda R^2}{3}}\right)\,R$,
i.e.,
\begin{equation}
RT\geq
\frac{\sqrt{1-\frac{\Lambda R^2}{3}}-\sqrt{V(R)}}
{\pi \frac{r_{+2}^2}{R^2}}\,,
\quad\quad\quad 0\leq\Lambda R^2<1\,.
\label{FHSdSdominates2SLR2<1}
\end{equation}
We see that Eq.~\eqref{FHSdSdominatesSLR2<1}
is an implicit equation,
because $r_{+2}=r_{+2}(R,T)$.
For each $\Lambda R^2$ in the interval above, and for each
$RT$ one gets an $r_{+2}$, which can then be
put into Eq.~\eqref{FHSdSdominates2SLR2<1}
to see whether 
the Schwarzschild-de Sitter phase dominates
over the de Sitter phase or not.
In the case it dominates then a black hole can
nucleate thermodynamically from hot
de Sitter space.

For $\Lambda R^2\ll 1$ we can find some interesting
numbers.
One finds equality between the actions, i.e., that
$F_{\rm SdS}=F_{\rm HdS}$, see
Eqs.~\eqref{FHSdSdominatesSLR2<1} and \eqref{FHSdSdominates2SLR2<1},
when $RT=(RT)_{\rm eq}$ with
\begin{equation}
(RT)_{\rm eq}=\frac{27}{32\pi}\left(
1-\frac{13}{486}\,\Lambda R^2\right)\,,
\quad\quad\quad
\Lambda R^2\ll 1\,,
\label{FHSdSdominatesSLR2<<1finaleq}
\end{equation}
valid in first order, as all equations in the discussion
below 
will be valid in this order.
It is of interest to put this value in
decimal notation, i.e., 
$(RT)_{\rm eq}=0.269\left(
1-0.027\Lambda R^2\right)$, approximately.
For $\frac{r_{+2\,\rm eq}}{R}$ one has in this case that
\begin{equation}
\frac{r_{+2\,\rm eq}}{R}=\frac89\left(
1+\frac{77}{729}\Lambda R^2
\right)\,,
\quad\quad\quad
\Lambda R^2\ll 1\,.
\label{r+RFHSdSdominatesSLR2<<1finaleq}
\end{equation}
In decimal notation, this can be put
as
$\frac{r_{+2\,\rm eq}}{R}=0.889\left(
1+0.106\Lambda R^2
\right)$,
approximately.
Thus,
$F_{\rm SdS}\leq F_{\rm HdS}$, see
Eqs.~\eqref{FHSdSdominatesSLR2<1} and
\eqref{FHSdSdominates2SLR2<1},
when
\begin{equation}
RT\geq (RT)_{\rm eq},,
\label{FHSdSdominatesSLR2<<1final}
\end{equation}
and 
when 
\begin{equation}
\frac{r_{+2}}{R}\geq \frac{r_{+2\,\rm eq}}{R}\,.
\label{r+RFHSdSdominatesSLR2<<1final}
\end{equation}
So, when the inequalities 
given in Eqs.~\eqref{FHSdSdominatesSLR2<<1final}
and \eqref{r+RFHSdSdominatesSLR2<<1final}
hold, then the black hole phase dominates,
but nevertheless
the hot de Sitter phase has some
probability of turning up. 

There is another radius of interest here,
which although not strictly
thermodynamic, it appears through dynamical
arguments, and is important in this discussion
of phases and phase transitions.
For matter or energy enclosed in a box,
which one can consider that it configures a star,
there is a mass, or energy,
above which the star cannot
support its self gravity and tends to collapse.
This is called the Buchdahl limit
which for spaces with positive cosmological constant
has been calculated in 
\cite{andreassonbohmer}.
Here one should envisage this limit as giving,
for a given $R$ fixed,  the mass
$m_{\rm Buch}$  above which the
energy within the system is so large that the
system collapses.
For a given $R$ and $\Lambda$, $m_{\rm Buch}$ is
$\frac{m_{\rm Buch}}{R}=\frac29
+\frac29\sqrt{1+3\Lambda R^2}
-\frac{\Lambda R^2}{3}$.
Since $m=\frac{r_+}{2}\left(1-\frac{\Lambda r_+^2}{3}\right)$,
see Eq.~\eqref{V=0euc} one has
the Buchdahl limit is
given by the equation
$\frac{r_{+\rm Buch}}{R}\left(1-
\left(\frac{r_{+\rm Buch}}{R}\right)^2
\frac{\Lambda R^2 }{3}
\right)=\frac49
+\frac49\sqrt{1+3\Lambda R^2}
-\frac{2\Lambda R^2}{3}
$,
which is a cubic equation
for $\frac{r_{+\rm Buch}}{R}$ 
that can
in principle be solved.
Given $\frac{r_{+\rm Buch}}{R}$
one can then work out what is the
temperature 
$(RT)_{\rm Buch}$ that yields the
related black hole with radius
$\frac{r_{+2}}{R}$.
Here we are dealing with small $\Lambda R^2$.
in this case one gets 
\begin{equation}
(RT)_{\rm Buch}=\frac{27}{32\pi}\left(
1+\frac{985}{486}\,\Lambda R^2\right)\,,
\quad\quad\quad
\Lambda R^2\ll 1\,.
\label{Tbuch}
\end{equation}
One further has
\begin{equation}
\frac{r_{+\rm Buch}}{R}=\frac89\left(
1+\frac{64}{81}\,\Lambda R^2\right)\,,
\quad\quad\quad
\Lambda R^2\ll 1\,.
\label{r+BuchdahlLR2<<1}
\end{equation}
These two values can be put
in decimal notation as
$(RT)_{\rm Buch}=0.269\left(
1+2.027\,\Lambda R^2\right)$
and 
$
\frac{r_{+\rm Buch}}{R}=0.889\left(
1+0.790\,\Lambda R_{\rm Buch}^2\right)$,
approximately.
There is collapse when for a
given $\Lambda$ and a given $R$ and $T$ one has 
\begin{equation}
RT\geq(RT)_{\rm Buch}\,.
\label{Tbuchnew}
\end{equation}
and 
\begin{equation}
\frac{r_{+2}}{R}
\geq
\frac{r_{+\rm Buch}}{R}
\label{r+BuchdahlLR2<<1condition}
\end{equation}
So, there is collapse if for a
given $\Lambda$, $R$ and $T$ Eqs.~\eqref{Tbuchnew}
and \eqref{r+BuchdahlLR2<<1condition}
are obeyed.
Now, interestingly enough, comparing
Eq.~\eqref{FHSdSdominatesSLR2<<1finaleq}
with 
\eqref{Tbuch}
and 
Eq.~\eqref{r+RFHSdSdominatesSLR2<<1finaleq}
with 
\eqref{r+BuchdahlLR2<<1}
we see that 
\begin{equation}
(RT)_{\rm Buch}>(RT)_{\rm eq}\,.
\label{RTcomparebuchnew}
\end{equation}
and 
\begin{equation}
\frac{r_{+\rm Buch}}{R}>
\frac{r_{+\,\rm eq}}{R}\,.
\label{comparisonr+2BuchdahlLR2<<1}
\end{equation}
Thus, one has
that for sufficiently high
temperatures the black hole is a dominant phase
but not the unique, hot de Sitter might pop up,
and for even higher temperatures then 
the black hole is the unique phase as the system
tends to collapse.
A comment is in order. The Buchdahl bound applies to a
self-gravitating mechanical system consisting of a ball of radius $R$
containing matter.  For a fixed $R$, the bound determines the maximum
value of the gravitational radius $r_{+2}$, i.e., the maximum mass or
energy within $R$, above which the system collapses. The system we are
working with is a thermodynamic system, with a boundary that has
radius $R$ and temperature $T$ fixed.  In the approximation we are
using, the system contains no matter with a black hole appearing as a
result of thermodynamically imposed data in the ensemble, not as a
result of a dynamic process. Nevertheless, one can think in going to
the next order of approximation, where now the system contains lumps of
energy or particles.  In this case, for fixed $R$, there is a maximum
value for the energy within $R$, above which the gravitational radius
$r_{+2}$ is higher than the value permitted by the Buchdahl bound, and
one can infer that the system must collapse. This reasoning is
plausible, however it comes from dynamical arguments, and as such is
outside the thermodynamic approach we are using.

So, we have the following picture for
fixed and tiny $\Lambda R^2$.
For $0\leq RT
<\frac{\sqrt{27}}{8\pi}
\left( 1-\frac{5}{54}\,\Lambda R^2\right)
$, see Eq.~\eqref{nosolutionslimit}
there is only hot de Sitter space.
For $\frac{\sqrt{27}}{8\pi}
\left( 1-\frac{5}{54}\,\Lambda R^2\right)
\leq RT <\frac{27}{32\pi}\left(
1-\frac{13}{486}\,\Lambda R^2\right)$,
see Eqs.~\eqref{solutionslimit} and
\eqref{FHSdSdominatesSLR2<<1finaleq}, hot de Sitter space is a phase
that dominates over the Schwarzschild-de Sitter black hole phase,
where the black hole is the large one, the small one being unstable is
of no interest in this context.
For
$\frac{27}{32\pi}\left(
1-\frac{13}{486}\,\Lambda R^2\right)= RT$, 
the Schwarzschild-de Sitter black hole
and the pure de Sitter phases coexist equally,
see Eq.~\eqref{FHSdSdominatesSLR2<<1finaleq}.
For
$\frac{27}{32\pi}\left(
1-\frac{13}{486}\,\Lambda R^2\right)< RT
<\frac{27}{32\pi}\left(
1+\frac{985}{486}\,\Lambda R^2\right)$,
the Schwarzschild-de Sitter black hole phase
dominates over the pure de Sitter phase.
For
$\frac{27}{32\pi}\left(
1+\frac{985}{486}\,\Lambda R^2\right)\leq RT
<\infty$,
see Eqs.~\eqref{Tbuch}
and \eqref{RTcomparebuchnew},
there is only 
the Schwarzschild-de Sitter black hole phase, 
the system suffers total gravitational collapse.
Note that in the phase transition
from hot de Sitter to
the Schwarzschild-de Sitter black hole phase
there is topology change, since here the Euclidean topology
of hot de Sitter is $S^1\times R^3$,
and the Euclidean topology of
the
Schwarzschild-de Sitter black hole is $R^2\times S^2$.

The case $\Lambda R^2=0$ is a
particular case of the $\Lambda R^2\ll1$
case just shown.
Nevertheless, there are interesting aspects
worth  mentioning.
In this case the
thermodynamic phases are 
hot flat space and the Schwarzschild
black hole.
The Schwarzschild
black hole
phase dominates, or the two phases coexist equally,
when
$F_{\rm Schw}\leq F_{\rm HFS}$. We can take it
directly to be 
$RT\geq
\frac{1-\sqrt{V(R)}}
{\pi \frac{r_{+2}^2}{R^2}}$, where
here $V(R)=1-\frac{r_{+2}}{R}$.
From the calculations above 
one finds that for
$0\leq RT
<\frac{\sqrt{27}}{8\pi}
$
there is only hot flat space.
For $\frac{\sqrt{27}}{8\pi}
\leq RT <\frac{27}{32\pi}$, hot de Sitter space is a phase that
dominates over the Schwarzschild-de Sitter black hole phase.  For
$\frac{27}{32\pi}= RT$, the Schwarzschild-de Sitter black hole and the
pure de Sitter phases coexist equally.
Now, note that 
$\frac{r_{+2}}{R}=\frac89$ 
is the radius where the two actions,
for hot flat and Schwarzschild spaces, have the
same value, which is zero in this case \cite{y86}.
But this value is in fact equal to 
 the Buchdahl
limiting radius in general relativity, as was found
for dimensions $d\geq4$ in
\cite{Andre:2020,Andre:2021,fernandeslemos}.
The fact that the thermodynamic existence
of a black hole phase and the
Buchdal limit coincide in this case
is an interesting and
unexpected property, and
it can help to compare both
processes,
thermodynamic an dynamic,
of forming a black hole.
Then, for slightly
higher temperatures, one can infer
that when the Schwarzschild black hole phase
dominates, it actually has sufficient energy to
collapse itself to a black hole, i.e.,
when the two phases, hot
flat space and Schwarzschild black hole,
start to coexist, the black hole phase actually
dominates completely, since the system has sufficient
energy to collapse to a black hole.
Thus, here, for
$\frac{27}{32\pi}< RT
<\infty$,
the Schwarzschild black hole phase
should be the only phase that exists, 
as the system must suffer total gravitational collapse.

For generic $\Lambda R^2$ in the range $\Lambda R^2<1$ a similar
analysis can be made, but we do not do it here.

\subsection{Comments on the $\Lambda R^2<1$ case}

We note that although the analysis made in
Eqs.~\eqref{nosolutionslimit}-\eqref{rcsolutionslimit} is valid for a
very small cosmological constant $\Lambda R^2$, and the analysis made
in
Eqs.~\eqref{solutionsLambdaR2<near1}-\eqref{rcsolutionsLambdaR2<near1}
is valid for a cosmological constant $\Lambda R^2$ near unity from
below, one can have a good idea of the behavior of the solutions as
$\Lambda R^2$ is increased from zero up to a value less than
unity. This is also done with the help with the results of the high
temperature limit Eqs.~\eqref{TH+1}-\eqref{CAr+2}.

For $\Lambda R^2=0$, i.e.,
for zero cosmological constant,
we recover York's results for a heat reservoir in a
Schwarzschild black hole space. In this case, when
$RT<\frac{\sqrt{27}}{8\pi}$ there are no
black hole solutions, only hot flat
space, and when $RT\geq\frac{\sqrt{27}}{8\pi}$ there are two
solutions, the small black hole $r_{+1}$ which is unstable,
and the large black hole
$r_{+2}$, which is stable, the two solutions are the same when
$RT=\frac{\sqrt{27}}{8\pi}$ with horizon radius
$\frac{r_{+1}}{R}=\frac{r_{+2}}{R}=\frac23$.
Still for $\Lambda R^2=0$,
when $RT$ is very high,
i.e., the temperature $T$ of the heat reservoir
is very high, then
$\frac{r_{+1}}{R}$ is small tending to zero and $\frac{r_{+2}}{R}$ is
large tending to one. The behavior
of the $\Lambda R^2=0$ can be heuristically
explained through the thermal wavelength of the radiation $\lambda$
and the size of the reservoir $R$.
For small $T$, one has that the corresponding thermal wavelength
$\lambda\equiv\frac{1}{T}$ is high relative to $R$.  In fact, $RT$
small means $\frac{R}{\lambda}$ small, i.e., $\frac{\lambda}{R}$ high,
so that the wavelength of the
thermal energy packets is stuck to the walls
of the reservoir, and the packets cannot collapse to form a black
hole.  For higher $T$, $\lambda$ is small
relative to $R$.  Thus, $RT$
large means $\frac{R}{\lambda}$ large, i.e., $\frac{\lambda}{R}$ low,
so that the wavelength of the thermal packets is sufficiently
small, the packets are free inside the
reservoir, and eventually collapse to form a black hole.

For $\Lambda R^2$ fixed and tiny we can spell the results as well.
Now the space of black hole solutions is over a two-dimensional
domain, specifically, $RT$ and $\Lambda R^2$.
When
$RT$ is less than some number, which itself is smaller than
$\frac{\sqrt{27}}{8\pi}$, then there are no solutions, only hot de
Sitter space, and when $RT$ is larger than this same number, which
itself is smaller than $\frac{\sqrt{27}}{8\pi}$, then there are two
solutions, the small black hole $r_{+1}$, which is unstable, and the
large black hole $r_{+2}$, which is stable.
When $RT$ is very high, i.e., when the temperature $T$ of the heat
reservoir is very high, then $\frac{r_{+1}}{R}$ is small tending to
zero and $\frac{r_{+2}}{R}$ is large tending to one.
The two solutions are the same solution when
$RT=\frac{\sqrt{27}}{8\pi} \left( 1-\frac{5}{54}\,\Lambda R^2\right)$,
$\Lambda R^2$ being the fixed and tiny value, with coincident horizon
radius given by $\frac{r_{+1}}{R}=\frac{r_{+2}}{R}=\frac23
\left(1+\frac{17}{81}\Lambda R^2\right)$.
Moreover, from the result that the coincident horizon radius has
the value just given, one can deduce that as one goes along
increasing $\Lambda$, specifically, increasing $\Lambda R^2$, for
fixed $RT$, the radii $\frac{r_{+1}}{R}$ and $\frac{r_{+2}}{R}$
increase.
This behavior for spaces with tiny $\Lambda R^2$ can be heuristically
explained through the thermal wavelength,
the size of the reservoir, and the cosmological length.
For small $T$, one has that the corresponding thermal wavelength
$\lambda\equiv\frac{1}{T}$ is high relative to $R$.  In fact, $RT$
small means $\frac{R}{\lambda}$ small, i.e., $\frac{\lambda}{R}$ high,
so that the wavelength of the thermal
energy packets is stuck to the walls
of the reservoir, and the packets cannot collapse to form a black
hole. 
But now, due to the new cosmological
length scale $\ell$ set by $\Lambda$,
$\ell=\frac{1}{\sqrt\Lambda}$, the space
inside the reservoir is more curved and, so to
speak,  a bit larger, and thus
a lower $T$, i.e., a higher $\lambda$, is
allowed so that they are free to collapse inside
the reservoir and form a black hole in this case.

For $\Lambda R^2$ fixed, not tiny and less than one, we can deduce
several results.  Solutions $\frac{r_{+1}}{R}$ and $\frac{r_{+2}}{R}$
start to appear at a certain $RT$ which is ever decreasing as $\Lambda
R^2$ is increasing, and when $\Lambda R^2$ is close to one then one
finds that $RT$ is close to and a bit higher than $\frac1{2\pi}$.
Thus, when $RT$ is small one can find black hole solutions for spaces
with cosmological constant near one, but there are no black holes for
spaces with any other lower cosmological constant.
Moreover, for $RT$ close to $\frac1{2\pi}$, then $\frac{r_{+1}}{R}$
and $\frac{r_{+2}}{R}$ are very near one and merge at $\frac1{2\pi}$.
In addition, for fixed $RT$ as one goes along increasing $\Lambda$,
i.e., increasing $\Lambda R^2$, then the radii $\frac{r_{+1}}{R}$ and
$\frac{r_{+2}}{R}$ increase. The
solution $\frac{r_{+1}}{R}$ is the small solution and the solution
$\frac{r_{+2}}{R}$ is the large solution that for $\Lambda R^2$ close
to one yields $\frac{r_{+2}}{R}$ close to one.
When $RT$ is very high, i.e., when the temperature $T$ of the heat
reservoir is very high, then $\frac{r_{+1}}{R}$ is small tending to
zero and $\frac{r_{+2}}{R}$ is large tending to one.
This behavior of $\Lambda R^2$ fixed, not tiny and less than one, can
be heuristically explained through the thermal wavelength,
the size of the reservoir, and the cosmological length.  As
the temperature gets lower and lower, the associated thermal
wavelength $\lambda=\frac1T$ gets higher and higher, and to have black
hole solutions the space needs to be more curved, and so larger, to
accommodate those wavelengths $\lambda$ and allow the corresponding
thermal energy packets
to collapse. For very low temperatures, in the limit
$RT\to\frac1{2\pi}$, i.e., $\frac{\lambda}{R}=2\pi$,
energy packets can
only build a black hole horizon for sufficiently high $\Lambda$, i.e.,
for $\Lambda R^2\to 1$.  This first black hole that appears at
$RT\to\frac1{2\pi}$ and $\Lambda R^2\to 1$ has large horizon radius
given by $\frac{r_{+1}}{R}=\frac{r_{+2}}{R}\to1$.
For
higher $T$, i.e., higher $RT$,
then the wavelength of the energy packets
is sufficiently small that allows
for black hole solutions $\frac{r_{+1}}{R}$ and $\frac{r_{+2}}{R}$.

To sum up, in the $\Lambda R^2<1$ case, for sufficiently high
temperatures there are two solutions, the
small mass branch with black
hole horizon radius $r_{+1}$
which is unstable, and the massive branch with
black hole horizon radius which is stable, a fact the holds for any
temperature $RT$ and any $\Lambda R^2<1$.  This is similar to what
happens in
the thermodynamics of
pure Schwarzschild, i.e., $\Lambda R^2=0$.  
As $\Lambda
R^2$ is increased from zero, black holes can form with less and less
temperatures $RT$, and for $\Lambda R^2$ near one, can form black
holes with the least temperature, namely, $RT$
tending to $\frac{1}{2\pi}$.

The case with $\Lambda R^2=1$ precisely has to be dealt as a separate case,
as an intermediate value case for $\Lambda R^2$. As we will show it
reserves interesting surprises.

\section{Thermodynamics of Schwarzschild-de Sitter black holes in the
canonical ensemble: Intermediate value of
the cosmological constant, $\Lambda R^2=1$, the Nariai
universe inside the heat reservoir}
\label{Rlamda2=1}

\subsection{Solutions and the Euclidean metric}

We treat here the intermediate positive cosmological constant, $\Lambda
R^2=1$, problem. Again, we put the boundary of a spherical cavity with
a black hole in a positive cosmological constant background inside a
heat reservoir, at some radius $R$, where it is also specified a fixed
temperature $T$, see Fig.~\ref{schwdesitterreservoir1} anew.  
The intermediate value
of the cosmological constant is precisely
\begin{equation}
\Lambda R^2=1\,.
\label{range=1}
\end{equation}
For this value of $\Lambda R^2$, and for fixed $RT$ one can
find solutions of Eq.~\eqref{tbh2} for black hole horizon radii $r_+$
analytically.  However, when $\Lambda R^2=1$ the problem has to be
treated with care.  There are still two solutions, $r_{+1}$ and
$r_{+2}$.

The solution $r_{+1}$, the small black hole
solution, can be taken
directly from
Eq.~\eqref{tbh2} putting $\Lambda R^2=1$
which is then 
$4\pi RT=
\frac{1}{ \frac{r_+}{R}}
\frac{
1-\left(\frac{r_+}{R}\right)^2}
{
\sqrt{1-\frac{r_+}{R}}\,
\sqrt{\left(1-\frac13\left(1+
\frac{r_+}{R}+\left(\frac{r_+}{R}\right)^2\right)\right)}
}$. This equation can be transformed into a
quartic equation in $\frac{r_+}{R}$
yielding then the solution $r_{+1}$, the small and
unstable solution.

The solution $r_{+2}$, the large black hole solution, needs special
attention.
One cannot simply put $\Lambda R^2=1$ into Eq.~\eqref{tbh2},
i.e.,
$4\pi RT= \frac{1}{ \frac{r_+}{R}} \frac{ 1-\Lambda
R^2\left(\frac{r_+}{R}\right)^2} { \sqrt{1-\frac{r_+}{R}}\,
\sqrt{\left(1-\frac{\Lambda R^2}{3}\left(1+
\frac{r_+}{R}+\left(\frac{r_+}{R}\right)^2\right)\right)} }$,
which gives directly the solution
$\frac{r_{+2}}{R}=1$ with $RT=\frac{1}{2\pi}$.  The correct way is to
take the limit $\Lambda R^2\to1$ and $\frac{r_{+2}}{R}\to1$.  Then,
$RT$ can have a broad range of values.  One then also finds that, for
$\frac{r_{+2}}{R}\to1$, the cosmological radius solution can be taken
from Eq.~\eqref{rcofr+Lambdaagain} to give $\frac{r_{\rm c2}}{R}\to1$.
This limit takes us to the Nariai universe, which we now
find.  The Nariai universe is to be seen as Schwarzschild-de Sitter
black hole with maximal mass, it is an extremal case.

The Nariai solution can be found from the Schwarzschild-de Sitter
solution in the limit that the two horizons $r_{+2}$ and $r_{\rm c2}$
coincide, see also Appendix~\ref{appendix:sdsnbasics}.  Here, we have
a heat reservoir at $R$ that acts as a reservoir for the inside
region, the region containing a black hole. This heat reservoir, at
$R$, is in between $r_{+2}$ and $r_{\rm c2}$, and thus the limit we
want to take is such that $r_{+2}$, $R$, and $r_{\rm c2}$ coincide,
see Appendix~\ref{nariailimitapp} for detail.  We drop the subscript
{\footnotesize 2} in the following analysis.
Now, if we do $\frac{r_+}{R}\to1$, then Eq.~\eqref{tn},
$T=\frac{T^{\rm H}_+}{\sqrt{V(R)}}$, together with
Eq.~\eqref{meteucnew}, $V(r)=\left(1-\frac{r_+}{r}\right)
\left(1-\frac{\Lambda r^2}{3}\left(1+ \frac{r_+}{r} +
\left(\frac{r_+}{r}\right)^2\right)\right)$, gives at face value that
the heat reservoir is at very high temperature $T$.  But, there is a
way to have $T$ finite with $\frac{r_+}{R}\to 1$.  From Eq.~\eqref{tn}
we see that if we do $\frac{r_+}{R}\to 1$ concomitantly with $T^{\rm
H}_+\to 0$ then $T$ is finite.  Since $T^{\rm H}_+=\frac{1}{4\pi
r_+}\left(1-\Lambda r_+^2\right) $, see Eq.~\eqref{tbh}, $T^{\rm
H}_+\to 0$ means $1-\Lambda r_+^2\to0$, but since $\frac{r_+}{R}\to 1$
this also means $1-\Lambda R^2\to0$.  In brief, in this limit we have
$\frac{r_+}{R}\to 1$ and $\Lambda R^2\to1$, both from below and both
of the same infinitesimal order.
Then, Eq.~\eqref{tn}, $T=\frac{T^{\rm
H}_+}{\sqrt{V(R)}}$, gives in this limit that
$T=\frac{1}{2\pi R}
\frac{ \left(1-\frac{r_+}{R}\right)+\left(1-\sqrt{\Lambda R^2}\right)}
{\sqrt{1-\frac{r_+}{R}}
\sqrt{\left(1-\frac{r_+}{R}\right)+2\left(1-\sqrt{\Lambda R^2}\right)}}$.
Since $1-\sqrt{\Lambda R^2}$ and $1-\frac{r_+}{R}$
are infinitesimal in this limit,
we see that $T$ is finite and can have a
range of values depending on the precise infinitesimal values of
$1-\sqrt{\Lambda R^2}$ and $1-\frac{r_+}{R}$.
One more thing. We have deduced that in this limit
$r_+\to R\to \frac{1}{\sqrt\Lambda}$, so that 
from
Eq.~\eqref{rcofr+Lambdaagain} one has
$r_{\rm c}\to \frac{1}{\sqrt\Lambda}$, and since
$\frac{1}{\sqrt\Lambda}\to R$, then $r_{\rm c}\to R$.
So, in the limit we have $r_+=R=r_{\rm c}$.
To see
that this is the Nariai limit with a reservoir $R$ in the middle, we
have to do some work on the original line element,
Eqs.~\eqref{met1euc+} and \eqref{meteuc}.  We
present the results below, see 
Appendix \ref{nariailimitapp} for a detailed derivation.

Let us start. 
The reservoir temperature $T$ is also
the local Tolman temperature $T$ at $R$
and has an associated expression given by
$
T=\frac{T^{\rm H}_+}{\sqrt{V(R)}}
$,
with $T^{\rm H}_+$ being the
tiny Hawking temperature
as we have just found.
Expanding the metric potential $V(r)$
of Eq.~\eqref{meteuc} near $r_+$ in a Taylor series
gives 
$
V(r)= 4\pi T^{\rm H}_+(r-r_+)-\frac{1}{R^2}\left(r-r_+\right)^2$,
plus higher order terms.
Make now the transformations $(t,r)\to(\bar t,
z)$ as
$
r-r_+=4\pi T^{\rm H}_+R^2
\sin^2\left(\frac12{\rm arcos}(\frac{{z}}{R})\right)$
and $t=\frac{\bar t}{2\pi T^{\rm H}_+R}$
with
$0\leq t\leq\frac{1}{T^{\rm H}_+}$
corresponding to
$0\leq {\bar t}\leq2\pi R$
and 
$r_+\leq r\leq R$ corresponding to $-R\leq z\leq Z$.
Then, since $V(r)$ is actually a $V(r,r_+)$, we have here
$V(r)=V(r,r_+)=V(r-r_+) =V(z)=(2\pi T^{\rm H}_+ R)^2\sin^2
\left({\rm arcos}(\frac{{ z}}{R})\right)
=(2\pi T^{\rm H}_+ R)^2\left(1-\frac{z^2}{R^2}\right)$.
From the original
Schwarzschild-de Sitter line element, Eqs.~\eqref{met1euc+} and
\eqref{meteuc}, together with Eq.~\eqref{V=0euc},
and dropping the bar in $\bar t$ which is now meaningless,
we obtain then the Nariai line element,
i.e.,
\begin{align}
ds^2=V(z)
\,dt^2+&\dfrac{dz^2}{V(z)}+ R^2\left(d\theta^2+\sin^2\theta\,d\phi^2
\right)\,,\nonumber\\
&0\leq t\leq2\pi R\,,\quad
-R<z<Z\,,
\label{nariai0newz}
\end{align}
where $Z$ is now the
heat reservoir boundary
in the $z$-direction, the 
other coordinates
are in the range
$0\leq \theta\leq
\pi$, $0\leq\phi<2\pi$, and
the metric potential $V(z)$ being given by
\begin{equation}
V(z)=1-\frac{z^2}{R^2} \,.
\label{Vnariai0newz}
\end{equation}
The line element given in Eqs.~\eqref{nariai0newz}
and \eqref{Vnariai0newz}
corresponds to the Nariai universe, which
can be seen to be decomposable into
a two-dimensional de Sitter
space times a sphere.
So, the ensemble with its boundary
data, $T$ and $R$, 
provide automatically the range of
coordinates of the solution.
Note also that
the range of values
for the heat reservoir boundary $Z$
is $-R\leq Z\leq R$.
From 
Eq.~\eqref{Vnariai0newz}
we see that there are two horizons,
one is $z_+=-R$, the other is
$z_{\rm c}=R$, 
but the subscripts now are
just names, since the two horizons are of the same type.
The topology of the Nariai universe
in Euclideanized form
is $R^2 \times S^2$ and
its boundary has $S^1 \times S^2$ topology,
where the $S^1$ subspace has proper length
$\frac1T$.

\begin{figure}[b]
\includegraphics*[scale=0.9]{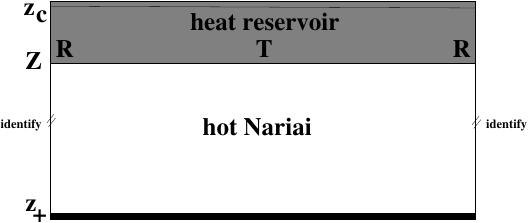}
\caption{
A drawing of a Nariai horizon $z_+$ within a heat reservoir at
temperature $T$, with cylindrical radius $R$, and situated at $Z$.
The cosmological horizon $z_{\rm c}$ is situated beyond the heat
reservoir.  The Euclideanized space and its boundary have $R^2 \times
S^2$ and $S^1 \times S^2$ topologies, respectively, where the $S^1$
subspace with proper length $\beta=\frac1T$ is not displayed.  See
text for more details.
}
\label{nariai1bh}
\end{figure}

Now, the temperature $T$ is 
\begin{equation}
T=\frac{T^{\rm H}_+}{\sqrt{V(Z)}}\,,
\label{tnnariai}
\end{equation}
with $T^{\rm H}_+$ being the Hawking temperature given by
$
T^{\rm H}_+=\frac{\kappa}{2\pi }
$,  and
$\kappa$ being
the surface gravity of the black hole
horizon.  For the metric
(\ref{nariai0newz}) one has $\kappa =\frac{1}{2}{V^{\prime}(z_+)}$,
and
so the Hawking temperature for the + horizon is
$T^{\rm H}_+=\frac{1}{4\pi}
\left(\frac{dV}{dz}\right)_{z_+}$. Using Eq.~(\ref{nariai0newz})
we get
\begin{equation}
T^{\rm H}_+=\frac{1}{2\pi R}.  \label{Tnariaibh2}
\end{equation}
As well, from Eq.~\eqref{Vnariai0newz} we have
that the metric potential at the heat reservoir boundary $Z$ is
\begin{equation}
V(Z)=1-\frac{Z^2}{R^2} \,.
\label{Vnariai0newZ}
\end{equation}
So, Eqs.~\eqref{tnnariai}-\eqref{Vnariai0newZ}
give
that the reservoir temperature
is given by
$T=\frac{1}{2\pi R \sqrt{1-\frac{Z^2}{R^2}}}$,
where $Z$ is the boundary
on  the $z$ coordinate, see
Fig.~\ref{nariai1bh} for
a representation of the Nariai universe in a heat reservoir.
For a given $T$ at the boundary, for the ensemble, there are two
solutions of this equation, in general. Namely, one solution is for
$Z$ between $-R$ and 0 and the other for $Z$ between 0 and $R$.  These
two solutions yield different physical situations of course, as the
reservoir boundary $Z$ is put in different positions relatively to
$z_+$. Note that in Schwarzschild-de Sitter space, the two solutions
were for the horizon $r_+$, one small $r_{+1}$ the other large
$r_{+2}$, both relative to the reservoir $R$.  Here, the two solutions
are not for the horizons but instead for the boundary $Z$, one $Z_1$,
the other $Z_2$, with
\begin{equation}
Z_1=-R\sqrt{1-\frac{1}{(2\pi RT)^2}} \,,
\label{Z1}
\end{equation}
and
\begin{equation}
Z_2=R\sqrt{1-\frac{1}{(2\pi RT)^2}} \,,
\label{Z2}
\end{equation}
now both relative to the horizon $z_+$.  Within the two choices for
the boundary, $Z_1$ or $Z_2$, we can pick the one we wish.  In
addition, since $z_+$ and $z_{\rm c}$ are indistinguishable, in the
sense they are of the same type, we can also interchange $z_+$ with
$z_{\rm c}$, in which case the situation would be the same.  The
boundary has always area radius $R$, which together with $T$, forms
the data for the canonical ensemble.

From Eqs.~\eqref{Z1} and \eqref{Z2} 
we see that for
\begin{equation}
RT<\frac1{2\pi} \,,
\label{nariainosolutionnew}
\end{equation}
there are no solutions for $Z_1$ or $Z_2$, so in this case the
boundary $Z$ does not exist, a reservoir does not exist, one cannot
define a temperature $T$ anywhere, and so there is no thermodynamic
Nariai solution.  Presumably one has simply hot de Sitter space inside
$R$. In decimal notation Eq.~\eqref{nariainosolutionnew} is
$RT<0.159$, approximately.

From Eqs.~\eqref{Z1} and \eqref{Z2} 
we see that for
\begin{equation}
RT\geq\frac1{2\pi} \,,
\label{nariaisolutionnew}
\end{equation}
there are two Nariai solutions, one with one horizon
$z_+=-R$ and boundary $Z_1$, the other with one horizon
$z_+=-R$ and
boundary $Z_2$, both boundaries can be picked up.
When the equality sign holds in Eq.~\eqref{nariaisolutionnew}
there is one solution
with $Z_1=Z_2=0$, so in this case the
boundary $Z$ pops up in the middle, 
at $Z=0$, and so $-R\leq z\leq0$.
In this case, 
The reservoir is at $Z=0$, has radius $R$,
and the horizon is at $z_+=-R$.

The generic Tolman temperature
formula in the Nariai space is
$
T(z)=\frac{1}{2\pi R \sqrt{1-\frac{z^2}{R^2}}}$
for $ -R\leq z<Z
\label{tempznariainew}$,
with
the reservoir temperature $T$ being expressed
as $T\equiv T(Z)$.
So, $T(z=-R)=\infty$ as expected since $z=-R$ is a
horizon, it is the horizon $z_+$.
Increasing $z$ from $-R$ one sees that $T(z)$ decreases and stops if one
picks $Z_1$, and if one picks $Z_2$
it decreases up to $z=0$, and then increases back up to $Z_2$.
In case $Z_2=R$, then 
$T(Z_2=R)=\infty$ as is expected since $z=R$
is a horizon, it is the horizon $z_{\rm c}$.
Clearly, the horizons are given by 
$z_+=-R$ and $z_{\rm c}=R$,
so they do not depend on $T$. The dependence on  $T$
is transferred to the boundary $Z$, so the structure has changed
from that of the Schwarzschild-de Sitter.

We now list the most relevant general formulas for the thermodynamics
of Nariai. These can be taken directly from the equations provided in
Sec.~\ref{generalresults}.  The action $I$ is now
\begin{equation}
I=\beta R -\pi R^2.
\label{actionbhnariai+}
\end{equation}
The action and the free energy are related by
$I=\beta F$, so
$
F=R -T\pi R^2\,.
\label{freeenergybhnariai+}$
Now, $F=E - TS$,
so the thermodynamic or quasilocal energy here
also the thermal energy at $R$ is 
\begin{equation}
E=R.
\label{Ebhnariai+}
\end{equation}
The entropy is
\begin{equation}
S=\pi R^2\,,
\label{entropybhnariai+}
\end{equation}
and is independent of $z_+$.  The heat capacity
$C_A=\left(\frac{dE}{dT}\right)_A$ is 
\begin{equation}
C_A=0\,,
\label{CAnariai}
\end{equation}
so there is neutral thermodynamic equilibrium in the Nariai universe.
That $C_A=0$ can be taken
directly from Eq.~\eqref{Ebhnariai+} which shows
that the energy $E$ has no dependence on the
temperature $T$.

\subsection{High-temperature limit}

For the value of the cosmological constant considered
in this section, $\Lambda R^2=1$, one can
work out the limit in which $RT$ goes to infinity, see
Eqs.~\eqref{Z1} and \eqref{Z2}. Since $R$ is the quantity
that we consider as the gauge,
$RT$ going to infinity is the same in this context as $T$ going to
infinity.
Let us then find explicit results by taking the limit
of high temperature.

For very high $T$, or very high $TR$,
one has from Eq.\eqref{Z1} the possibility
\begin{equation}
\frac{Z_1}{R}=- 1+\frac12\frac1{(2\pi RT)^2}\,,
\label{Z1highT}
\end{equation}
with 
$\frac12\frac1{(2\pi RT)^2}\ll1$.
In this case $Z_1$ is very near the horizon
$z_+=-R$ and one finds
that the space  in-between the horizon
and the reservoir is essentially a Rindler space.
To see this, note that in this limit $1-z^2\ll1$ for $0\leq z\leq
Z_1$. So with a 
$\bar z$ coordinate defined by $1-z^2={\bar z}^2$ one
obtains from
Eqs.~\eqref{nariai0newz} and \eqref{Vnariai0newz}
the line element
$ds^2={\bar z}^2 dt^2+ d{\bar z}^2
+ R^2\left(d\theta^2+\sin^2\theta\,d\phi^2\right)$, i.e.,
the Euclidean Rindler line element. 
For very high $T$, or very high $TR$,
one has from Eq.\eqref{Z2} the other possibility
\begin{equation}
\frac{Z_2}{R}=1-\frac12\frac1{(2\pi RT)^2}\,,
\label{Z2highT}
\end{equation}
with 
$\frac12\frac1{(2\pi RT)^2}\ll1$.
In this case 
the boundary $Z_2$ is very near the horizon $z_{\rm c}=R$,
but since $-R\leq z\leq Z_2<z_{\rm c}$
the space in-between the horizon
and the reservoir is not generically a Rindler
space, it approximates the Rindler line element 
only within the region
near $Z_2$.

\subsection{Thermodynamic phases and phase transitions between hot
Nariai and hot de Sitter in the $\Lambda R^2=1$ case}

We now work out thermodynamic phases and phase transitions
for the $\Lambda R^2=1$ case.  The
discussion is valid for the thermodynamically stable
black hole, the black hole $r_{+2}$, since the
unstable one $r_{+1}$ has at most a fleeting existence
and could not count for a phase.
Here the black hole $r_{+2}$ is in fact a Nariai universe.

From Eq.~\eqref{actionbhnariai+} we see
that the action $I$ for Nariai is $I_{\rm Nariai}=\beta R-\pi R^2$,
where we have used
that in Nariai one has $r_{+2}=R$.
So
the free energy 
$F=\frac{I}\beta=IT$
for a hot Nariai phase
is 
\begin{equation}
F_{\rm Nariai}= R-\pi R^2 T\,,
\quad\quad\quad
\Lambda R^2=1\,.
\label{FHSdSLR2Nariai=1}
\end{equation}
Another phase that might exist is hot de Sitter,
in which case $r_+=0$,
$V(R)=1-\frac{\Lambda R^2}{3}=\frac23$ as $\Lambda R^2=1$,
and the action is 
$I_{\rm HdS}=\beta R\left(1-\sqrt{\frac23}\right)$.
The free
energy is then
\begin{equation}
F_{\rm HdS}=\left(1-\sqrt{\frac23}\right)\,R\,,
\quad\quad\quad
\Lambda R^2=1\,.
\label{FHdSLR2=1}
\end{equation}
In the canonical ensemble the phase that has lowest $F$
is the phase that dominates.
So, the Nariai universe dominates
over hot de Sitter space, or the two phases coexist equally,
when 
\begin{equation}
F_{\rm Nariai}\leq F_{\rm HdS}\,,
\quad\quad\quad
\Lambda R^2=1\,.
\label{FHSdSdominatesSLR2=1}
\end{equation}
One finds equality between the two actions when 
$R-\pi R^2T=\left(1-\sqrt{\frac23}\right) R$,
i.e.,
when
\begin{equation}
(RT)_{\rm eq}=
\frac{\sqrt2}{\sqrt3\,\pi}\,,
\quad\quad\quad
\Lambda R^2=1\,.
\label{FHSdSdominatesSLR2=1equiv}
\end{equation}
In decimal notation, this is 
$(RT)_{\rm eq}= 0.260$, approximately.
So Nariai prevails over hot de Sitter,
or the two phases coexist equally,
when
\begin{equation}
RT\geq\frac{\sqrt2}{\sqrt3\,\pi}\,,
\quad\quad\quad
\Lambda R^2=1\,.
\label{FHSdSdominatesSLR2=1final}
\end{equation}

Recall
that for $RT<\frac1{2\pi}$,
there are no Nariai solutions only hot
de Sitter, see Eq.~\eqref{nariainosolutionnew},
and for $RT\geq\frac1{2\pi}$, see Eq.~\eqref{nariaisolutionnew},
two possible cases pop up,
the unstable black hole which is of no interest here, and
the Nariai universe which
is neutrally stable and of interest here.  So,
for $\Lambda R^2=1$ we have the following picture.
For 
$0\leq RT<\frac1{2\pi}$
hot
de Sitter is mandatory.
For $\frac1{2\pi}\leq RT<\frac{\sqrt2}{\sqrt3\,\pi}$
hot de
Sitter prevails as a thermodynamic phase over Nariai,
so that if the phase is
a Nariai one, it will probably
transition to a hot the Sitter phase.
For $\frac{\sqrt2}{\sqrt3\,\pi}=RT$
hot
de Sitter and Nariai coexist as thermodynamic phases.
For $\frac{\sqrt2}{\sqrt3\,\pi}
\leq RT<\infty$
Nariai prevails as a
thermodynamic phase over hot de Sitter.
Note that in the phase transition
from hot de Sitter to hot Nariai or from hot Nariai to hot de Sitter
there is topology change, since the Euclidean topology
of hot de Sitter is $S^1\times R^3$,
and the Euclidean topology of Nariai is $R^2\times S^2$.

\subsection{Comments on the $\Lambda R^2=1$ case}

It is really interesting that the resulting metric inside the heat
reservoir is described by the Nariai metric.  The procedure of
obtaining it in our context is completely different from the usual
procedure.  The heat reservoir radius $R$ and the temperature $T$ play
a crucial role here, and so the limit to the the Nariai
universe is naturally related to thermodynamics. As we have just seen,
the Nariai solution is of utmost importance in any analysis of the
Schwarzschild-de Sitter space in the canonical ensemble.

A feature of great importance in the overall picture is the minimum
temperature $T$, i.e., $RT$, of the ensemble above which there are
Nariai solutions for a given $\Lambda$, i.e., $\Lambda R^2$.
For
$RT<\frac1{2\pi}$ there are no black hole solutions whatsoever for any
$\Lambda R^2$, specifically, there are no small
black hole solutions with horizon radius
$\frac{r_{+1}}{R}$ neither
Nariai universes with $\frac{r_{+2}}{R}=1$.
Indeed, from the equations
found above for the heat reservoir boundary of the
Nariai universe, namely, 
$Z_1=-R\sqrt{1-\frac{1}{(2\pi RT)^2}}$
and
$Z_2=R\sqrt{1-\frac{1}{(2\pi RT)^2}}$,
we see that for
$
RT<\frac1{2\pi}$,
there are no solutions for $Z_1$ or $Z_2$. So in this case the
boundary $Z$ does not exist, a reservoir does not exist, one cannot
define a temperature $T$ anywhere, and so there is no thermodynamic
Nariai solution.  Presumably one has simply hot de Sitter space inside
$R$.  The reason is clear if one thinks in thermal wavelengths.  For
very small temperatures, the associated thermal wavelength is very
long and there is no boundary $Z$ that can accommodate such
corresponding thermal energy packets. 
For $RT=\frac1{2\pi}$ there is one solution only, it
has $\Lambda R^2=1$ precisely. There are no
solutions at this temperature
of any other $\Lambda R^2$.  So, a
Nariai solution is the first solution to pop up as one increases the
temperature from absolute zero.  As one increases $RT$ above
$\frac1{2\pi}$, solutions with $\Lambda R^2$ different from one start
to appear, first for $\Lambda R^2$ near one, than for $\Lambda R^2$
far from one as $RT$ is further increased as we have
discussed in the previous section for
$\Lambda R^2<1$.

The high temperature limit for the Nariai universe, i.e., for $\Lambda
R^2=1$, connects with the cases $\Lambda R^2<1$.  Indeed, when
$\Lambda R^2<1$, for $RT$ very large, there are two solutions, the
very small one $\frac{r_{+1}}{R}$ tending to zero, and the very large
one $\frac{r_{+2}}{R}$ tending to one. When $\Lambda R^2=1$ only the
very large one, $\frac{r_{+2}}{R}$, satisfies the condition
$\frac{r_{+2}}{R}=1$, necessary to get a Nariai universe.

It remains to be found the spectrum of solutions for $\Lambda
R^2>1$. We turn to this problem now, and it happens that there are
unexpected results.

\section{Thermodynamics of Schwarzschild-de Sitter black holes in the
canonical ensemble: Large values of the cosmological constant,
$\Lambda R^2>1$}
\label{Rlamda2>1}

\subsection{Solutions}

We treat here the large positive cosmological constant, $\Lambda
R^2>1$, problem. Again, we put the boundary of a spherical cavity with
a black hole in a positive cosmological constant background inside a
heat reservoir, at some radius $R$, where it is also specified a fixed
temperature $T$, see Fig.~\ref{schwdesitterreservoir1} anew.  Large
positive cosmological constant means exactly in this context that
\begin{equation}
1<\Lambda R^2\leq3\,.
\label{range>1}
\end{equation}
Within this range, for fixed $RT$ and generic $\Lambda R^2$, we can
draw the result that looking at Eq.~\eqref{tbh2} we can ascertain with
surprise that for $\Lambda R^2>1$ there are solutions with
$\frac{r_+}{R}<1$.  Nonetheless, it is hard to find analytic solutions of
Eq.~\eqref{tbh2} for black hole horizon radii $r_+$.
However, for $\Lambda R^2$ very near one from above, one can make some
progress.

For $\Lambda R^2$ a tiny bit larger than one, i.e., for 
$(\Lambda R^2-1)\ll 1$,
there are no black hole solutions 
for
\begin{equation}
\hskip -0.2cm
R T<\frac{1}{2\pi}
\left( 1+
\left(\frac{3}{8}\,\left(\Lambda R^2
\hskip -0.07cm
-
\hskip -0.07cm
1\right)\right)^\frac13\right),
\;\;
(\Lambda R^2
\hskip -0.07cm
-
\hskip -0.07cm
1)\ll 1\,,
\label{solutionslimit<LambdaR2>1}
\end{equation}
only hot de-Sitter space, valid in this order
of approximation, as all equations in this context below
are valid to this order.
Still for small $\Lambda R^2 -1$, i.e., for 
$(\Lambda R^2-1)\ll 1$, there
are two black hole solutions 
for
\begin{equation}
R T\geq\frac{1}{2\pi}
\left( 1+
\left(\frac{3}{8}\,\left(\Lambda R^2
\hskip -0.07cm
-
\hskip -0.07cm
1\right)\right)^\frac13\right),
\;\;
(\Lambda R^2
\hskip -0.07cm
-
\hskip -0.07cm
1)\ll 1\,,
\label{solutionslimit>LambdaR2>1}
\end{equation}
One of the two solutions is the small black
hole $r_{+1}(R,\Lambda,T)$,
and the other solution is the large black hole
$r_{+2}(R,\Lambda,T)$.
The plus sign inside the parenthesis in
Eq.~\eqref{solutionslimit>LambdaR2>1}
is what one expects really.
The two solutions merge
into one sole solution when 
the equality sign in
Eq.~\eqref{solutionslimit}
holds.
In this case the coincident
double solution has horizon radius
given by
\begin{equation}
\hskip -0.2cm
\frac{r_{+1}}{R}=\frac{r_{+2}}{R}= 1-\left(\frac38\,
(\Lambda R^2-1)^2\right)^{\frac13}\hskip -0.2cm,
\;\;
(\Lambda R^2
\hskip -0.07cm
-
\hskip -0.07cm
1)\ll 1\,.
\label{r+solutionslimitLambdaR2>1}
\end{equation}
The corresponding
cosmological radius can then be found
from Eq.~\eqref{rcofr+Lambdaagain}
to be given by 
\begin{equation}
\hskip -0.2cm
\frac{r_{\rm c1} }{R}=\frac{r_{\rm c2}}{R}=
1+
\left(\frac38\,
(\Lambda R^2-1)^2\right)^{\frac13}\hskip -0.2cm,
\;\;
(\Lambda R^2
\hskip -0.07cm
-
\hskip -0.07cm
1)\ll 1\,.
\label{rcsolutionslimitLambdaR2>1}
\end{equation}

One can work out in this order in $\Lambda R^2-1$
the action $I$, the energy $E$, the
entropy $S$, and the heat capacity $C_A$, given
through Eqs.~\eqref{actionbh}-\eqref{heatcapacity}.
Apart from the entropy $S=4\pi r_+^2$
for each of the two black hole solutions,
the other quantities would not be
particularly illuminating.
An instance where these quantities
can be worked out, in
particular the heat capacity $C_A$, is
the hight temperature limit to
which we now turn.

\subsection{High temperature limit: Analytical solution}

For the range of high values of the cosmological constant considered
in this section, $1<\Lambda R^2\leq3$, one can find solutions in
the limit in which $RT$ goes to infinity, see
Eqs.~\eqref{tn} and \eqref{tbh2}. Since $R$ is the quantity that we
consider as the gauge, $RT$ going to infinity is the same in this
context as $T$ going to infinity.  Let us then find explicit
results by taking the limit of high temperature. In this case the
equations can be solved.

For a given reservoir temperature $T$ there are two black hole
solutions, the small black hole solution $r_{+1}$ and the large black
hole solution $r_{+2}$.  We set $T$ fixed but very high, in the sense
$T\rightarrow \infty$. From Eq.~\eqref{tn} there are two
possibilities.  Either $T^{\rm H}_+\rightarrow \infty$ which
corresponds to the small black hole solution having a very small
$r_{+1}$, or $V(R)\rightarrow 0$ which corresponds to the large black
hole solution $r_{+2}$ but now not approaching the reservoir
radius in this range
of $\Lambda R^2$.
Let us work one solution at a time for a fixed very high value of $T$,
$T\rightarrow \infty$. Here we present the expressions at
zeroth order, not displaying the corrections in $\frac1T$.

\vskip 0.3cm
\noindent
The first solution for a very high heat reservoir temperature,
$T\to\infty$, is $r_+=r_{+1}\rightarrow 0$ with $T^{\rm
H}_+\rightarrow \infty$.  It is clear from Eqs.~(\ref{tbh}), together
with Eqs.~\eqref{tn} and \eqref{tbh2}, that this requires that the
black hole solution is of the form $r_+=r_{+1}\to 0$. So, in this
limit one has again $T^{\rm H}_{+1}= \frac{1}{4\pi r_{+1}}$, and then
from Eq.~\eqref{tbh2} one finds the small black hole $r_{+1}$ solution
to be of the form $\frac{r_{+1}}{R}= \frac{1}{4\pi RT\sqrt{1-\frac{
\Lambda R^2}{3}}}$.  The expression inside the square root is clearly
positive.  So, in the $T\to\infty$ limit we have,
\begin{equation}
\frac{r_{+1}}{R}= 0\,,
\label{r+1TinfiniteLambdaR2>1}
\end{equation}
plus higher order corrections.  As a by-product, we also find from
Eq.~\eqref{V=0euc} that in this limit one has $m_1=\frac{r_{+1}}{2}$.
For the heat capacity $C_A$, given in Eq.~\eqref{heatcapacity}, i.e.,
$C_A=\left(\frac{dE}{dT}\right)_A$, equivalently,
$C_A=\left(\frac{dE}{dT}\right)_R$, we find from Eq.~\eqref{Ebh} that
$ C_{A_{+1}}=\frac{1}{2\sqrt{V(R)}}\left(\frac{dr_{+1}}{dT}\right)_R$,
which upon using Eq.~\eqref{r+1Tinfinite} yields
$C_{A_{+1}}=-\frac{1}{8\pi T^2\left(1-\frac{\Lambda R^2}{3}\right)}
\leq0$, so in the limit
\begin{equation}
C_{A_{+1}}=0_-\,,
\end{equation}
here $0_-$ means that $C_{A_{+1}}$ tends to zero from negative values,
so that $C_{A_{+1}}$ is nonpositive.  Having negative heat capacity,
the small black hole $r_{+1}$ solution is thus unstable.

\vskip 0.3cm
\noindent The second solution
for  a very high heat reservoir temperature,
$T\to\infty$,
is some
$r_{+2}$
but now not necessarily near $R$.
Indeed, from Eq.~\eqref{tbh2},
i.e.,
$4\pi RT=
\frac{1}{ \frac{r_+}{R}}
\frac{
1-\Lambda R^2\left(\frac{r_+}{R}\right)^2}
{
\sqrt{1-\frac{r_+}{R}}\,
\sqrt{\left(1-\frac{\Lambda R^2}{3}\left(1+
\frac{r_+}{R}+\left(\frac{r_+}{R}\right)^2\right)\right)}
}$,
one sees
that, when $\Lambda R^2>1$, $RT\to\infty$
for 
$1-\frac{\Lambda R^2}{3}\left(1+
\frac{r_{+2}}{R}+\left(\frac{r_{+2}}{R}\right)^2\right)=0$. This is
a quadratic for $\frac{r_{+2}}{R}$ and it has as
solution
\begin{equation}
\frac{r_{+2}}{R}=\frac12\left(
\sqrt{
\frac{3}{\Lambda R^2}
}
\sqrt{4-\Lambda R^2}-1
\right)\,.
\label{r+2overRLambdaR2>1}
\end{equation}
This is the curve traced by $\frac{r_{+2}}{R}$
as a function of  $\Lambda R^2$ when $RT=\infty$.
So, from Eq.~\eqref{r+2overRLambdaR2>1} we find that
when $\Lambda R^2=1$ one has $\frac{r_{+2}}{R}=1$
as it should, and 
when $\Lambda R^2=3$ one has $\frac{r_{+2}}{R}=0$.
In this latter case, the solution $\frac{r_{+2}}{R}$
joins the small black hole solution  $\frac{r_{+1}}{R}=0$,
so that $\Lambda R^2=3$
is the turning point of $RT=\infty$,
$\frac{r_{+1}}{R}=\frac{r_{+2}}{R}=0$.
For the
heat capacity $C_A$, given in Eq.~\eqref{heatcapacity}, i.e.,
$C_A=\left(\frac{dE}{dT}\right)_A$, equivalently,
$C_A=\left(\frac{dE}{dT}\right)_R$, we
find that
$C_{A_{+2}}=
\frac{\sqrt{\Lambda R^2}}{8\pi T^2}
\left( 2+ \sqrt{\Lambda R^2}-\right.$ $\left.\sqrt{12-3\Lambda R^2}
\right)
\left(2-\sqrt{\Lambda R^2}+
\sqrt{12-3\Lambda R^2}\right)
\geq0$,
which after reworking
can also be written as
$C_{A_{+2}}=
\frac{\sqrt{\Lambda R^2}}{4\pi T^2}
\left( \Lambda R^2
+\sqrt{\Lambda R^2}\sqrt{12-3\Lambda R^2}
-4\right)
\geq0$, so in the infinite temperature
limit
\begin{equation}
C_{A_{+2}}=0_+\,,
\label{CA+2again}
\end{equation}
where $0_+$
means that $C_{A_{+2}}$ tends to zero from positive values, 
so that $C_{A_{+2}}$ is essentially positive, and
the large black hole $r_{+2}$
solution is stable.
For $\Lambda R^2=3$ the heat capacity is
$C_{A_{+2}}=\frac{\sqrt{3}}{\pi T^2}$,
and in the infinite temperature
limit one recovers Eq.~\eqref{CA+2again}.
The cosmological radius can also be found 
from Eq.~\eqref{r+2overRLambdaR2>1}, yielding
for any $\Lambda R^2$ in the range in question that
\begin{equation}
\frac{r_{\rm c2}}{R}=1\,,
\label{rc2overRLambdaR2>1}
\end{equation}
where Eq.~\eqref{rcofr+Lambdaagain}
has been used.
Note from the Tolman
formula that the whole region between
$r_{+2}$ and $R$ is at infinite
temperature.
Indeed, $T(r)=\frac{T}{V(R)}$, for $\frac{r_{+2}}{R}
\leq \frac{r}{R}\leq 1$.
Since $T=\infty$ and $V(R)$ is finite
one has that $T(r)$ is infinite in the region.
The 
temperature at $r$ normalized to the
heat reservoir temperature $T$ is $\frac{T(r)}{T}=
\frac1{V(R)}$ which is finite everywhere
except at $r_{+2}$ where it is infinite.
So the radius $r_{+2}$ yields a
doubly infinite temperature.
For
$\Lambda R^2=3$, the space is pure de Sitter
at infinite temperature with $\Lambda R^2=3$ from
the reservoir at $R$ up to the center
where there is a singular black hole
with horizon radius given by 
$\frac{r_{+2}}{R}=0$.


\subsection{Thermodynamic phases and phase transitions between hot
Schwarzschild-de Sitter and hot de Sitter in the $\Lambda R^2>1$ case}

We now mention thermodynamic phases and phase transitions
for the $\Lambda R^2>1$ case.  The
discussion is valid for the thermodynamically stable
black hole, the black hole $r_{+2}$, since the
unstable one $r_{+1}$ has at most a brief existence
that could not count as a phase.
From Eq.~\eqref{actionbh}
we get that the action $I$
for a hot Schwarzschild-de Sitter $r_{+2}$  phase
is 
$I_{\rm SdS}=\beta R\left(1-\sqrt{V(R)}\right) -\pi  r_{+2}^2$,
where 
from Eq.~\eqref{meteucnewR} we have $V(R)=\left(1-\frac{r_{+2}}{R}\right)
\left(1-\frac{\Lambda R^2}{3}\left(1+
\frac{r_{+2}}{R}+\left(\frac{r_{+2}}{R}\right)^2\right)\right)$, with
$1<\Lambda R^2\leq3$ here.
The free energy
$F=\frac{I}\beta=IT$ for hot Schwarzschild-de Sitter is
then 
\begin{equation}
F_{\rm SdS}= \left(1-\sqrt{V(R)}\right)\,R -\pi T r_{+2}^2\,,
\quad\quad\quad 1<\Lambda R^2\leq3\,,
\label{FHSdSLR2>1}
\end{equation}
Another phase that might exist is hot de Sitter,
in which case $r_+=0$,
$V(R)=1-\frac{\Lambda R^2}{3}$
and the action is 
$I_{\rm HdS}=\beta R\left(1-\sqrt{1-\frac{\Lambda R^2}{3}}\right)$.
The free
energy is then
\begin{equation}
F_{\rm HdS}=\left(1-\sqrt{1-\frac{\Lambda R^2}{3}}\right)\,R\,,
\quad\quad\quad 1<\Lambda R^2\leq3\,.
\label{FHdSLR2>1}
\end{equation}
In the canonical ensemble the phase that has lowest $F$
is the phase that dominates.
So, the hot Schwarzschild-de Sitter black hole dominates
over hot de Sitter, or the two phases coexist equally,
when 
\begin{equation}
F_{\rm SdS}\leq F_{\rm HdS}\,,
\quad\quad\quad 1<\Lambda R^2\leq3\,,
\label{FHSdSdominatesSLR2>1}
\end{equation}
i.e., 
$\left(1-\sqrt{V(R)}\right)\,R -\pi T r_{+2}^2\leq
\left(1-\sqrt{1-\frac{\Lambda R^2}{3}}\right)\,R$,
i.e.,
\begin{equation}
RT\geq
\frac{\sqrt{1-\frac{\Lambda R^2}{3}}-\sqrt{V(R)}}
{\pi \frac{r_+^2}{R^2}}\,,
\quad\quad\quad 1<\Lambda R^2\leq3\,.
\label{FHSdSdominates2SLR2>1}
\end{equation}
As before,
Eq.~\eqref{FHSdSdominatesSLR2<1}
is an implicit equation
because $r_{+2}=r_{+2}(R,T)$.
For each $\Lambda R^2$ in the interval above, and for each
$RT$ one gets an $r_{+2}$, which can then be
put into the expression just found
to see whether 
the Schwarzschild-de Sitter phase dominates
over the de Sitter phase or not.
In the case it dominates than a black hole can
nucleate thermodynamically from hot
de Sitter space.

Here, we just comment on the
limiting case, $\Lambda R^2=3$, which can be done directly.
In this case
the only solution is $T=\infty$,
with $r_+=0$, i.e., de Sitter space with
a singularity at the center. 
Thus, as $T\to\infty$,  one has
$F_{\rm SdS}\to R$ from below.
The de Sitter free energy for
$\Lambda R^2=3$ is 
$F_{\rm SdS}= R$, exactly.
Although both free energies are equal to $R$ in the limit,
one free energy tends to zero, the other is identically
zero.
So for $\Lambda R^2\to3$, $F_{\rm SdS}\leq F_{\rm HdS}$
and one can say that singular Schwarzschild
de Sitter phase
prevails.
In the limit both phases have the same free
energy and coexist in the ensemble in equal quantities. 
The difference between the two
phases is that one has a singular black
hole at the
center, i.e., a naked massless
singularity, and the other does not.
Note that in the phase transition
from hot de Sitter to
the Schwarzschild-de Sitter black hole phase,
and vice-versa, 
there is topology change, since the Euclidean topology
of hot de Sitter is $S^1\times R^3$,
and the Euclidean topology of
the
Schwarzschild-de Sitter black hole is $R^2\times S^2$.

\subsection{Comments on the $\Lambda R^2>1$ case}

We note that although the analysis made for
Eqs.~\eqref{solutionslimit<LambdaR2>1}-\eqref{rcsolutionslimitLambdaR2>1}
is valid for a very small value of $\Lambda R^2-1$, one can have a
good idea of the behavior of the solutions as $\Lambda R^2$ is
increased up to the value 3. This is also done with the help
of the results of the high temperature limit
Eqs.~\eqref{r+1TinfiniteLambdaR2>1}-\eqref{rc2overRLambdaR2>1}.

For $\Lambda R^2$ near one from above, we recover the previous result
that for $RT<\frac{1}{2\pi}$ there are no solutions.  So
$RT=\frac{1}{2\pi}$ is the minimum temperature to have solutions at
all.  As in the case $\Lambda R^2<1$ which has solutions for $RT$
higher than $\frac{1}{2\pi}$, the case $\Lambda R^2>1$ also has
solutions for $RT$ higher than $\frac{1}{2\pi}$.  For fixed $RT$
greater than the minimum value, i.e., $RT\geq\frac{1}{2\pi}$, two
solutions exist up to a maximum value of $\Lambda R^2$, where at this
value the two solutions merge, $r_{+1}=r_{+2}$.

For $RT=\infty$ one could find the exact dependence of
$\frac{r_{+2}}{R}$ in terms of $\Lambda R^2$.
Here, the maximum value is
$\Lambda R^2=3$,
where the solutions merge with horizon radius
$r_{+1}=r_{+2}=0$.
This behavior for
$RT=\infty$ can be
heuristically explained through the thermal wavelength,
the size of the reservoir, and the cosmological length.  $T$
going to infinity means that the associated thermal wavelength is zero
and so the small black hole as radius $r_{+1}=0$,
i.e., a black hole solution of zero size can be formed.
The understanding of the large black hole
with horizon radius $r_{+2}$ very small when compared to $R$,
indeed 
tending to zero, is here not so straightforward.
The cosmological scale $\ell\equiv\frac1{\sqrt\Lambda}$ has now
the minimum possible value, $\ell=\frac{R}{\sqrt3}$. $T$
going to infinity implies
that the associated thermal wavelength is vanishingly small,
and the result implies that this wavelength
only fits within the scale allowed by $\ell$ so that 
$r_{+2}$ is also vanishingly small.

\section{Diagrams for the Schwarzschild-de Sitter and Nariai
thermodynamic solutions and analysis}
\label{diagramsandanalysis}

\subsection{Diagrams for the Schwarzschild-de Sitter and Nariai
thermodynamic solutions}

\subsubsection{Preliminaries}

We now draw some diagrams that help in the understanding of the
thermodynamic solution
of the Schwarzschild-de Sitter and Nariai horizons
in a cavity.

There are two different sets of diagrams.  The first set contains six
diagrams. In each diagram, it is plotted, for a fixed value of $4\pi RT$,
the values of $\frac{r_+}{R}$ that are solution of the thermodynamic
problem, as a function of $\sqrt{\Lambda R^2}$,
see Fig.~\ref{figureyofx}.   The second set
contains also six diagrams. In each diagram, it is plotted, for a
fixed value of $\Lambda R^2$, the values of $\frac{r_+}{R}$ that are
solution of the thermodynamic problem, as a function of $4\pi RT$, see
Fig.~\ref{figureyofw}.
We use the variable
$4\pi RT$ rather than $RT$ because it is in a sense more natural
in this analysis.


\subsubsection{Diagrams with $RT$ fixed}

\begin{widetext}

\begin{figure}[h]
\centering
  \subfloat[\label{twohorizonswithRT=1/2pi}]
  {\includegraphics[width=.31\textwidth]{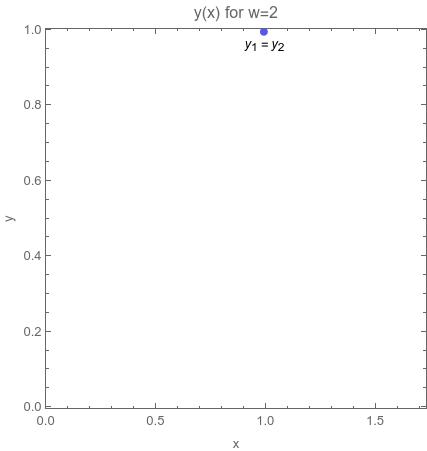}} 
  \hspace{5mm}
  \subfloat[\label{twohorizonswithRT=0.19}]
  {\includegraphics[width=.31\textwidth]{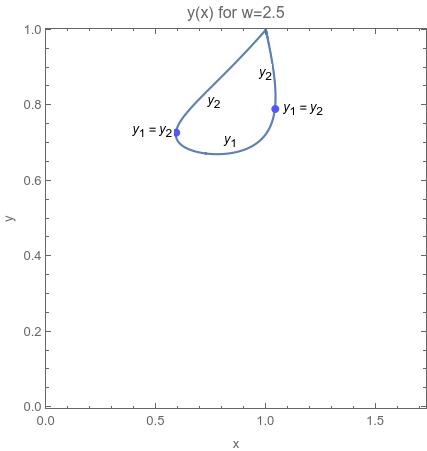}}
  \hspace{5mm}
  \subfloat[\label{twohorizonswithRT=srqt27/8pi}]
  {\includegraphics[width=.31\textwidth]{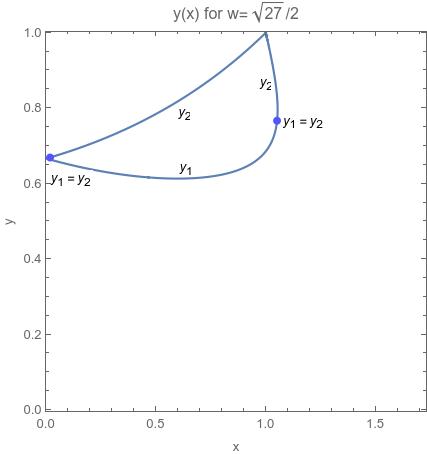}}
  \hspace{5mm}
  \subfloat[\label{twohorizonswithRT=0.23}]
  {\includegraphics[width=.31\textwidth]{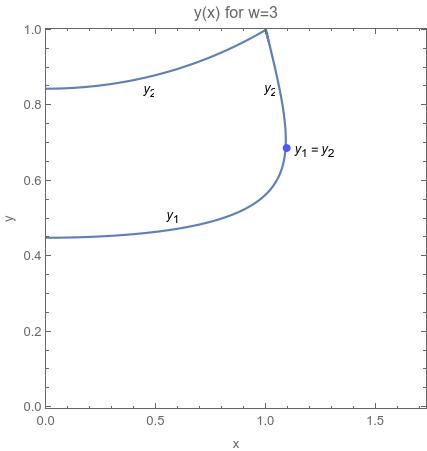}}
  \hspace{5mm}
  \subfloat[\label{twohorizonswithRT=10}]
  {\includegraphics[width=.31\textwidth]{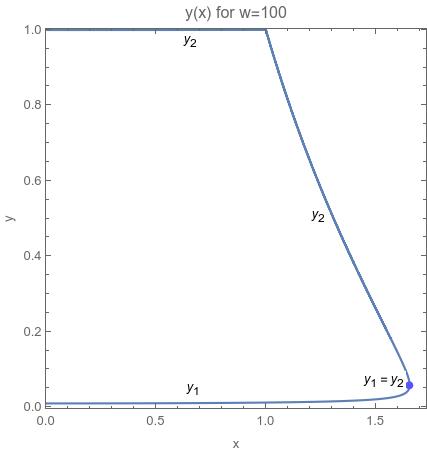}}
  \hspace{5mm}
  \subfloat[\label{twohorizonswithRT=1000}]
  {\includegraphics[width=.31\textwidth]{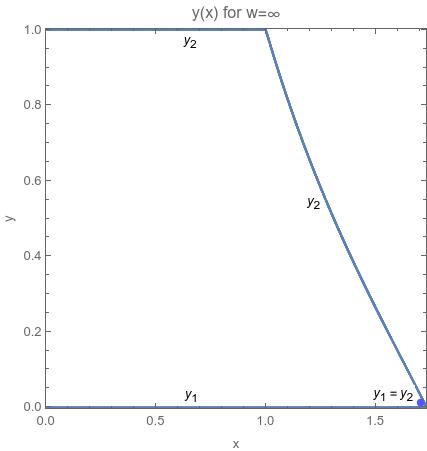}}
\caption{
Plots of $\frac{r_+}{R}$ as a function of $\sqrt{\Lambda R^2}$ for six
different values of $4\pi RT$.  For plotting purposes it is defined
$x\equiv \sqrt{\Lambda R^2}$, $y\equiv \frac{r_+}{R}$,
and
$w\equiv 4\pi RT$.
(a)
A plot of the horizon solution
for
temperature $w=2$, i.e., $RT=\frac1{2\pi}=0.16$,
the later equality being approximate.
The only solution is the Nariai solution with $x=1$ and 
$y_1=y_2=1$. 
(b)
A plot of the two horizon solutions $y_1$ and $y_2$ for temperature
$w=2.5$, i.e., $RT=0.20$ approximately.
(c)
A plot of the two horizon solutions $y_1$ and $y_2$ for
temperature
$w=\frac{\sqrt{27}}{2}=2.60$, i.e.,
$RT=\frac{\sqrt{27}}{8\pi}=0.21$,
the decimal equalities being approximate.  
(d)
A plot of the two horizon solutions $y_1$ and $y_2$ for
temperature $w=3$, i.e., $RT =0.24$ approximately.
(e)
A plot of the two horizon solutions $y_1$ and $y_2$ for temperature
$w=100$, i.e., $RT =8.0$ approximately.
(f)
A plot of the two horizon solutions $y_1$ and $y_2$ for
temperature  $w=10000$, i.e.,  $RT=796$ approximately.
Note that $10000\to\infty$  in this context.
See text for details.
}
\label{figureyofx}
\end{figure}
\end{widetext}

The first set of six  diagrams is shown in Fig.~\ref{figureyofx}. It gives
a snapshot for each $4\pi RT$ of how the black hole
horizon radii $\frac{r_+}{R}$
behave
in relation to $\sqrt{\Lambda R^2}$.


%
The case $0\leq 4\pi RT<2$ is not represented since there
are no black hole solutions.
One has either absolute zero pure de Sitter space when
$4\pi RT=0$, and hot, say, de Sitter space when $0< 4\pi RT<2$.
For a heat reservoir radius
$R$ of the order of the size of a neutron,
typical temperatures would be of the order of $10^{11}\,$K,
so the term hot even when $4\pi RT<2$
can be considered as appropriate.

%
The case $4\pi RT=2$, i.e., $RT=\frac1{2\pi}=0.159$, the last
equality being approximate, see
Fig.~\ref{twohorizonswithRT=1/2pi},
is the case with minimum temperature that yields a
solution $\frac{r_+}{R}$.  This first solution is a solution with
$\Lambda R^2=1$, and no other $\Lambda R^2$. It has
$\frac{r_{+1}}{R}=\frac{r_{+2}}{R}=1$, and no other radius. It is a
Nariai solution, the coldest one.  Indeed, it is the first solution
that pops out when $4\pi RT$, i.e., $T$, increases from zero. This means
that Nariai is easier to manufacture thermodynamically than
Schwarzschild-de Sitter.  The corresponding cosmological horizon
radius obey
$\frac{r_{\rm c1}}{R}= \frac{r_{\rm c2}}{R}=1$.


%
The case $4\pi RT=2.5$, i.e.,
$RT=0.199$ approximately, see Fig.~\ref{twohorizonswithRT=0.19}, shows that
for each $\Lambda R^2$ there are two solutions, the small one
$\frac{r_{+1}}{R}$, unstable,
and the large one $\frac{r_{+2}}{R}$, stable, but these
solutions only exist for a narrow range of the abscissas, namely,
$0.60\leq\Lambda R^2\leq 1.10$, where the numbers are approximate,
i.e., there are solutions for $\Lambda R^2<1$ and for $\Lambda R^2>1$,
but all solutions are still near $\Lambda R^2=1$. The Nariai solution
is also included in this case, now with a temperature higher than the
previous case.  The cosmological horizon radii $\frac{r_{\rm c1}}{R}$
and $\frac{r_{\rm c2}}{R}$ can then be found directly from the
corresponding horizon radii.


%
The case $4\pi RT=\frac{\sqrt{27}}{2}=2.598$, i.e.,
$RT=\frac{\sqrt{27}}{8\pi}=0.207$, the equalities
in decimal notation being
approximate, see
Fig.~\ref{twohorizonswithRT=srqt27/8pi}],
displays for the first time a solution with zero
cosmological constant, $\Lambda R^2=0$, which is the pure Schwarzschild
solution, the one with zero cosmological constant, found first by
York. This solution is a coincident horizon solution with
$\frac{r_{+1}}{R}=\frac{r_{+2}}{R}=\frac23$.
For other larger $\Lambda R^2$ there are two solutions, the small one
$\frac{r_{+1}}{R}$, unstable,
and the large one $\frac{r_{+2}}{R}$, stable, and these
solutions only exist for some range of the abscissas, namely,
$0\leq\Lambda R^2\leq 1.12$, where the latter
number is approximate.
The Nariai solution is also included in this case, now with a
temperature higher than the previous case. The cosmological horizon
radii $\frac{r_{\rm c1}}{R}$ and $\frac{r_{\rm c2}}{R}$ can then be
found directly from the corresponding horizon radii using
the appropriate equation.



%
The case $4\pi RT=3$, i.e.,
$RT=0.239$ approximately, see Fig.~\ref{twohorizonswithRT=0.23},
shows that for zero cosmological constant, 
$\Lambda R^2=0$, there are
two pure Schwarzschild solutions
with $\frac{r_{+1}}{R}<\frac23$ and $\frac{r_{+2}}{R}>\frac23$, the
first unstable, the second stable.
For other larger $\Lambda R^2$ there are also
two solutions, the small one
$\frac{r_{+1}}{R}$, unstable,
and the large one $\frac{r_{+2}}{R}$, stable, and these
solutions only exist for some range of the abscissas, namely,
$0\leq\Lambda R^2\leq 1.20$, the latter number being approximate.
The Nariai solution
is also included in this case, now with a temperature higher than the
previous case. 
The cosmological horizon radii $\frac{r_{\rm
c1}}{R}$ and $\frac{r_{\rm c2}}{R}$ can then be found directly
from the corresponding horizon radii.


%
The case $4\pi RT=100$, i.e.,
$RT=7.958$ approximately,
see Fig.~\ref{twohorizonswithRT=10},
is a case where the temperature is high, but not divergingly high.
For zero
cosmological constant,  $\Lambda R^2=0$, the two pure
Schwarzschild solutions are,  one
with $\frac{r_{+1}}{R}$
now very small approaching zero, which is unstable, and the
other $\frac{r_{+2}}{R}$  now large and approaching one,
which is stable.
For other larger $\Lambda R^2$ there are also
two solutions, the small one approaching zero
$\frac{r_{+1}}{R}$, unstable,
and the large one  approaching one
$\frac{r_{+2}}{R}$, stable, and these
solutions exist for a larger range of the abscissas, namely,
$0\leq\Lambda R^2\leq 2.70$, the latter number
being approximate. The Nariai solution
is also included in this case, now with a temperature higher than the
previous case. 
The cosmological horizon radii $\frac{r_{\rm
c1}}{R}$ and $\frac{r_{\rm c2}}{R}$ can then be found directly
from the corresponding horizon radii. 


%
The case $4\pi RT=\infty$,
precisely $4\pi RT=10000$, i.e.,  $RT=795.8$ approximately,
see Fig.~\ref{twohorizonswithRT=1000}, displays the maximum spectrum of
solutions. This case has been analyzed above in some detail and exact
expressions for $\frac{r_{+1}}{R}$ and $\frac{r_{+2}}{R}$ have been
found.  There are solutions in the range $0\leq\Lambda R^2\leq 3$.
The small horizon solution has $\frac{r_{+1}}{R}=0$ all the way
and is unstable.  The
large horizon solution has $\frac{r_{+2}}{R}=1$ up to $\Lambda R^2=1$,
and then decreases to zero at $\Lambda R^2=3$ where it joins the
$\frac{r_{+1}}{R}$ solution, and it is
a stable solution. The
Nariai solution is also included in this case, now with a temperature
tending to infinity, $4\pi RT\to\infty$. 
The cosmological horizon radii
$\frac{r_{\rm c1}}{R}$ and $\frac{r_{\rm c2}}{R}$ can then be found
directly from the corresponding horizon radii and there are exact
expressions for them.

\subsubsection{Diagrams with $\Lambda R^2$ fixed}

\begin{widetext}

\begin{figure}[h]
\centering
  \subfloat[\label{yofw1xsquareequal0}]
  {\includegraphics[width=.28\textwidth]{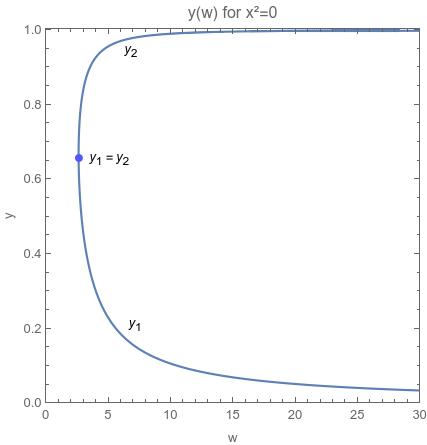}} 
  \hspace{0mm}
  \subfloat[\label{yofw2xequal08}]
  {\includegraphics[width=.28\textwidth]{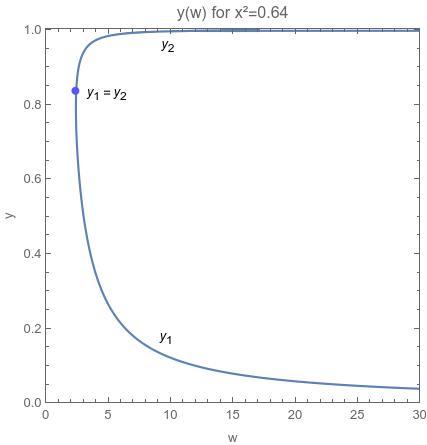}}
  \hspace{0mm}
  \subfloat[\label{yofw3xequal1}]
  {\includegraphics[width=.28\textwidth]{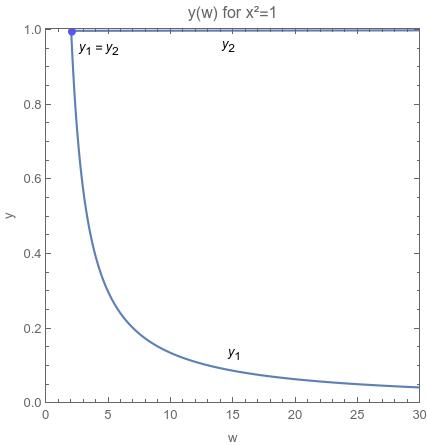}}
  \hspace{0mm}
  \subfloat[\label{yofw4xequal11}]
  {\includegraphics[width=.28\textwidth]{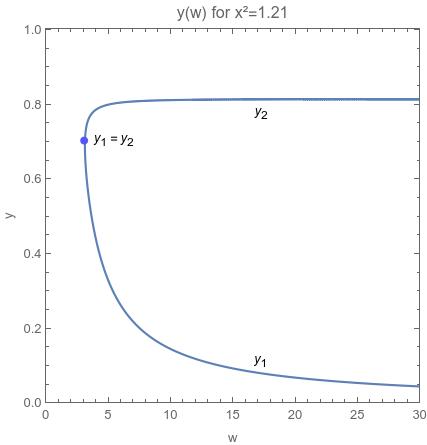}}
  \hspace{0mm}
  \subfloat[\label{yofw5xequal15}]
  {\includegraphics[width=.28\textwidth]{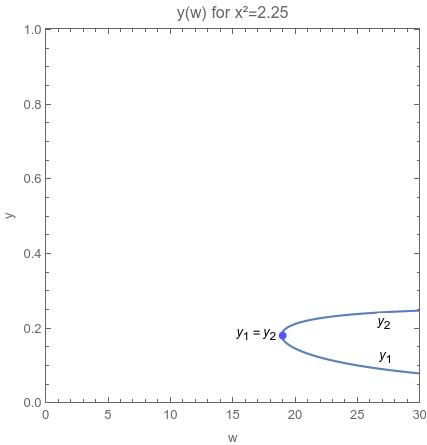}}
  \hspace{0mm}
  \subfloat[\label{yofw6xequalsqrt3}]
  {\includegraphics[width=.28\textwidth]{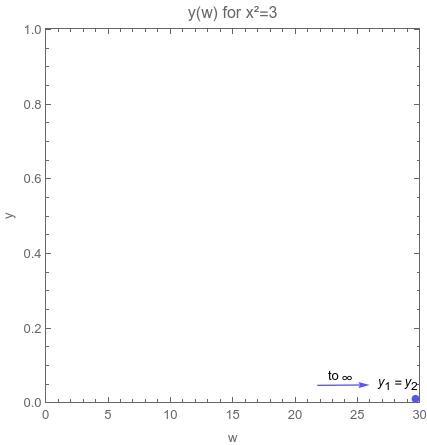}}
\caption{
Plots of $\frac{r_+}{R}$ as a function of $4\pi RT$ for six different
values of $\Lambda R^2$.  For plotting purposes it is defined
$x\equiv\sqrt{\Lambda R^2}$,
$y\equiv\frac{r_+}{R}$, and $w\equiv 4\pi RT$.
(a)
A plot of the two horizon solutions $y_1$ and $y_2$
for $x^2=0$. At temperature
$w=\frac{\sqrt{27}}{2}$, i.e.,
$RT=\frac{\sqrt{27}}{8\pi}$, the coincident solution $y_1=y_2=\frac23$
appears.  For larger $w$, $y_1$ tends to zero and $y_2$ tends to one.
(b)
A plot of the two horizon solutions $y_1$ and $y_2$ for
$x^2=0.64$.
(c)
A plot of the two horizon solutions $y_1$ and $y_2$
for $x^2=1$.
At temperature
$w=2$, i.e., $RT=\frac{1}{2\pi}$,
the solution $y_2=1$ appears, which in this case is
a coincident solution, indeed  $y_1=y_2=1$. It is a Nariai
solution. 
The solution $y_2=1$ for
any higher temperature $w$ is Nariai.
(d)
A plot of the two horizon solutions $y_1$ and $y_2$ for
$x^2=1.21$.
(e)
A plot of the two horizon solutions $y_1$ and $y_2$ for
$x^2=2.25$. 
(f)
A plot of the two horizon solutions $y_1$ and $y_2$ for
$x^2=3$.  Here, there is only
the coincident solution,
with values $y_1=y_2=0$ and temperature $w=\infty$.
See text for details.
}
\label{figureyofw}
\end{figure}

\end{widetext}

The second set of six diagrams is shown in Fig.~\ref{figureyofw}.  It
gives a snaphot for each $\Lambda R^2$ of how the black hole horizon
radii $\frac{r_{+}}{R}$ behave in relation to $4\pi RT$.


%
The case $\Lambda R^2=0$, see Fig.~\ref{yofw1xsquareequal0}, is the case
with minimum $\Lambda R^2$ in this context.  For this case,
in the range $0\leq
4\pi RT<\frac{\sqrt{27}}{2}$, there are no radii $\frac{r_{+}}{R}$ that
are solution of the thermodynamic problem,
so presumably the inside of the reservoir is
filled with hot de Sitter space.  At
$4\pi RT=\frac{\sqrt{27}}{2}$ the coincident solution $\frac{r_{+1}}{R}=
\frac{r_{+2}}{R}=\frac23$ appears.  For larger $4\pi RT$,
$\frac{r_{+1}}{R}$ tends to zero and $\frac{r_{+2}}{R}$ tends to
one. Since $\Lambda R^2=0$ means zero cosmological constant, i.e., 
$\Lambda=0$, this case is York's solution, specifically,
the pure Schwarzchild
case.


%
The case $\Lambda R^2=0.64$, see Fig.~\ref{yofw2xequal08}, is a
case with an intermediate value of  $\Lambda R^2$.  At some definite
value of 
$4\pi RT$, smaller
than the value of the previous case,
the coincident solution $\frac{r_{+1}}{R}= \frac{r_{+2}}{R}$
appears.  For larger $4\pi RT$, $\frac{r_{+1}}{R}$ tends to zero and
$\frac{r_{+2}}{R}$ tends to one.


%
The case $\Lambda R^2=1$, see Fig.~\ref{yofw3xequal1}, is the case
where the Nariai universe exists.
For this case,
in the range $0\leq
4\pi RT<2$, there are no radii $\frac{r_{+}}{R}$ that
are solution of the thermodynamic problem,
so presumably the inside of the reservoir is
filled with hot de Sitter space. 
At $4\pi RT=2$ the coincident solution
appears with $\frac{r_{+1}}{R}= \frac{r_{+2}}{R}=1$. This is
the coldest Nariai solution.
For larger $4\pi RT$, $\frac{r_{+1}}{R}$
tends to zero and $\frac{r_{+2}}{R}$ tends to one, indeed
the solution
$\frac{r_{+2}}{R}=1$ for any $4\pi RT\geq2$ is
a hot Nariai universe.


%
The case $\Lambda R^2=1.21$, see Fig.~\ref{yofw4xequal11}, is a
case typical of  $\Lambda R^2>1$.  At some definite $4\pi RT$ the
coincident solution $\frac{r_{+1}}{R}= \frac{r_{+2}}{R}$ appears.  For
larger $4\pi RT$, $\frac{r_{+1}}{R}$ tends to zero and $ \frac{r_{+2}}{R}$
tends to some value value that is less than one.


%
The case $\Lambda R^2=2.25$, see Fig.~\ref{yofw5xequal15}, is also a
case typical of  $\Lambda R^2>1$, but the
new features are more evident.  At some definite $4\pi RT$ the
coincident solution $\frac{r_{+1}}{R}= \frac{r_{+2}}{R}$ appears, now
with greater $4\pi RT$ than the previous case.  For larger $4\pi RT$,
$\frac{r_{+1}}{R}$ tends to zero and $ \frac{r_{+2}}{R}$ tends to some
value value less than one. This
value less than one decreases rapidly with increasing $\Lambda R^2$.

%
The case $\Lambda R^2=3$, see Fig.~\ref{yofw6xequalsqrt3}, is the last
possible case, is a limit case.  In this case, there is only one
solution, which is the coincident solution $\frac{r_{+1}}{R}=
\frac{r_{+2}}{R}=0$.  In the plot this solution is represented as a
point in the highest shown $4\pi RT$, which is meant
to be $4\pi RT=\infty$.
There is divergently hot de Sitter space in the cavity at radius
$R$, apart from a central singularity at zero radius,
$\frac{r_{+1}}{R}= \frac{r_{+2}}{R}=0$.


%
%

%
%

\subsubsection{Additions}

It is important to make additional comments to the 
the plots that have been displayed
in Figs.~\ref{figureyofx}
and \ref{figureyofw}. 
In what follows the discussion will be qualitative.

Stacking with interpolation 
Figs.~\ref{twohorizonswithRT=1/2pi}-\ref{twohorizonswithRT=1000},
one can glimpse the correctness of 
Figs.~\ref{yofw1xsquareequal0}-\ref{yofw6xequalsqrt3},
and stacking with interpolation 
Figs.~\ref{yofw1xsquareequal0}-\ref{yofw6xequalsqrt3},
one  can glimpse, in turn, the correctness of 
Figs.~\ref{twohorizonswithRT=1/2pi}-\ref{twohorizonswithRT=1000}.

One striking feature, that can be deduced from the plots, is that the
space of black hole horizon radius solutions is enlarged as the
reservoir temperature $T$, or rather $4\pi RT$, is increased. In fact, for
very low temperatures there are no solutions for any $\Lambda$, or
rather, for
any $\Lambda R^2$.  At the temperature $4\pi RT=2$ there is
only one solution, the coldest
possible Nariai universe.  For higher $4\pi RT$ there are
solutions for some values of $\Lambda R^2$, but not all.  For
instance, for $4\pi RT=\frac{\sqrt{27}}{2}$, there appears a solution
with $\Lambda=0$, i.e., $\Lambda R^2=0$, and there are solutions up to
$\Lambda R^2=1.12$, approximately, so that the range of $\Lambda R^2$
is $0\leq\Lambda R^2\leq1.12$ for this temperature.  Finally, for
infinite temperature, $4\pi RT=\infty$, there are solutions for all
possible $\Lambda R^2$, namely $0\leq\Lambda R^2\leq3$. To understand
other features of the solutions it is perhaps advisable to
incorporate the case $\Lambda R^2=1$ in both the low and
high cosmological constant
cases and thus 
divide
the whole range into
$0\leq\Lambda R^2\leq1$ and $1\leq\Lambda R^2\leq3$.

In the range $0\leq\Lambda R^2\leq1$, now with the help of
Figs.~\ref{figureyofx} and \ref{figureyofw}, one can summarize the
qualitative explanation for the reason of why
black hole solutions with lower
cosmological constant appear only for higher $T$, i.e., higher $RT$.
Thus, let us start with
$\Lambda=0$, so that the cosmological length scale
$\ell=\frac1{\sqrt{\Lambda}}$ is infinite, 
$\ell=\infty$.
In this case there is no coupling of
this length scale with the other two,
$\lambda=\frac1T$ and $R$.
In this situation, we see that
for low $T$, high $\lambda$, one has $\lambda\gg R$, so that, since
the thermal wavelength is very large compared to the
reservoir radius $R$, then this wavelength
is stuck to the reservoir
and
the corresponding energy
cannot collapse to form a black hole
in any circumstances.
When $T$ is sufficiently  increased, i.e., 
$RT$ is larger than about one, the wavelength is
sufficiently small,
and the corresponding energy 
can travel freely inside the reservoir
and can collapse,
so that formation of black holes is possible.  The value
$4\pi RT=\frac{\sqrt{27}}{2}$ divides no black hole from black hole
solutions.
Now, let us do $\ell$ finite. This is the third scale.  For low enough
$\ell$, say a bit larger than $R$, the space inside the reservoir gets
higher curvature due to the high cosmological constant, and so in some
manner this inner space has more length along the radius, so that
although the reservoir area radius is still $R$, the
radial length is large, and so the volume is also
larger.  This means that
energy packets
with higher $\lambda$, relatively to the cases with lower
$\ell$, can continue to travel freely in the inside and can form black
holes.
The limiting situation is when $R$ and $\ell$ are equal,
i.e., $\Lambda R^2=1$, or $\frac{R^2}{l^2}=1$, so that
energy packets with the highest possible $\lambda$, actually
$\lambda=2\pi R$, can give
a solution, which in this case is a Nariai solution.

In the range $1\leq\Lambda R^2\leq3$, now with the help of
Figs.~\ref{figureyofx} and \ref{figureyofw}, one can also s give a
qualitative explanation for the reason of why now black hole solutions
with higher cosmological constant, i.e., higher $\Lambda R^2$, appear
only for higher $T$, i.e., higher $RT$.  Within this range, the
cosmological constant is very high, i.e., the cosmological length
scale $\ell$ is very short, and so determines and dominates the
processes.  Indeed, now $\ell<R$.
Let us now start with the situation when $R$ and $\ell$
are equal, i.e., $\Lambda R^2=1$, or $\frac{R^2}{l^2}=1$, so that the
energy packets with the highest possible $\lambda$, actually
$\lambda=2\pi R$, can give a solution, a Nariai solution in this case.
Still, for $\frac{R^2}{l^2}=1$ and higher temperatures, i.e,
lower wavelengths $\lambda$, there is smaller unstable
$r_{+1}$ black hole solution and the Nariai solution.
For $\frac{R^2}{l^2}$ larger than one,
i.e., $\ell$ a bit smaller,
the temperatures have to
be a  higher, and so the
wavelength $\lambda$ of the energy
packets has to be a bit smaller, to
have the two black black hole solutions, as in the $\Lambda R^2<1$
case. The small $r_{+1}$ solution forms in the same manner.
The new feature is with the large black hole solution
$r_{+2}$. Now, for fixed $\Lambda R^2<1$, the
$r_{+2}$ solution is always less then $R$ even when
$T$ is very large. This means that
the corresponding small wavelengths $\lambda$ now
are constrained by the scale $\ell$ so that the interplay
is between $r_{+2}$, $\lambda$, and $\ell$ and not anymore
between $r_{+2}$, $\lambda$, and $R$.  
The final case is when $\ell=\frac{R}{\sqrt3}$. The space
inside is pure de Sitter, except for a singular black horizon at the
center, with the energy packets having a zero thermal wavelength
$\lambda=0$, with the temperature $T$ of the
reservoir being infinite. Only those $\lambda=0$
energy packets can collapse
to form a black hole, packets with a higher $\lambda$, corresponding
to a lower reservoir $T$ cannot fit to the
scale set by $\ell$.

Another characteristic radius that is part of the Schwarzschild-de
Sitter solution is the cosmological radius $r_{\rm c}$. In the setting
we are working, where the heat reservoir is for the inside that
harbors a possible black hole, the cosmological radius $r_{\rm c}$ has
only a secondary role. This characteristic radius $r_{\rm c}$
can be calculated
once the black hole horizon radius is found on thermodynamic grounds.

\subsection{
Mathematical analysis of the plots: Black hole horizons}

\subsubsection{Nomenclature}

We now obtain through a mathematical
analysis some important features displayed in the
plots above, Figs.~\ref{figureyofx} and \ref{figureyofw}. We repeat
here Eq.~\eqref{tbh2}, i.e., $4\pi RT= \frac{1}{ \frac{r_+}{R}} \frac{
1-\Lambda R^2\left(\frac{r_+}{R}\right)^2} { \sqrt{1-\frac{r_+}{R}}\,
\sqrt{\left(1-\frac{\Lambda R^2}{3}\left(1+
\frac{r_+}{R}+\left(\frac{r_+}{R}\right)^2\right)\right)} }$. The
natural variables without units are
$\Lambda R^2$ and
$\frac{r_+}{R}$. In this context it is
perhaps preferable to work with 
$\sqrt{\Lambda R^2}$ rather than with $\Lambda R^2$,
so to shorten the notation we define the
variables $x$ and $y$ as
\begin{equation}
x\equiv\sqrt{\Lambda R^2}\,,
\label{xdef}
\end{equation}
\begin{equation}
y\equiv\frac{r_+}{R}\,,
\label{ydef}
\end{equation}
with the range being $0\leq x\leq\sqrt3$, or
 $0\leq x^2\leq3$, and
$0\leq y\leq1$. In these variables, 
Eq.~\eqref{tbh2} is
$4\pi TR=
\frac{1-x^2y^2}{y\sqrt{1-y}
\sqrt{1-\frac{x^2}{3}(1+y+y^2)}}
$.
Define in addition the variable $w$ as 
\begin{equation}
w\equiv4\pi RT\,.
\label{zdef}
\end{equation}
Then, with these definitions Eq.~\eqref{tbh2} is
\begin{equation}
w=
\frac{1-x^2y^2}{y\sqrt{1-y}
\sqrt{1-\frac{x^2}{3}(1+y+y^2)}}\,,
\label{zofxy}
\end{equation}
with $2\leq w<\infty$. Solutions exist only for $w\geq2$.

Now, for a fixed temperature $T$, more properly
for a fixed $RT$, i.e., fixed $w$, one
has $dw=0$, and so
$\frac{dy}{dx}=-\frac{\frac{\partial w}{\partial x}}
{\frac{\partial w}{\partial y}}$.
After some calculations we obtain
\begin{equation}
\frac{dy}{dx}=-\frac{2xy(1-y)Q(y)}{3R(y)}\,,
\label{dydx}
\end{equation}
where
\begin{equation}
Q(y)\equiv(1+y+y^2)(1+x^2y^2)-6y^2\,,
\label{Qy}
\end{equation}
and
\begin{equation}
R(y)\equiv-\frac{2}{3}(1+x^2y^2)(1-y)[3-
x^2(1+y+y^2)]+(1-x^2y^2)^2y\,.
\label{Ry}
\end{equation}
In addition, we need in the analysis
$\frac{\partial w}{\partial y}$.
Obtain from Eq.~\eqref{zofxy} that
\begin{equation}
\frac{\partial w}{\partial y}=
\frac{\sqrt{3} {S}}{2y^2(1-y)^{3/2}
\Bigl[ 3-x^2(1+y+y^2)\Bigr]^{3/2}}
\label{wy}
\end{equation}
where we have defined 
\begin{equation}
{S}(x,y)=x^{4}[2y^{2}(1-y^{3})+
3y^{5}]+2x^{2}(-y^{3}-3y^{2}+1)-6+9y\,.
\label{defSbar}
\end{equation}

We have seen that the point $\Lambda R^2=1$, i.e.,
$x^2=1$, is important, as it gives the
Nariai solution. 
So, let us consider below
the limit when $x\rightarrow 1$
from below and from above.
We will see now that the result depends
on how the limit is taken. We recall that for each
$\Lambda R^2$ there are two solutions, $r_{+1}$,
the small solution,
and $r_{+2}$, the large solution,
which change as $RT$ is changed,
i.e., for each $x$, there are $y_1$
and $y_2$, which change as $w$ is changed.

\subsubsection{Analysis}
There are general
results here that we can mention. Let us
analyzed the three
ranges
separately, namely, the regimes $x^2<1$, $x^2=1$, and $x^2>1$.

\vskip 0.5cm

\noindent
{\bf $x^2{\mathbf <1}$:}

\noindent
This range of $x^2$ is specifically
$0\leq x^2<1$. In this range of $x$, the range of
$y$ is
\begin{equation}
0\leq y<1\,.
\label{0y1}
\end{equation}
We now find $y_1$, then $y_2$, and then we analyze
the coincident solutions $y_1=y_2$.

First we find $y_1$, i.e., $r_{+1}$, when $x$
varies within this range, $0\leq x^2<1$.
For that, we fix $y<1$ and move along the positive
$x$
direction.
In the space
$(x,y)$ the corresponding point moves along
a horizontal line. Then, from Eq.~\eqref{zofxy}
one finds
$w=
\frac{1-x^2y^2}{y\sqrt{1-y}
\sqrt{1-\frac{x^2}{3}(1+y+y^2)}}$,
i.e., $y$ obeys an equation of the type 
\begin{align}
y_1\sqrt{1-y_1}\sqrt{1-\frac{x^2}{3}(1+y_1+y_1^2)}\,&w=1-x^2y_1^2\,,
\nonumber\\
&
0\leq x^2<1\,,
\label{y1}
\end{align}
for a given $x$ fixed. 
One of the solutions of this equation
is $y_1$, with $w$ fixed and $w$
with a value obeying $w>2$,
and with $y_1(x)<1$ always. The limit
of $x\to1$ for $y_1$ is smooth, see below.
Another property is that 
for any $x$, $w\rightarrow \infty $ when $y_1\rightarrow 0$.
This is the solution $y_1=0$, i.e., $\frac{r_{+1}}{R}=0$, when the
temperature goes to infinity.
Another interesting property is the point where
$\frac{dy_1}{dx}=0$, if there is one.
This is when $r_{+1}$ attains a minimum value
in relation to $\Lambda R^2$. This is the root of equation
$Q(y)=0$, and happens
for the small solution, i.e., $y_1$.
From Eq.~\eqref{Qy},
$Q(y_1)=0$ gives
\begin{equation}
x^2=\frac{5y_1^2-y_1-1}{y_1^2(1+y_1+y_1^2)}\,.
\label{smallsolutiony1}
\end{equation}
Now, the lowest $y_1$ is given when $x=0$ by the solution of
$5y_1^2-y_1-1=0$, which is $\frac{1+\sqrt{21}}{10}$.
The highest $y_1$ is $y_1\to1$ which yields $x\to1$.
Thus, $\frac{dy_1}{dx}=0$ in the range
$0\leq x^2<1$ happens in the range
\begin{equation}
\frac{1+\sqrt{21}}{10}\leq y_1<1\,,
\end{equation}
i.e., 
$0.56\leq y_1<1$,
the first number in the left inequality being
approximate, with the range of $x$ being $0\leq x^2<1$.
The corresponding range of $w$
from Eq.~\eqref{zofxy}
is 
$
2< w\leq \frac{10\sqrt{10}}
{(1+\sqrt{21})\sqrt{9-\sqrt{21}}}
$,
or in round numbers
$
2<w\leq  2.70
$,
the last number being approximate.
Why there is an $x$ for which $\frac{dy}{dx}=0$
in the $y_1$ solution is
not clear on physical, heuristic, terms, but
possibly 
is a nonlinear
interplay between the
length scale $\ell\equiv\frac1{\sqrt\Lambda}$
and the length scales $R$ and $\lambda=\frac1T$.
Further properties for the small back hole depend on the specific $x$
and the specific $w$ that one picks, but there is nothing else
more general that we can mention.

Second we find $y_2$, i.e., $r_{+2}$, when $x$
varies within this range, $0\leq x^2<1$.
From Eq.~\eqref{zofxy},
one has
\begin{align}
y_2\sqrt{1-y_2}\sqrt{1-\frac{x^2}{3}(1+y_2+y_2^2)}\,&w=1-x^2y_2^2\,,
\nonumber\\
&
0\leq x^2<1\,,
\label{y2new}
\end{align}
for a given $x$ fixed, and 
one finds that there is a solution $y_2(x)$ for each $x$.
The maximum of the curve is at $x=1$ and $y_2=1$,
and at this point
the derivative can be any of those found above,
so it is not well defined.
It is important to find 
the behavior 
and the properties of $y_2$
when $x$ is near 1, i.e., $x\to 1$ with  $y_2\to 1$.
A direct property is that for any $x$,
$w\to\infty$ when $y_2\to1$.
This is the solution $y_2=1$, i.e., $\frac{r_{+2}}{R}=1$, when the
temperature goes to infinity.
Other important properties
are related to the derivative of $y_2$,
in particular, $\frac{dy_2}{dx}$
for  $x\to 1$. The calculations of
$\frac{dy_2}{dx}$ for  $x\to 1$ are done in
detail in the Appendix
\ref{fomulassectionvi}, here we state the result,
namely,
\begin{equation}
\frac{dy_2}{dx}
=\frac{w}{\sqrt{w^2-4}}-1\,,\quad\quad w\geq2\,,
\quad\quad x\to1\,.
\label{dydxnew}
\end{equation}
Therefore, taking the two limits for the temperature,
i.e., for $w$, we have 
\begin{align}
\frac{dy_2}{dx}\to \infty \quad{\rm for} \quad
w\to 2\,,\quad\quad\quad  
\frac{dy_2}{dx}\to 0 \quad
&w\to\infty\,,
\nonumber\\
& x\to1\,.
\label{lim1}
\end{align}
Thus, the derivative $\frac{dy_2}{dx}$
obeys $\frac{dy_2}{dx}\geq0$, and can 
have values
from 0 to infinity
when one approaches $x=1$ from 
$x<1$.
Another
possible interesting property is the point where
$\frac{dy_2}{dx}=0$, if there is one.
For the solution $y_2$
one finds that there is no point with zero derivative.

Now we find $y_1=y_2$ for $w$ fixed, i.e., the point $x$
where $y_1=y_2$.
There can
exist two points at which $\frac{dy}{dx}=\infty $
for fixed $w$, one for $x<1$, the other for $x>1$.
Here we work out $x<1$.
It is given by the root of equation $R(y)=0$, see Eq.~\eqref{Ry},
and it defines
the bifurcation point where $y_1=y_2$ for $w={\rm constant}$.
The equation is
$\frac{2}{3}(1+x^2y^2)(1-y)[3-
x^2(1+y+y^2)]-(1-x^2y^2)^2y=0$, which can be put
in the form
$
x^4 \left[ \frac23y^2(1-y)(1+y+y^2)+y^5\right]
-x^2\left[2y^2-\frac23(1-y)(1+y+y^2)
\right]-(2-3y)=0
$. This is a quadratic in $x^2$
of the form $x^4a(y)-x^2b(y)-c(y)=0$,
where $a(y)=\frac23y^2(1-y)(1+y+y^2)+y^5$,
$b(y)=2y^2-\frac23(1-y)(1+y+y^2)$, and 
$c(y)=2-3y$.
The solution is
\begin{equation}
x^2=\frac{b(y)-\sqrt{b^2(y)+4a(y)c(y)}}
{2a(y)}\,,\quad\quad\quad 0< x^2\leq 1
\,,
\label{Ry=0}
\end{equation}
Inverting Eq.~\eqref{Ry=0} one finds $y(x)$,
i.e., the solution $y_1=y_2$.
For $0\leq x^2<1$ there are solutions
for $y>\frac23$, which from Eq.~\eqref{zofxy} means in turn
$w<\frac{\sqrt{27}}{2}$, i.e., 
for $RT<\frac{\sqrt{27}}{8\pi}$, with $RT=\frac{\sqrt{27}}{8\pi}$
at $x=0$ being marginal. For higher $w$, i.e.,
higher $RT$,
there are no solutions, $y_1$ and $y_2$
never coincide for any $x$ in this range
$0\leq x^2<1$.

Now, we find $y_1=y_2$, the coincident solution
for fixed $x$, and $w$ varying.
This is when $\frac{\partial w}{\partial y}=0$, see Eq.~\eqref{wy},
i.e., ${S}(x,y)=0$.
From Eq.~\eqref{defSbar} we have then that the
bifurcation points where $y_1=y_2$ for $x={\rm constant}$
are given by the equation
$
x^{4}[2y^{2}(1-y^{3})+3y^{5}]+2x^{2}(-y^{3}-3y^{2}+1)-6+9y=0
\label{defSbar=0}
$,
i.e., 
$
x^{4}[2y^2(1-y^3)+3y^5]-x^{2}[2(y^3+3y^2-1)]-3(2-3y)=0
$. This is a quadratic in $x^2$, which can be written
as $x^4d(y)-x^2e(y)-f(y)=0$, with
$d(y)=2y^2+y^5$,
$e(y)=2(y^3+3y^2-1)$,
$f(y)=3(2-3y)$, and with solution
\begin{equation}
x^2=\frac{e(y)-
\sqrt{e^2(y)+ 4d(y)f(y)}
}
{2d(y)}\,,\quad\quad\quad 0< x^2\leq 1
\,.
\label{more=0}
\end{equation}
Inverting Eq.~\eqref{more=0}
one has $y(x)$, for which $y_1=y_2$.
To make progress
and find an analytical result we take $x\to 1$ and $y\to 1$.
The calculations are involved and we leave them to
the Appendix
\ref{fomulassectionvi}, the
result is 
that the coincident solution takes the form
\begin{equation}
y_1=y_2= 1-\left(\frac38 \left(1-x^2\right)^2\right)^\frac13\,,
\label{solutionapp0}
\end{equation}
for $(1-x^2)\ll1$.  Then, returning to the original variables one has
$\frac{r_{+1}}{R}=\frac{r_{+2}}{R}= 1-\left(\frac32\, (1-\Lambda
R^2)^2\right)^{\frac13}$ for $1-\Lambda R^2$ tiny, which is
Eq.~\eqref{solutioninitial1}.

\vskip 0.5cm
\noindent
{\bf $x^2{\mathbf =1}$:}

\noindent
This specific value of $x$, $x=1$, is important
as it hides a considerable structure, indeed, the
whole Nariai solution is inside it. 
For this value of $x$, the range of
$y$ is
\begin{equation}
0\leq y\leq1\,.
\label{0y1x=1}
\end{equation}
We now find $y_1$, then $y_2$, an then we analyze
the coincident solutions $y_1=y_2$.

First we find  $y_1$, i.e., $r_{+1}$,  when $x=1$.
For that, we fix $y<1$ and move along the positive
$x$
direction.
In the space
$(x,y)$ the corresponding point moves along
a horizontal line. Then, from Eq.~\eqref{zofxy}
one finds that for $x=1$ one has
$w=\frac{\sqrt{3}\sqrt{1-y}(1+y)}{y\sqrt{2-y-y^2}}
$.
Since $2-y-y^2=(1-y)(2+y)$, one has
\begin{equation}
\left(
y_1\sqrt{2+y_1}\right)w={\sqrt{3}(1+y_1)}\,,
\quad\quad  x^2=1\,,
\label{y1x=1}
\end{equation}
and, from Eq.~\eqref{wy},
one also finds $\frac{d w}{d y}<0$.
Now, Eq.~\eqref{y1x=1}
has one solution $y_1(x=1)$, with $w$ fixed and $w$
with a value obeying $w>2$.
For any $w$, with $w>2$ one has
$y_1(x=1)<1$ always.
It is clear that point $(x,y_1)=(1,y_1)$ lies on
the branch of the curve that
corresponds to the small root,
to the small black hole.

Second, we find   $y_2$, i.e., $r_{+2}$.
From Eq.~\eqref{zofxy}
one finds that for $x=1$ one has
\begin{equation}
\left(
y_2\sqrt{2+y_2}\right)w={\sqrt{3}(1+y_2)}\,,
\quad\quad  x^2=1\,.
\label{y2x=1}
\end{equation}
For $x=1$, the large black hole
always has $y_2=1$ but $w$ can be any as we have seen.
So, Eq.~\eqref{y2x=1} in the context of $y_2$
is deceiving. 
Indeed, if we put $y_2=1$ into 
Eq.~\eqref{y1x=1} one finds $w=2$, i.e., $RT=\frac1{2\pi}$.
But we know, from
our calculation above that at $x=1$,
$w$ can be any, indeed $2\leq w<\infty$.
So, Eq.~\eqref{y2x=1} only gives one of the
infinite number of solutions, which all correspond
to the Nariai universe. Since this has been and will be further
discussed
we refrain to take further comments.

Now we find $y_1=y_2$ for $w$ fixed. Here
$x$ is fixed,  $x=1$,
i.e., $x^2=1$. 
The solution is 
\begin{equation}
x^2=y_1=y_2=1\,,
\label{y1=y2=x=1}
\end{equation}
with $w=2$.
Then, in the original variables
we have $\Lambda R^2=\frac{r_{+1}}{R}=\frac{r_{+2}}{R}=1$
with $RT=\frac1{2\pi}$. It is the first Nariai
universe solution as far as increasing temperature
goes.

Now we find $y_1=y_2$ for $x$ fixed
and $w$ varying.
It is given by Eq.~\eqref{y1=y2=x=1}
since $w$ is fixed as we saw.

\vskip 0.5cm
\noindent
{\bf $x^2{\mathbf >1}$:}

\noindent

\noindent
This range of $x^2$ is specifically
$1<x^2\leq 3$. In this range of $x$, the range of
$y$ is
\begin{equation}
0<y\leq y_{\rm e}\,,\quad\quad\quad\quad
y_{\rm e}=-\frac{1}{2}+\sqrt{\frac{3}{x^2}-
\frac{3}{4}}\,,
\label{y0}
\end{equation}
where $y_{\rm e}=y_{\rm e}(x)$ is the
edge, or maximum, value that $y$ can have
for each $x$ in this range.
The expression for $y_{\rm e}$ is given by the 
root of the equation
$
y^2+y+1-\frac{3}{x^2}=0$, see Eq.~\eqref{zofxy},
with solution  
$
y_{\rm e}=-\frac{1}{2}+\sqrt{\frac{3}{x^2}-
\frac{3}{4}}
$.
Note that
$y_{\rm e}=-\frac{1}{2}+\sqrt{\frac{3}{x^2}-
\frac{3}{4}}<\frac{1}{x}\leq1$.
When $y\rightarrow y_{\rm e}$ we have from Eq.~\eqref{zofxy}
that 
$w= \frac{1-x^2y_{\rm e}^2}{y_{\rm e}\sqrt{1-y_{\rm e}}\sqrt{
(y_{\rm e}-y)(y_{\ast}+y)}}$
 with
$y_{\ast}\equiv\frac{1}{2}+\sqrt{3}
\sqrt{\frac{1}{x^2}-\frac{1}{4}}$.
Since $y_{\ast}=y_{\rm e}+1$ this can also be written as
$
w= \frac{1-x^2y_{\rm e}^2}{y_{\rm e}\sqrt{1-y_{\rm e}}\sqrt{
(y_{\rm e}-y)(y_{\rm e}+y+1)}}
$.
So, for each $x$ in the range 
$1<x^2\leq 3$, one finds
$y_{\rm e}$ from Eq.~\eqref{y0}, and then one
finds the corresponding $w$, i.e.,
the corresponding $RT$, which for $y\rightarrow y_{\rm e}$
is $w\to\infty$.
We now find $y_1$, then $y_2$, and then we analyze
the coincident solutions $y_1=y_2$.

First we find  $y_1$, i.e., $r_{+1}$, when $x$
varies within the range we are working,
namely, $1< x^2\leq3$.
For that, we give some $w$, then we fix
$y<y_{\rm e}$ and move along the positive
$x$
direction.
In the space
$(x,y)$ the corresponding point moves along
a horizontal line. 
Then, from Eq.~\eqref{zofxy}
one finds
$w=
\frac{1-x^2y^2}{y\sqrt{1-y}
\sqrt{1-\frac{x^2}{3}(1+y+y^2)}}$,
i.e., $y$ obeys an equation of the type
\begin{align}
y_1\sqrt{1-y_1}\sqrt{1-\frac{x^2}{3}(1+y_1+y_1^2)}\,&w=1-x^2y_1^2\,,
\nonumber\\
&
1< x^2\leq3\,,
\label{y1x>1}
\end{align}
for a given $x$ fixed. 
One of the solutions of this equation
is $y_1$, with $w$ fixed and $w$
with a value obeying $w>2$,
and with $y_1(x)<y_{\rm e}$ always. The limit of $y_1$ for
$x\to1$ is smooth.
Also,
for any $x$ in the range
$1< x^2\leq3$, when $w\rightarrow \infty$ one has $y_1\rightarrow 0$.
This is the solution $y_1=0$, i.e., $r_{+1}=0$, when the
temperature goes to infinity.
Another
possible interesting property is the point
where $\frac{dy_1}{dx}=0$.
From Eq.~\eqref{dydx} this happens when 
$Q(y_1)=0$. Then, from 
Eq.~\eqref{Qy}
one finds that
there are none in the range $1< x^2\leq3$.
Moreover, $y_1(x)$ is
a smooth curve, infinitely
differentiable, in the whole range 
$0\leq x^2\leq3$, there is no discontinuity in any derivative
at $x=1$.
Further properties for the small back hole depend on the specific $x$
and the specific $w$ that one picks, but there is nothing else
general that we can mention.

Second we find $y_2$, i.e., $r_{+2}$.
From Eq.~\eqref{zofxy},
one finds
\begin{align}
y_2\sqrt{1-y_2}\sqrt{1-\frac{x^2}{3}(1+y_2+y_2^2)}\,
&w=1-x^2y_2^2\,,
\nonumber\\
&
1< x^2\leq3\,,
\label{y2x>1}
\end{align}
and
one finds that there is a solution $y_2(x)$ for each $x$.
The maximum of the curve is at $x=1$, $y_2=1$,
and the derivative can be any of those derived above,
so it is not well defined in this sense.
It is important to discuss
the behavior 
and the properties of $y_2$, when $x\to 1$ from above.
A direct property is that for any $x$,
$w\to\infty$ when $y_2\to y_{\rm e}$.
This is the solution $y_2=y_{\rm e}$ when the
temperature goes to infinity.
Other important properties are related
to $\frac{dy_2}{dx}$,
in particular,
$\frac{dy_2}{dx}$ when $x\to1$ from above.
The calculations
for finding
$\frac{dy_2}{dx}$ when $x\to1$ from above,
are done in detail in the Appendix
\ref{fomulassectionvi},
here we state the result, namely
\begin{equation}
\frac{dy_2}{dx} =-\frac{w}{\sqrt{w^2-4}}-1\,,\quad\quad w\geq2\,,
\quad\quad x\to 1\,,
\label{dydxnewnew}
\end{equation}
where the limit is from above.
Therefore, taking the two limits for the temperature,
i.e., for $w$, we have 
%
%
\begin{align}
\frac{dy_2}{dx}
\to-\infty \;\;{\rm for} \;\;
w\to 2\,,
\quad
\frac{dy_2}{dx}&
\to -2
\;\;{\rm for} \;\;
w\to\infty\,,
\nonumber\\
&x\to 1\,,
\label{lim1new}
\end{align}
where the limit is
from above.
Thus, the derivative $\frac{dy_2}{dx}$
obeys $\frac{dy}{dx}\leq0$,
and it can be from $0$ to minus infinity
when one approaches
$x=1$ from
$x>1$.
We see again that the point $(x,y)=(1,1)$ is very rich,
in fact it corresponds to
the Nariai limit, a
universe full of structure as we have studied.
Another
possible interesting property is the point where
$\frac{dy_2}{dx}=0$, if there is one.
For the solution $y_2$
one finds that there is no point with zero derivative.

Now we find $y_1=y_2$ for $w$ fixed, i.e., the point $x$
where $y_1=y_2$.
There exist two points at which $\frac{dy}{dx}=\infty $
for fixed $w$, one for $x<1$, the other for $x>1$.
Here we work out $x>1$.
It is given by the root of equation $R(y)=0$, see Eq.~\eqref{Ry},
and it defines
the bifurcation points where $y_1=y_2$ for $w={\rm constant}$.
The equation is
$\frac{2}{3}(1+x^2y^2)(1-y)[3-
x^2(1+y+y^2)]-(1-x^2y^2)^2y=0$, which can be put
in the form
$
x^4 \left[ \frac23y^2(1-y)(1+y+y^2)+y^5\right]
-x^2\left[2y^2-\frac23(1-y)(1+y+y^2)
\right]-(2-3y)=0
$. This is a quadratic in $x^2$.
Writing it as $x^4a(y)-x^2b(y)-c(y)=0$,
where $a(y)=\frac23y^2(1-y)(1+y+y^2)+y^5$,
$b(y)=2y^2-\frac23(1-y)(1+y+y^2)$, and 
$c(y)=2-3y$,
the solution is
\begin{equation}
x^2=\frac{b(y)+\sqrt{b^2(y)+4a(y)c(y)}}
{2a(y)}\,,\quad\quad\quad 1< x^2\leq 3
\,.
\label{Ry=0x>1}
\end{equation}
Inverting Eq.~\eqref{Ry=0x>1} one obtains $y(x)$
for which $y_1=y_2$.
For $1< x^2\leq 3$ there are solutions
for $y>0$, i.e.,
$0\leq y<y_{\rm e}$,
which means in turn
$w>2$, i.e., 
for $2<w<\infty$.
For $w=\infty$, one has
$x^2=3$, and $y_1=y_2=0$.

Now, we find $y_1=y_2$, the coincident solution
for fixed $x$, and $w$ varying.
This is when $\frac{\partial w}{\partial y}=0$, see Eq.~\eqref{wy},
i.e., ${S}(x,y)=0$.
From Eq.~\eqref{defSbar} we have then that the
bifurcation points where $y_1=y_2$ for $x={\rm constant}$
are given by the equation
$
x^{4}[2y^{2}(1-y^{3})+3y^{5}]+2x^{2}(-y^{3}-3y^{2}+1)-6+9y=0
\label{defSbar=0x>1}
$,
i.e., 
$
x^{4}[2y^2(1-y^3)+3y^5]-x^{2}[2(y^3+3y^2-1)]-3(2-3y)=0
$. This is a quadratic in $x^2$, which can be written
as $x^4d(y)-x^2e(y)-f(y)=0$, with
$d(y)=y^5+2y^2>0$,
$e(y)=2(y^3+3y^2-1)$,
$f(y)=3(2-3y)$, and with solution
\begin{equation}
x^2=\frac{e(y)+
\sqrt{e^2(y)+ 4d(y)f(y)}
}
{2d(y)}\,,\quad\quad\quad 1< x^2\leq 3
\,.
\label{more=0x>1}
\end{equation}
One can show that the solution of
the quadratic equation with the minus sign before radical is
inconsistent with the condition $x>1$, so it is
not considered.
Inverting Eq.~\eqref{more=0x>1} one has $y(x)$
for which $y_1=y_2$.
To make progress and find an analytical result
we take $x\to 1$ and $y\to 1$.
The calculations are involved and we leave them to the
Appendix
\ref{fomulassectionvi},
the result is that the coincident solution takes the form
\begin{equation}
y_1=y_2= 1-\left(\frac38\left(x^2-1\right)^2\right)^\frac13\,,
\label{solutionapp11}
\end{equation}
for 
$x^2-1\ll1$.
Then, returning to the original variables one has,
$\frac{r_{+1}}{R}=\frac{r_{+2}}{R}= 1-\left(\frac38\,
(\Lambda R^2-1)^2\right)^{\frac13}$
for 
$\Lambda R^2-1$ tiny, which is
Eq.~\eqref{r+solutionslimitLambdaR2>1}.

\subsubsection{Synopsis}

When looking for roots $y=y(x,w)$
of the equation $w(x,y)=w$,
we have found
that, depending on the fixed
value of $w$, one can have
no roots, one root, or
two roots. 
The no root situation means that
the space inside the heat reservoir is
hot de Sitter space. 
The one root situation means
that there is one possible black
hole solution $y_1=y_2$
inside the reservoir.
The 
two roots situation means that 
there are two possible black
hole solutions 
inside the reservoir, one small $y_1$ and unstable,
the other large $y_2$ and stable. 
The case $x^2=1$ is important. 
When $x=1$ there is the sole solution
$y_1=y_2=1$ for $w=2$, but 
for any other $w$, $w>2$,  there are two solutions,
one is $y_1$, corresponding
to the small black hole, the other
is $y_2=1$,
corresponding to
the larger black holes, which
for $x^2=1$ have transubstantiated
into the Nariai cosmological universes
with all the allowed values of $w$, $w\geq2$.

\subsection{Further analysis: Cosmological horizons}

In the Schwarzschild-de Sitter space 
there is also the radius of the cosmological horizon $r_{\rm c}$.
However, since the radius of
the heat reservoir $R$ is inside $r_{\rm c}$, this latter has no
role in the thermodynamics. The cosmological horizon radius,
is just a parameter, which has only a coadjuvant role in the
whole setting. It is found once $r_+$ has been found from
the thermodynamics. 

The radius of the cosmological horizon $r_{\rm c}$ is the largest root
of the equation $V(r)=0$, see Eq.~\eqref{meteuc} for the expression of
$V(r)$.  As we have seen, it is related to the the black hole horizon
radius according to $r_{\rm c}^2+r_{\rm
c}r_++r_+^2-\frac{3}{\Lambda }=0$, whence $ r_{\rm c}=
-\frac{r_+}2+\frac{r_+}2\sqrt{\frac{12-3\Lambda r_+^2}{\Lambda
r_+^2}}$, see Eq.~\eqref{rcofr+Lambdaagain}, or $r_{\rm
c}=-\frac{r_+}{2}+\sqrt{\frac{3}{\Lambda }-\frac34 r_+^2}$.
Let us define 
\begin{equation}
u\equiv \frac{r_{\rm c}}{R}\,,
\label{udef}
\end{equation}
as the cosmological radius in units of $R$.
Then, from Eq.~\eqref{rcofr+Lambdaagain}
we have that $u$ is given in terms of $x$ and $y$
of Eqs.~\eqref{xdef} and \eqref{ydef}, respectively, 
by
\begin{equation}
u=-\frac{y}{2}+\sqrt{\frac{3}{x^2}-\frac{3}{4}y^2}\,.
\label{uequation}
\end{equation}
Calculating $\frac{du}{dy}$
one finds 
\begin{equation}
\frac{du}{dy}\leq0,
\label{dudy}
\end{equation}
i.e., $\frac{dr_{\rm c}}{dr_+}\leq0$.
This means that as
the horizon radius $r_+$ increases the
cosmological horizon decreases, in general.
We restrict here to calculate
some of the properties of the cosmological
radius $u$.

We want to calculate
the properties of $u$ at the neighborhood of the
point $x=1$, $y=1$, $u=1$, and find $\frac{du}{dx}$.
Since this pertains to the large black hole,
we use, as usual, the subscript 2 to refer
to that solution, see
Appendix~\ref{fomulassectionvi} for details.
One finds that
\begin{equation}
\left(\frac{du_2}{dx}\right)_{x=1_-}=
\left(\frac{dy_2}{dx}\right)_{x=1_+}
\label{change1}
\end{equation}
and 
\begin{equation}
\left(\frac{du_2}{dx}\right)_{x=1_+}=
\left(\frac{dy_2}{dx}\right)_{x=1_-}
\label{change2}
\end{equation}
where $1_-$ and $1_+$ mean that one is taking the
limit $x\to1$ from below, i.e., $x<1$,
and from above, 
i.e., $x>1$, respectively.
Using Eqs.~\eqref{dydxnew},
\eqref{dydxnewnew},
\eqref{change1}, and \eqref{change2} 
we can deduce some further
specific properties of the cosmological horizon.
They are that
$\left(\frac{du_2}{dx}\right)_{x=1_-}
\to-\infty$ for $w\to 2$,
and $\left(\frac{du_2}{dx}\right)_{x=1_+}\to -2$
for $w\to \infty$, and that
$\left(\frac{du_2}{dx}\right)_{x=1_-}
\to\infty$ for $w\to 2$,
and $\left(\frac{du_2}{dx}\right)_{x=1_+}\to 0$
for $w\to \infty$.

\begin{figure}[h]
\centering
{\includegraphics[width=.30\textwidth]{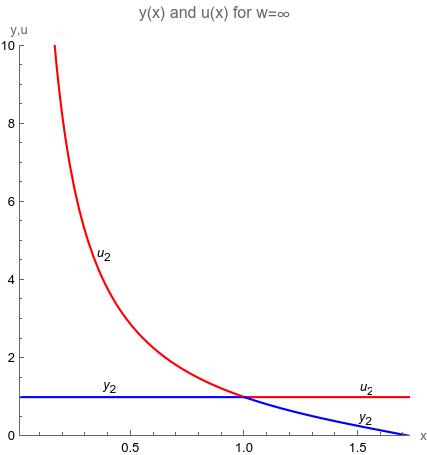}} 
\caption{
Plots of $\frac{r_{+2}}{R}$ and $\frac{r_{\rm c2}}{R}$ as a function
of $\sqrt{\Lambda R^2}$ for $RT=\infty$, i.e., essentially infinite
reservoir temperature.  For plotting purposes it is defined
$x\equiv\sqrt{\Lambda R^2}$, $y_2\equiv\frac{r_{+2}}{R}$,
$u_2\equiv\frac{r_{\rm c2}}{R}$, and $w\equiv 4\pi RT$. So this is the case
$w=\infty$.  Note that $y_2$ and $u_2$ exchange character at the
bifurcation point $x=1$.  See text for details.
}
\label{yofxuofxgloballabel}
\end{figure}

This interchange of equations
as displayed in Eqs.~\eqref{change1}
and \eqref{change2},
is
mostly clearly
seen in the case $w=\infty$, i.e., $RT\to \infty$.
Indeed, the case $w\to\infty$ can be solved exactly
as we have seen in Eq.~\eqref{uequation}.
Thus, for $0\leq x^2\leq 1$, one has $y_2=1$ and $u_2=
-\frac{1}{2}+\sqrt{\frac{3}{x^2}-
\frac{3}{4}}$, whereas for
$1\leq x^2\leq 3$, $y_2=
-\frac{1}{2}+\sqrt{\frac{3}{x^2}-
\frac{3}{4}}$ and $u_2=1$, see Fig.~\ref{yofxuofxgloballabel}.

In brief, the black hole horizon equation for
$\Lambda R^2<1$ turns into the cosmological horizon equation for
$\Lambda R^2>1$, and the cosmological horizon equation
for $\Lambda R^2<1$ turns into the black hole horizon equation for
$\Lambda R^2>1$. 

\vskip 8cm

\newpage

\section{Conclusions}
\label{conc}

We have taken on the problem of understanding the Schwarzschild-de
Sitter black hole thermodynamically.  For that we have put the black
hole in a cavity of radius $R$
surrounded by a heat reservoir and
kept at temperature $T$. This structure allowed the thermodynamic
problem to be solved within the canonical formalism and the Euclidean
path integral to quantum gravity.  Moreover, it led naturally into two
other spaces, namely, hot de Sitter and Nariai cosmological spaces,
besides the initial Schwarzschild-de Sitter black hole.

There are new results. One is that the range of the relevant parameter
$\Lambda R^2$ is extendable up to 3, i.e., $0\leq\Lambda R^2\leq3$.
It was also found that to properly treat the problem one has to divide
this range into three ranges, $0\leq \Lambda R^2<1$, $\Lambda R^2=1$,
$1< \Lambda R^2\leq 3$.

On the lower side of $\Lambda R^2$, $0\leq \Lambda R^2<1$, York's
solution for pure Schwarzschild is automatically incorporated when
$\Lambda R^2=0$, appearing first for $RT=\frac{\sqrt{27}}{8\pi}$, with
a coincident black hole horizon radius $r_{+1}=r_{+2}=\frac23 R$.  For
higher $\Lambda R^2$ the coincident black hole horizon radius gets
increased values for some lower $RT$.  A heuristic understanding of
this behavior has been given.  Changing the values and $\Lambda R^2$
and $RT$ one obtains either two thermodynamics solutions $r_{+1}$
small and $r_{+2}$ large, the first thermodynamically unstable and the
second stable, or no solutions in which case one is in the presence of
hot de Sitter.

At the intermediate value of $\Lambda R^2$, $\Lambda R^2=1$, the small
$r_{+1}$ unstable black hole exists. More interestingly, the large
$r_{+2}$ black hole is now the solution $r_{+2}=R$ and opens out into
a spectrum of a beautiful set of Nariai universes that can have all
temperatures in the range $\frac{1}{2\pi}\leq RT\leq\infty$.  The
ensemble data $R$ and $T$ now determine the location of the boundary
$Z$ in the Nariai universe rather than the black hole radius
solution. The Nariai universe is thermodynamically neutrally stable.

On the higher side of $\Lambda R^2$, $1<\Lambda R^2\leq 3$, unexpected
black hole solutions also arise.  The small $r_{+1}$ unstable black
hole exists without changing character.  The large $r_{+2}$ stable
solutions interchange the role of black hole $r_{+2}$ and cosmological
$r_{\rm c2}$ horizons, and the maximum value $r_{+2}$ can have is
attained for infinite temperature, i.e., $RT=\infty$, and is less than
one changing as $\Lambda R^2$ changes up to 3.  The case at the end of
the range, $\Lambda R^2=3$, only exists for infinite temperature, and
represents a reservoir filled with de Sitter space inside, except at
the center, where there is a black hole with zero horizon radius
$r_+=0$, i.e., a naked singularity.

Another important result is that increasing the temperature from zero,
i.e., increasing $RT$ from zero, one finds that the first solution
that appears for the whole range $0\leq\Lambda R^2\leq3$ has
temperature $RT=\frac1{2\pi}$ and is a solution with $\Lambda R^2=1$,
and radius $r_{+1}=r_{+2}=R$, i.e., it is the coincident Nariai
solution.  As one increases $RT$ solutions with lower and higher
$\Lambda R^2$ peal out.

Yet another interesting finding is related to phase transitions.  In
the $\Lambda R^2<1$ case, in particular for $\Lambda R^2\ll1$, the
possible thermodynamic phases that appear as one increases the
temperature from zero are hot de Sitter only, hot de Sitter favored in
relation to the Schwarzschild-de Sitter black hole, hot de Sitter
coexisting equally with the Schwarzschild-de Sitter black hole, the
Schwarzschild-de Sitter black hole favored in relation to hot de
Sitter, and the Schwarzschild-de Sitter black hole alone.  This latter
case comes out when one considers a high enough temperature so that
there is sufficient energy in the cavity to surpass the Buchdahl bound
and  presumably
the system collapses.  There are topology changes when the system
performs a phase transition from hot de Sitter to Schwarzschild-de
Sitter and vice versa as it is allowed in this formalism, since it is
in a semiclassical approximation to quantum gravity, and in quantum
gravity topology changes of the psce can happen.
In the $\Lambda R^2=1$ case, i.e., the Nariai universe, one has that
the possible thermodynamic phases that appear as one increases the
temperature from zero are hot de Sitter only, hot de Sitter favored in
relation to the Nariai universe, hot de Sitter coexisting equally with
the Nariai universe, the Nariai universe favored in relation to hot de
Sitter. Here there is no phase with only the Nariai universe.  There
are topology changes when the system performs a phase transition from
hot de Sitter to the Nariai universe and vice versa as it is allowed
in this formalism.
In the $\Lambda R^2>1$ case, phase transitions
between hot de Sitter and the Schwarzschild-de Sitter black hole,
can also be explored.

Thus, we have given a full
thermodynamic description of the Schwarzschild-de Sitter
black hole space in
a finite size cavity.

\begin{acknowledgments}
We thank conversations with Tiago Fernandes and Francisco Gandum
on the Euclidean path integral approach
to quantum gravity.
We acknowledge financial support from Funda\c c\~ao para a Ci\^encia e
Tecnologia - FCT through the project~No.~UIDB/00099/2020.
\end{acknowledgments}

\appendix

\renewcommand\thefigure{\thesection\arabic{figure}}

\section{The Schwarzschild-de Sitter and
Nariai spacetimes: Basics}
\label{appendix:sdsnbasics}

\setcounter{figure}{0}

\subsection{The Schwarzschild-de Sitter spacetime}

\begin{figure}[h]
\includegraphics*[scale=0.7]{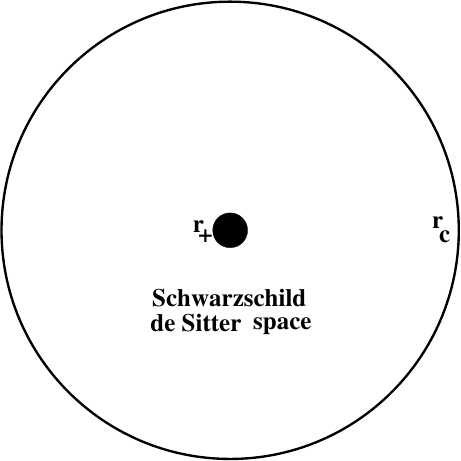}
\caption{
A drawing of the black hole with its event horizon $r_+$ and of the
cosmological horizon $r_{\rm c}$ in the Schwarzschild-de Sitter
spacetime. The topology of the 3-space is $R^3$.
}
\label{schwdesitterspace}
\end{figure}

The line element of
the Schwarzschild-de Sitter spacetime in spherical
coordinates $(t,r,\theta,\phi)$ is given by
\begin{equation}
ds^2=-V(r) \,
dt^2+\frac{dr^2}{V(r)}+r^2(d\theta^2+ \sin^2\theta
\,d\phi^2),
\label{met1}
\end{equation}
where
metric potential $V(r)$ has the form
\begin{equation}
V(r)=1-\frac{2m}{r}-\frac{\Lambda r^2}{3},
\label{met}
\end{equation}
with $m$ being the spacetime mass and $\Lambda$ the cosmological
constant which we consider positive, $\Lambda >0$,
see also Fig.~\ref{schwdesitterspace}.
The coordinate ranges are $-\infty< t<\infty$,
$r_+<r<r_{\rm c}$, $0\leq\theta\leq\pi$, and $0\leq\phi<2\pi$, where
$r_+$ and $r_{\rm c}$ are the black hole and cosmological
horizons of the spacetime, respectively.
The spacetime has  topology
$R^4$.
These coordinates can be further extended, e.g.,
through
a Kruskal-Szekeres extension, but it is not necessary here to do so.

The equation
$V(r)=0$
is
\begin{equation}
r-2m-\frac{\Lambda r^3}{3}=0,
\label{V=0}
\end{equation}
which 
can be written as
$m=\frac{r}{2}\left(1-\frac{\Lambda r^2}{3}\right)$.
Note that Eq.~\eqref{V=0}
has at most two positive real roots. When it has roots,
one
root corresponds to the black hole horizon
$r_+$, with  
\begin{equation}
r_+=r_+(m,\Lambda)\,.
\label{rooth}
\end{equation}
The radius of the black hole horizon obeys the inequality 
$
0\leq r_+\leq
\frac{1}{\sqrt{\Lambda}}
$.
The explicit form of $r_+(m,\Lambda),$ 
can be given since Eq.~\eqref{V=0} is a cubic equation
for $r$, but it is cumbersome and there is no need
to present it explicitly.
Given $r_+$, then
$V(r)$ in Eq.~\eqref{met}
can be written as 
$V(r)=1-\frac{r_+\left(1-\frac{\Lambda r_+^2}{3}\right)}{r}-
\frac{\Lambda r^2}{3}$
or
$V(r)
=1-\frac{r_+}{r}
-\frac{\Lambda}{3r}\left(r^3-r_+^3\right)$, where use of Eq.~\eqref{V=0}
has been made.
The other root
corresponds to the cosmological horizon $r_{\rm c}$, 
\begin{equation}
r_{\rm c}=r_{\rm c}(m,\Lambda),
\label{rootc}
\end{equation}
with $r_{\rm c}\geq r_+$.
The radius of the cosmological horizon obeys the inequality 
$
\frac{1}{\sqrt{\Lambda}}
\leq r_{\rm c}\leq
\sqrt{\frac{3}{\Lambda}}
$.
The explicit form of $r_{\rm c}(m,\Lambda),$ 
can be given since Eq.~\eqref{V=0} is a cubic equation
for $r$, but it is cumbersome and there is no need
to present it explicitly.

On the other hand, since there are two roots 
of Eq.~\eqref{V=0}, $r_+$ and
$r_{\rm c}$, one can write 
\begin{equation}
V(r)=\frac{\Lambda}{3r}
(r-r_+)(r_{\rm c}-r)(r+r_++r_{\rm c})\,,
\label{Vofrofrcr+}
\end{equation}
with 
\begin{equation}
r_{\rm c}^2+r_{\rm c}r_++r_+^2=\frac{3}{\Lambda}\,,
\label{forLambda}
\end{equation}
and
\begin{equation}
r_{\rm c}r_+(r_{\rm c}+r_+)=\frac{6m}{\Lambda}\,.
\label{form}
\end{equation}
Thus $\Lambda$ and $m$ can be swapped for $r_+$ and $r_{\rm c}$.
Moreover, Eq.~\eqref{forLambda} can be written as $r_{\rm c}^2+
r_{\rm c}r_++r_+^2-\frac{3}{\Lambda}=0$ which is a quadratic either
for $r_{\rm c}$ in terms of $r_+$ or vice versa. The solution is
\begin{equation}
r_{\rm c}=
-\frac{r_+}2+\frac{r_+}2
\sqrt{\frac{12-3\Lambda r_+^2}{\Lambda r_+^2}}\,,
\label{rcofr+Lambda}
\end{equation}
or $r_{\rm c}=
-\frac{r_+}2+\sqrt{\frac{3}{\Lambda}-\frac34 r_+^2}$.
Of course, the equation $r_{\rm
c}^2+r_{\rm c}r_++r_+^2-\frac{3}{\Lambda}=0$ is also a quadratic for
$r_+$ which gives $r_+$ in terms of $r_{\rm c}$ in the same form of
Eq.~\eqref{rcofr+Lambda} with the roles reversed.
Another way of obtaining this is that
since there are two solutions of Eq.~\eqref{V=0}, $r_+$ and
$r_{\rm c}$, one has from Eq.~\eqref{V=0} that $r_{\rm
c}-2m-\frac{\Lambda r_{\rm c}^3}{3}=0$ and $r_+-2m-\frac{\Lambda
r_+^3}{3}=0$.  Subtracting one equation from the other one eliminates
$2m$ to get $(r_{\rm c}-r_+) -\frac{\Lambda}{3} \left( r_{\rm
c}^3-r_+^3\right)=0$.  Since $\left( r_{\rm c}^3-r_+^3\right)= (r_{\rm
c}-r_+)(r_{\rm c}^2+r_{\rm c}r_++r_+^2)$, one finds that
$1-\frac{\Lambda}{3}(r_{\rm c}^2+r_{\rm c}r_++r_+^2)=0$, i.e., $r_{\rm
c}^2+r_{\rm c}r_++r_+^2-\frac{3}{\Lambda}=0$.  One can then write the
solution of $r_{\rm c}$ in term of $r_+$ as
$r_{\rm c}= -\frac{r_+}2+\sqrt{\frac{3}{\Lambda}-\frac34 r_+^2}$,
which is Eq.~\eqref{rcofr+Lambda}.  In addition, Eq.~\eqref{form} with
the help of Eq.~\eqref{forLambda} can be written as $r_{\rm c}^2+
r_{\rm c}r_+-\frac{2mr_+^2}{r_+-2m}=0$, which is a quadratic either
for $r_{\rm c}$ in terms of $r_+$ or vice versa.  The solution is
\begin{equation}
r_{\rm c}=
-\frac{r_+}2
+\frac{r_+}2\sqrt{\frac{r_++6m}{r_+-2m}}\,,
\label{rcofr+m}
\end{equation}
or $r_{\rm c}= \frac{r_+}2\left(
\sqrt{\frac{r_++6m}{r_+-2m}}-1\right)$.
Another way of obtaining this is that
since there are two solutions of Eq.~\eqref{V=0}, $r_+$ and
$r_{\rm c}$, one has from Eq.~\eqref{V=0} that
$r_{\rm
c}r_+^3-2mr_+^3-\frac{\Lambda r_{\rm c}^3r_+^3}{3}=0$ and
$r_+r_{\rm c}^3-2mr_{\rm c}^3-\frac{\Lambda
r_+^3r_{\rm c}^3}{3}=0$.  Subtracting one equation from
the other one eliminates
$\Lambda$ to get
$-r_+r_{\rm c}(r_{\rm c}^2-r_+^2)+2m(r_{\rm c}^3-r_+^3)=0$.
Thus,
$-r_+r_{\rm c}(r_{\rm c}-r_+)(r_{\rm c}+r_+)+2m
(r_{\rm c}-r_+)
(r_{\rm c}^2+r_{\rm c}r_+
+r_+^2)=0$, i.e.,
$-r_+r_{\rm c}(r_{\rm c}+r_+)+2m
(r_{\rm c}^2+r_{\rm c}r_+
+r_+^2)=0$. So, $r_{\rm c}^2+
r_{\rm c}r_+-\frac{2mr_+^2}{r_+-2m}=0$, with solution 
$r_{\rm c}=
-\frac{r_+}2
+\frac{r_+}2\sqrt{\frac{r_++6m}{r_+-2m}}$
which is Eq.~\eqref{rcofr+m}.

Note that $r_{\rm c}=r_+$ for $r_{\rm
c}=r_+=\frac{1}{\sqrt\Lambda}=3m\label{extr2x}$, and this happens when
$9m^2\Lambda =1\label{extr}$. In this limit one can have either an
extremal Schwarzschild-de Sitter spacetime, where the two regions off
the extremal horizon, the inside and the outside regions, are time
dependent, or the Nariai spacetime where the topology of the space
changes, see below.  For $9m^2\Lambda >1$ there are no horizons, the
spacetime is asymptotically de Sitter, with a massive naked
singularity at the center.

\subsection{The Nariai spacetime}

\begin{figure}[h]
\includegraphics*[scale=0.9]{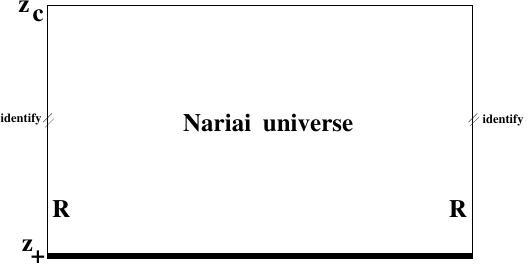}
\caption{
A drawing of the Nariai universe with its event horizon $z_+$ and its
cosmological horizon $z_{\rm c}$.  The cylindrical character of the
Nariai space is made clear after identifying the two vertical end lines
of the diagram.  The radius $R$ is the radius of the cylinder.  The
labelling of the two horizons $z_+$ and $z_{\rm c}$ is convention as
they are of the same type. The topology of the
3-space is $R\times S^2$.
}
\label{nariaispace}
\end{figure}

The line element of
the Nariai spacetime in spherical
coordinates $(t,{z},\theta,\phi)$,  
is given by
\begin{equation}
ds^2=-V({z}) \,
dt^2+\frac{d{z}^2}{V({z})}+
\frac1\Lambda\left(d\theta^2+ \sin^2\theta
\,d\phi^2\right),
\label{metneriai1}
\end{equation}
where
the metric potential $V({z})$ has the form
\begin{equation}
V({z})=1-\Lambda {z}^2,
\label{metnariai}
\end{equation}
with $\Lambda$ being the cosmological
constant which we consider positive, $\Lambda >0$,
see also Fig.~\ref{nariaispace}.
The coordinate ranges are $-\infty< t<\infty$,
${z}_+<{z}<{z}_{\rm c}$,
where
${z}_+$ and ${z}_{\rm c}$ are the horizons of the spacetime,
$0\leq\theta\leq\pi$, and $0\leq\phi<2\pi$.
Note that the spacetime has topology
$R^2\times S^2$.
These coordinates can be further extended, e.g.,
through
a Kruskal-Szekeres extension, but it is not necessary here to do so.

The equation
$V({z})=0$ is
\begin{equation}
1=\Lambda {z}^2.
\label{V=0nariai}
\end{equation}
In general Eq.~\eqref{V=0nariai}
has two  roots. One
root corresponds to the black hole horizon
${z}_+$, by convention, with  
\begin{equation}
{z}_+=-\frac{1}{\sqrt\Lambda}\,.
\label{roothnariai}
\end{equation}
The other root
corresponds to the cosmological horizon ${z}_{\rm c}$,
by convention, with
\begin{equation}
{z}_{\rm c}=+\frac{1}{\sqrt\Lambda},
\label{rootcnariai}
\end{equation}
with ${z}_{\rm c}\geq {z}_+$.

Note that when 
\begin{equation}
\Lambda =\infty\,,
\label{extrnariai}
\end{equation}
both roots coincide, 
\begin{equation}
{z}_+={z}_{\rm c}=0,
\label{extr2xnariai}
\end{equation}
and in this extremal limit the spacetime disappears.

Note also that when
$\Lambda =0$,
the $z_+$ and $z_{\rm c}$ have roots
\begin{equation}
z_+=-\infty\,,\quad
z_{\rm c}=+\infty\,,
\label{minknariai}
\end{equation}
and in this limit there are 
no horizons, one is in the presence
of simply a Minkowski space
with two coordinates possibly  wrapped
around, i.e., one can have
a torus, a cylinder, or an infinite
plane. 
This can be seen from the angular part of
Eq.~\eqref{metneriai1}
by doing $\theta=\sqrt{\Lambda}\,x$, 
$\phi=\sqrt{\Lambda}\,y$, and then doing
$\Lambda\to0$ to give 
$ds^2=-
dt^2+dz^2
+dx^2+dy^2$,
where the range of the coordinates
$x$ and $y$ is
$0\leq x\leq x_1$ and 
$0\leq y\leq y_1$, respectively, with 
$x_1$ and $y_1$ having any value one chooses from
a finite value to infinite.

\section{The Nariai limit
from the Schwarzschild-de Sitter
space in a thermodynamic setting}
\label{nariailimitapp}

\setcounter{figure}{0}

\begin{widetext}

\begin{figure}[h]
\includegraphics*[scale=0.9]{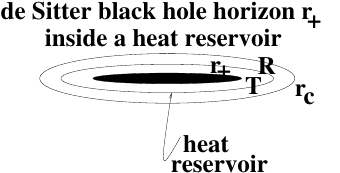}
\hskip 3cm
\includegraphics*[scale=0.9]{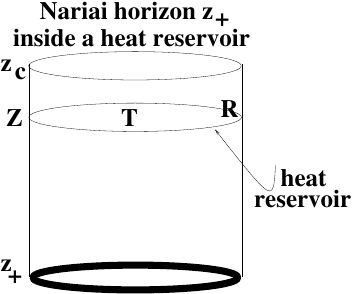}
\caption{
Drawings of a black hole horizon inside a heat reservoir in the
Schwarzschild-de Sitter black hole space on the left,
and of the Nariai
universe with one of its horizons
inside a heat reservoir
on the right.  Specifically, on the left the drawing
shows a slant view of a $t={\rm constant}$ and
$\theta={\rm constant}$ space
of the black hole horizon
$r_+$ inside a heat reservoir at temperature $T$ and radius $R$ in the
Schwarzschild-de Sitter geometry. Outside $R$ there is the
cosmological horizon $r_{\rm c}$.  The Euclideanized space and its
boundary have $R^2 \times S^2$ and $S^1 \times S^2$ topologies,
respectively, with the $S^1$ subspace having proper length
$\beta=\frac1T$.  On the right, the drawing
shows a slant view of a $t={\rm constant}$ and
$\theta={\rm constant}$ space
of 
the horizon $z_+$ inside a heat reservoir at
temperature $T$, with cylindrical radius $R$, and situated at $Z$ in
the Nariai universe geometry.  Outside $Z$ there is the cosmological
horizon $z_{\rm c}$.  The Euclideanized space and its boundary have
$R^2 \times S^2$ and $S^1 \times S^2$ topologies, respectively, with
the $S^1$ subspace having proper length $\beta=\frac1T$.
In pictorial terms it is clear how the Schwarzschild-de
Sitter black hole space originates the Nariai universe, with the slant
view of the Schwarzschild-de Sitter space helping in the visualization
of the process. Indeed, if the two Schwarzschild-de Sitter different
horizon radii, $r_+$ and $r_{\rm c}$, are squeezed into the heat
reservoir radius $R$, then the two horizons pinch off to form the
Nariai universe with the heat reservoir still at temperature $T$,
with cylindrical radius $R$,
and situated now at some $Z$ 
in-between
the two displaced distinct horizons $z_+$ and $z_{\rm c}$.  See text
for more details.
}
\label{nariai1bhapp}
\end{figure}

\end{widetext}

In order to understand the limiting thermodynamic process of obtaining
a Nariai space from a Schwarzschild-de Sitter space in the limit
$\Lambda R^2=1$, $\frac{r_+}{R}=1$, and $\frac{r_{\rm c}}{R}=1$, it is
useful to resort to Fig.~\ref{nariai1bhapp}.

To be complete we give again the Schwarzschild-de Sitter Euclidean
line element
\begin{align}
ds^2=V(r) \,
dt^2+&\frac{dr^2}{V(r)}+r^2(d\theta^2+ \sin^2\theta
\,d\phi^2),\nonumber\\
&0\leq t<\beta^{\rm H}_+\,, \, r_+\leq r\leq R\,,
\label{appmet1euc+}
\end{align}
where the metric potential $V(r)$
has the form
\begin{equation}
V(r)=1-\frac{2m}{r}-\frac{\Lambda r^2}{3}\,.
\label{appmeteuc}
\end{equation}
Note  that
the range of coordinates of Euclidean time
$t$ is $0\leq t<\beta^{\rm H}_+$, where
$\beta^{\rm H}_+$ is the period
of the coordinate such that the line element
given by Eqs.~\eqref{appmet1euc+} and \eqref{appmeteuc}
has no conical singularities.
The relation with the Hawking temperature $T^{\rm H}_+$
is
$\beta^{\rm H}_+=\frac1{T^{\rm H}_+}$.
In addition,
due to the reservoir at radius $R$ the range
of coordinates of
$r$ is now 
$r_+\leq r\leq R$.
Note now that this is Euclidean space and
the topology of this space is
$R^2\times S^2$.
There are two horizon for $V(r)=0$, see Eq.~\eqref{appmeteuc}, 
the black hole horizon $r_+$ and the cosmological horizon $r_{\rm c}$.

The local Tolman temperature $T$
of the heat reservoir at $R$
has to be treated with care. It is
\begin{equation}
T=\frac{T^{\rm H}_+}{\sqrt{V(R)}}\,,
\label{apptn}
\end{equation}
with $T^{\rm H}_+$ being the Hawking temperature given by
$
T^{\rm H}_+=\frac{\kappa_+}{2\pi }
$,  and
$\kappa_+$ being
the surface gravity of the black hole
horizon and $V(R)$ is the potential at $R$.
For the line element of Eq.~\eqref{appmet1euc+} 
one has $\kappa_+ =\frac{1}{2}{V^{\prime}(r_{ +})}$,
where a prime means derivative with respect
to $r$, and
so the Hawking temperature for the black hole is $T^{
\rm H}_+=\frac{1}{4\pi} \left(\frac{dV}{dr}\right)_{r_{
+}}$. Using Eq.~\eqref{appmeteuc} we get
\begin{equation}
T^{\rm H}_+=
\frac{1}{4\pi r_+} \left(1-\Lambda r_+^2\right).
\label{apptbh}
\end{equation}
The potential at $R$ is
\begin{equation}
V(R)=1-\frac{2m}{R}-\frac{\Lambda R^2}{3}\,.
\label{appmeteucatR}
\end{equation}


The Nariai solution can be found
from the  Schwarzschild-de Sitter solution in the
limit that 
the two horizons 
$r_+$ and $r_{\rm c}$ coincide.
Here, we have a heat reservoir
at $R$ that acts for the inside region
which is a cavity with 
the black hole. This heat reservoir
at $R$ is in-between 
$r_+$ and $r_{\rm c}$, and
thus the limit is such that 
$r_+$, $R$, and $r_{\rm c}$ coincide.
Now, 
if we do 
$r_+\to R$ then
 Eq.~\eqref{apptn}, 
$T=\frac{T^{\rm H}_+}{\sqrt{V(R)}}$,
gives at face value that the heat reservoir
is at very high temperature since $V(R)\to0$.
But, there is a way to have $T$ finite
with 
$r_+\to R$, and this is 
to do concomitantly $T^{\rm H}_+\to
0$.
Since
$T^{\rm H}_+=\frac{1}{4\pi r_+}\left(1-\Lambda
r_+^2\right) $, see Eq.~\eqref{apptbh},
$T^{\rm H}_+\to
0$  means 
$1-\Lambda
r_+^2\to0$.
Since $r_+\to R$, we put,
\begin{equation}
\frac{r_+}{R}=1-\varepsilon \,,
\label{appRr+varep}
\end{equation}
$\varepsilon\ll1$,
which implies
\begin{equation}
\sqrt{\Lambda R^2}=1-\delta\,,
\label{delta}
\end{equation}
where
$\delta\equiv \frac12(1-\Lambda r_+^2)-\varepsilon$,
$\delta\ll1$, and $\frac{\varepsilon}{\delta}$
is of order one.
Thus, $R$ and
the length scale $\frac1{\sqrt\Lambda}$ are equal at zeroth order.
Since $r_+\to R\to \frac1{\sqrt\Lambda}$ then
from Eq.~\eqref{rcofr+Lambdaagain}
one has
$\frac{r_{\rm c}}{R}\to1$, i.e., one has
$r_{\rm c}\to \frac1{\sqrt\Lambda}$, and since 
$\frac1{\sqrt\Lambda}\to R$, then $r_{\rm c}\to R$.
From the expansion of Eqs.~\eqref{appRr+varep}
and \eqref{delta}
one has that 
Eq.~\eqref{apptbh} gives
\begin{equation}
T^{\rm H}_+
=\frac{\delta +\varepsilon }{2\pi R}\,,
\label{hawkingTtinydelta}
\end{equation}
in first order, as required. To
understand the behavior of $V(r)$ near $R$ in this limit
we use Eq.~\eqref{appmeteucatR}. It gives
\begin{equation}
V(R)=\varepsilon\left(\varepsilon+2\delta\right),
\label{Vofdeltavarepsilon}
\end{equation}
in first order.
Thus, Eq.~\eqref{apptn}, i.e., $T=\frac{T^{\rm H}_+}{\sqrt{V(R)}}$
gives
\begin{equation}
T=\frac{1}{2\pi R}\frac{\frac{\varepsilon}{\delta}+1}
{
\sqrt{\frac{\varepsilon}{\delta}\left(
\frac{\varepsilon}{\delta}+2\right)
}
}
\,,
\label{Tvarepsilondelta}
\end{equation}
which given a $RT$, i.e., given a $T$,
implies some $\frac{\varepsilon}{\delta}$, and
so is finite and consistent.

But we have not finished. The metric potential $V(R)$
given in Eq.~\eqref{appmet1euc+}
vanishes in these coordinates, and the line element 
of Eq.~\eqref{appmet1euc+} looses sense.
So, we have to pay attention to this limit indeed
with care.
It is the Nariai limit with a reservoir $R$
in the middle, see Fig.~\eqref{nariai1bhapp}, an
interesting case.
Expanding the metric potential $V(r)$ near $r_+$ in a Taylor series
gives $V(r)=\left(\frac{dV}{dr}\right)_{r=r_+}
\left(r-r_+\right)+\frac12 \left(\frac{d^2V}{dr^2}\right)_{r=r_+}
\left(r-r_+\right)^2$ plus higher order terms.  Recall that
$\left(\frac{dV}{dr}\right)_{r=r_+}=4 \pi T^{\rm H}_+$ and find
$\left(\frac{d^2V}{dr^2}\right)_{r=r_+}=-\frac{2}{r_+^2}$.  Now, in
this limit $r_+=R$, so
$\left(\frac{d^2V}{dr^2}\right)_{r=r_+}=-\frac{2}{R^2}$.  So,
\begin{equation}
V(r)= 4\pi T^{\rm H}_+(r-r_+)-\frac{1}{R^2}\left(r-r_+\right)^2\,,
\label{Vofrfornariai}
\end{equation}
plus higher order terms.  Make the
coordinate transformations $(t,r)\to(\bar t,
{\bar z})$ as
\begin{equation}
r-r_+=4\pi T^{\rm H}_+R^2
\sin^2\left(\frac12\frac{{\bar z}}{R}\right)\,,\quad\quad
t=\frac{\bar t}{2\pi T^{\rm H}_+R}\,,
\label{transf1}
\end{equation}
with $0\leq t\leq\frac{1}{T^{\rm H}_+}$
corresponding to
 $0\leq {\bar t}\leq2\pi R$, and
$r_+\leq r\leq R$ corresponding
to $0\leq\bar z\leq \bar Z$.
Then obtain from Eq.~\eqref{appmeteuc} that
\begin{equation}
V(r)=V(r-r_+) =V(\bar z)=(2\pi T^{\rm H}_+ R)^2\sin^2
\left(\frac{{\bar z}}{R}\right)\,,
\label{newVofr}
\end{equation}
and from Eq.~\eqref{appmet1euc+}
obtain the line element,
\begin{equation}
ds^2=\sin^2\left(\frac{{\bar z}}{R}\right)
\,d{\bar t}^2+d{\bar z}^2+ R^2d\Omega^2 \,,
\label{Nariai1appnewVofr}
\end{equation}
which is a form
of the Nariai line element.
Note now from Eq.~\eqref{newVofr} that
$
V(R)=V(R-r_+) =V(\bar Z)=(2\pi T^{\rm H}_+ R)^2\sin^2
\left(\frac{{\bar Z}}{R}\right)
$,
and so from Eq.~\eqref{apptn}, i.e.,
$
T=\frac{T^{\rm H}_+}{\sqrt{V(R)}}
$,
one has
\begin{equation}
T=\frac{1}{2\pi R\sin \left(\frac{{\bar Z}}{R}\right)} \,.
\label{NewTNariai1}
\end{equation}
This means that the ensemble boundary values $T$ and $R$ specify
automatically the maximum value for ${\bar z}$, namely ${\bar Z}$.
So, for the ensemble with boundary data one has
$0<t<2\pi R$, $0\leq {\bar z}\leq {\bar Z}$, $0\leq \theta\leq
\pi$, $0\leq\phi<2\pi$.  In turn $0\leq {\bar Z}\leq R \pi$.  Note
that ${\bar z}=0$ and ${\bar z}=R \pi$ are horizons.

We can continue further. 
Let us make another coordinate transformation,
$z=R\cos\frac{\bar z}{R}$, Then 
\begin{equation}
ds^2=V(z)
\,dt^2+\dfrac{dz^2}{V(z)}+ R^2d\Omega^2 \,,
\label{nariai0newzapp}
\end{equation}
where we have dropped the bar in $\bar t$.
Then, the metric potential $V(z)$ is 
\begin{equation}
V=1-\frac{z^2}{R^2} \,,
\label{Vofz}
\end{equation}
the one given in
Eq.~\eqref{metnariai}.
So, for the ensemble with boundary data one has
$0<t<2\pi R$, $-R<z<Z$,
$0\leq \theta\leq \pi$, $0\leq\phi<2\pi$.
In turn $-R\leq Z\leq R$.
Note that $z=-R$ and $z=R$ are horizons.
The line element given in Eq.~\eqref{nariai0newzapp}
and \eqref{Vofz}
corresponds to a two-dimensional de Sitter
space times a sphere,
the topology is $R^2\times S^2$. It has two horizons,
we label $z_+=-R$
and $z_{\rm c}=R$, but now they are
just labels, since the two horizons
have the same character.

We have that the
heat reservoir temperature
$T$ is now 
\begin{equation}
T=\frac{T^{\rm H}_{\rm +\, Nariai}}{\sqrt{V(Z)}}\,,
\label{apptnnariai}
\end{equation}
with $T^{\rm H}_{\rm +\, Nariai}$
being the Hawking temperature given by
$
T^{\rm H}_{\rm +\, Nariai}=\frac{\kappa_+}{2\pi}$, and
$\kappa_+$ being
the surface gravity of the black hole
horizon.  For the metric given in Eq.~(\ref{nariai0newzapp})
one has $\kappa_+ =\frac{1}{2}{V^{\prime}(z_+)}$,
and
so the Hawking temperature for + horizon is
$T^{\rm H}_{\rm +\, Nariai}=\frac{1}{4\pi}
\left(\frac{dV}{dz}\right)_{z_+}$. Using Eqs.~\eqref{nariai0newzapp}
and \eqref{Vofz}
we get
\begin{equation}
T^{\rm H}_{\rm +\, Nariai}=\frac{1}{2\pi R}.  \label{appTnariaibh2}
\end{equation}
So, from Eqs.~\eqref{apptnnariai} and \eqref{appTnariaibh2}
one finds
\begin{equation}
T=\frac{1}{2\pi R \sqrt{1-\frac{Z^2}{R^2}}}\,,
\label{appTnariaibh2explicit}
\end{equation}
where $Z$ is the boundary of $Z$.
For a given $T$ at the boundary, for the ensemble, there are two
solutions of Eq.~\eqref{appTnariaibh2explicit}
in general, namely,
one for $Z$ between $-R$ and 0 and the other
for $Z$ between 0 and $R$. These two solutions
may be thought of yielding
different physical situations. Note also 
that in Schwarzschild-de Sitter the two solutions
were for the horizon $r_+$, one small $r_{+1}$
the other large $r_{+2}$, here the two solutions
are not for the horizons but instead for the
boundary $Z$, one $Z_1$, the other
$Z_2$, with
\begin{equation}
Z_1=-R\sqrt{1-\frac{1}{(2\pi RT)^2}} \,,
\label{appZ1}
\end{equation}
and
\begin{equation}
Z_2=R\sqrt{1-\frac{1}{(2\pi RT)^2}} \,,
\label{appZ2}
\end{equation}
We are free to choose where to put the boundary,
we have two choices, either $Z_1$ or $Z_2$,
noting that $z_+$ and $z_{\rm c}$ can also
be exchanged if one wants. Exchanging
$Z_1$ with $Z_2$ and concomitantly
exchanging $z_+$ with $z_{\rm c}$ 
reverses to the original situation.
The boundary
is at $Z_1$ or $Z_2$, has
cylindrical radius $R$, and acts
as a reservoir for the inner region,
i.e.,
$z\leq Z_1$ or $z\leq Z_2$
depending on where we choose to put the boundary.
Note, in addition that the 
reservoir instead of being two-dimensional with
space topology $S^2$ as in the de Sitter space, it is
still two dimensional
but now with topology $R\times S^1$, i.e., it is a
cylinder.

From Eqs.~\eqref{appZ1} and \eqref{appZ2} 
we see that for
\begin{equation}
RT<\frac1{2\pi} \,,
\label{appnariainosolutionnew}
\end{equation}
there are no solutions for $Z_1$ or $Z_2$, so in this case the
boundary in $Z$ does not exist, and so there is no
Nariai solution for the
thermodynamic problem in the canonical ensemble.
The reason is clear if one thinks in thermal wavelengths,
indeed when the 
temperatures are relatively very small,
the associated thermal wavelengths are very
long and there is no boundary Z that can accommodate
them.

From Eqs.~\eqref{appZ1} and \eqref{appZ2} 
we see that for
\begin{equation}
RT\geq\frac1{2\pi} \,,
\label{appnariaisolutionnew}
\end{equation}
there are two Nariai solutions, one with one horizon
$z=-R$ and boundary $Z_1$, the other 
with with one horizon
$z=-R$ and boundary $Z_2$, both boundaries can be picked up.
When the equality sign holds 
there is one solution
with $Z_1=Z_2=0$, so in this case the
boundary in $Z$ pops up in the middle, 
at $z=0$, and so $-R\leq z\leq0$.
The reservoir is at $Z_1=Z_2=0$ and has
cylindrical radius $R$
and the horizon is at $z=-R$.

Three comments are in order.
The first comment is that the resulting
metric inside
the reservoir is
in this case described by the Nariai universe
as we have seen. The
procedure of obtaining the Nariai metric as a limit from the
Schwarzschild-de Sitter metric is known and has different versions,
e.g., see \cite{gp,diaslemoscardosonariaiinhigherd}. However, we have
done it here in a completely different manner and in a completely
different context that is related naturally to thermodynamics.  The
procedure we used is similar to that described in \cite{prd97} for the
extremal limit of nonextremal electrically charged black holes.  Here,
in Eq.~\eqref{Vofrfornariai}, the potential expansion takes the form
$V(r)= 4\pi T^{\rm H}_+(r-r_+)-\frac{1}{R^2}\left(r-r_+\right)^2$
instead of $V(r)= 4\pi T^{\rm H}_+(r-r_+)+
\frac{1}{R^2}\left(r-r_+\right)^2$, i.e., a plus instead of the minus
sign of \cite{prd97} appears in the second term of the expansion.  As
a result, we have arrived at the Nariai metric instead of
obtaining the
Bertotti-Robinson metric as in \cite{prd97}.
The second comment, is that in the above considerations, we have
discussed two completely different limits, one is the high-temperature
limit and the other is the extremal Nariai limit. However, they can
be combined if
the boundary $Z_1\to -R$. 
The third comment is that we have done the Nariai limit from below,
i.e., from $\Lambda R^2<1$.  If we do the limit $\Lambda R^2\to1$ from
above, i.e., from $\Lambda R^2>1$, we get the same result, namely, the
Nariai universe inside a heat reservoir in the canonical ensemble.

\section{Deduction of formulas of Sec.~\ref{diagramsandanalysis}}
\label{fomulassectionvi}

\vskip 0.5cm

Here, we deduce in detail some equations found in
Sec.~\ref{diagramsandanalysis}.
Specifically, we want to deduce
Eqs.~\eqref{dydxnew},
\eqref{solutionapp0}, \eqref{dydxnewnew},
and \eqref{solutionapp11}, and study
Eqs.~\eqref{change1}
and \eqref{change2}.

For the derivation of
Eqs.~\eqref{dydxnew},
\eqref{solutionapp0}, \eqref{dydxnewnew},
and \eqref{solutionapp11}
it is convenient to
shorten the notation and define the variables 
$\sqrt{\Lambda R^2}$ and $\frac{r_+}{R}$
as 
$x$ and $y$ variables, namely, 
\begin{equation}
x\equiv\sqrt{\Lambda R^2} \,,
\label{xdef1}
\end{equation}
\begin{equation}
y\equiv\frac{r_+}{R} \,.
\label{ydef1}
\end{equation}
With these definitions we want to find 
Eqs.~\eqref{dydxnew}
and ~\eqref{solutionapp0} of the $x^2<1$ case,
and Eqs.~\eqref{dydxnewnew}
and ~\eqref{solutionapp11} of the $x^2>1$ case.

\vskip 0.5cm

\noindent
{\bf $x^2{\mathbf <1}$:}

\noindent
Let us deduce Eq.~\eqref{dydxnew}.  We have to find
the behavior of $y_2$, i.e.,
$r_{+2}$, at $x^2$ near 1,
i.e., $\Lambda R^2$ near 1.
From Eq.~\eqref{zofxy}, or from Eq.~\eqref{y1}, one finds
that there is a solution $y_2(x)$ for each $x$.  The maximum of the
curve is at $x=1$ and $y_2=1$, and the derivative there is not well
defined.  So, it is important to discuss the behavior and the
properties of $y_2$ when $x\to 1$.  For that we define infinitesimal
quantities $\delta$ and $\varepsilon$, related to $x$ and $y$,
respectively, by
\begin{equation}
x=1-\delta\,,\quad\quad\quad y=1-\varepsilon\,,
\label{appdeltadef}
\end{equation}
with $\delta\ll1$ and $\varepsilon\ll1$, both positive, and with
$\frac{\varepsilon}{\delta}$ being a quantity of order one.  Indeed,
$\delta>0$ for $x<1$, and $\varepsilon\geq0$ always.
Then, from Eq.~\eqref{zofxy} one gets
\begin{equation}
w= \frac{2(\frac{\varepsilon}{\delta}+1)}{\sqrt{
\frac{\varepsilon}{\delta}
(\frac{\varepsilon}{\delta}+2)}}\,,
\label{appnewz}
\end{equation}
valid in first order, in the order we are working.  One sees that for
$w$ finite $\frac{\varepsilon}{\delta}$ is finite, so generically here
in this calculation $\varepsilon$ is
indeed of the order $\delta$.  From
Eq.~\eqref{dydx} one also gets
\begin{equation}
\frac{dy}{dx}= \frac{\varepsilon}{\delta}\,,
\label{appdydxmu}
\end{equation}
also valid in the order we are working.  We obtain from
Eq.~\eqref{appnewz} that $\frac{\varepsilon}{\delta}
=\frac{w}{\sqrt{w^2-4}}-1$ with $w\geq2$.  Thus, since here
$\frac{dy}{dx}=\frac{\varepsilon}{\delta}$, 
Eq.~\eqref{appdydxmu}, one finds  that
\begin{equation}
\frac{dy}{dx}
=\frac{w}{\sqrt{w^2-4}}-1\,,\quad\quad w\geq2\,,
\label{appdydxnew}
\end{equation}
which is Eq.~\eqref{dydxnew}, the equation we wanted to deduce.

Let us deduce now
Eq.~\eqref{solutionapp0}.
We want to find $y_1=y_2$, the coincident solution
for fixed $x$, and $w$ varying.
This is when $\frac{d w}{d y}=0$, see Eq.~\eqref{wy},
i.e., ${S}(x,y)=0$.
From Eq.~\eqref{defSbar} we have then that the
bifurcation points where $y_1=y_2$ for $x={\rm constant}$
are given by 
Eq.~\eqref{more=0}.
We are interested in finding $y_1=y_2$ when $x\to1$.
Instead of working with Eq.~\eqref{more=0}
we work with 
Eq.~\eqref{defSbar} which is more direct.
We find
from
$x=1-\delta$
of Eq.~\eqref{appdeltadef}, that Eq.~\eqref{defSbar} in this limit is
${S}=(y-1)^{3}(y^2+3y+4)-4\delta
(y-1)^2(y+1)(1+y+y^2)+4\delta^2(2y^2+y^{5})$.
In addition, from $y=1-\varepsilon$
of Eq.~\eqref{appdeltadef},
Eq.~\eqref{defSbar} becomes then
$
{S}=12\delta^2
-24\delta\varepsilon^2-8\varepsilon^3
\label{appnewdefS0}
$.
As we have seen, the coincident
root is given
when $\frac{d w}{dy}=0$, i.e.,
${S}=0$. Then, 
$12\delta^2
-24\delta\varepsilon^2-8\varepsilon^3=0$, i.e.,
$\delta^2
-2\delta\varepsilon^2-\frac23\varepsilon^3=0$.
This is a quadratic in $\delta$, with solution given by
$\delta=\sqrt{
\frac23\varepsilon^3+\varepsilon^4}+\varepsilon^2$, i.e.
$\delta
=\varepsilon^{\frac32}\sqrt{\frac23+
\varepsilon
}+\varepsilon^2
$.
Since $\varepsilon\ll1$ the dominant term gives
\begin{equation}
\delta=\left(\frac23\varepsilon^3\right)^\frac12\,.
\label{appsolutionvarepbardelta}
\end{equation}
Thus, $\delta$ goes with $\varepsilon^\frac32$
or, if one prefers,  $\varepsilon$ goes with $\delta^\frac23$,
indeed $\varepsilon=\left(\frac32\delta^2\right)^\frac13$.
Recall that 
$x=1-\delta$, and so $x^2=1-2\delta$ in this
approximation, and that 
$y=1-\varepsilon$.
Then, substituting $\delta$ for
$x^2$ and $\varepsilon$ for $y$
in Eq.~\eqref{appsolutionvarepbardelta},
one has
\begin{equation}
x^2=1-\left(\frac83(1-y)^3\right)^\frac12\,.
\label{appsolutionapp0inverted}
\end{equation}
We see that Eq.~\eqref{appsolutionapp0inverted} is
Eq.~\eqref{more=0} in the limit $x\to1$ and $y\to1$.
Inverting Eq.~\eqref{appsolutionapp0inverted} 
one has
that the coincident solution
takes the form
\begin{equation}
y_1=y_2= 1-\left(\frac38 \left(1-x^2\right)^2\right)^\frac13\,,
\label{appsolutionapp0}
\end{equation}
valid for $1-x^2\ll1$,
which is Eq.~\eqref{solutionapp0}, the equation we wanted to deduce.

\vskip 1.0cm

\noindent
{\bf $x^2{\mathbf >1}$:}

\noindent
Let us deduce Eq.~\eqref{dydxnewnew}.  We have to find $y_2$, i.e.,
$r_{+2}$. 
 From Eq.~\eqref{zofxy}, or from Eq.~\eqref{y1x>1}, one finds
that there is a solution $y_2(x)$ for each $x$.
The maximum of the
curve is at $x=1$ and $y_2=1$, and the derivative there it is not well
defined.  So, it is important to discuss the behavior and the
properties of $y_2$ when $x\to 1$ from above.
 For that we define infinitesimal
quantities $\bar\delta$ and $\bar\varepsilon$, related to $x$ and $y$,
respectively, by
\begin{equation}
x=1+{\bar\delta}\,,\quad\quad\quad y=1-{\bar\varepsilon}\,,
\label{appdeltadefx>1}
\end{equation}
with $\bar\delta\ll1$ and $\bar\varepsilon\ll1$, both positive, and with
$\frac{\bar\varepsilon}{\bar\delta}$ being a quantity of order one.
 Indeed,
$\bar\delta>0$ for $x>1$, and $\bar\varepsilon>0$ always.
Then, from Eq.~\eqref{zofxy}
one gets
\begin{equation}
w= \frac{2(\frac{{\bar\varepsilon}}{{\bar\delta}}-
1)}{\sqrt{\frac{{\bar\varepsilon}}{{\bar\delta}}
(\frac{{\bar\varepsilon}}{{\bar\delta}}-2)}}
\,,
\label{appnewz2}
\end{equation}
valid in first order, in the order we are working.
One sees that for $w$ finite $\frac{\bar\varepsilon}{\bar\delta}$ is
finite, so generically here in this calculation $\bar\varepsilon$ is
of the order of $\bar\delta$.
From Eq.~\eqref{dydx} one also gets
\begin{equation}
\frac{dy}{dx}=- \frac{{\bar\varepsilon}}{{\bar\delta}}\,,
\label{appdydxmu2}
\end{equation}
also valid in the order we are working. 
Since ${\bar\delta}>0$ and ${\bar\varepsilon}>0$, 
we obtain from Eq.~\eqref{appnewz2} 
$\frac{{\bar\varepsilon}}{{\bar\delta}}
=\frac{w}{\sqrt{w^2-4}}+1$ with $w\geq2$.
Thus, we have from
Eq.~\eqref{appdydxmu2} that
\begin{equation}
\frac{dy}{dx} =-\frac{w}{\sqrt{w^2-4}}-1\,,\quad\quad w\geq2\,,
\label{appdydxnewnew}
\end{equation}
which is Eq.~\eqref{dydxnewnew}, the equation we wanted to deduce.

Let us deduce now
Eq.~\eqref{solutionapp11}.
We want to find $y_1=y_2$, the coincident solution
for fixed $x$, and $w$ varying.
This is when $\frac{d w}{d y}=0$, see Eq.~\eqref{wy},
i.e., ${S}(x,y)=0$.
From Eq.~\eqref{defSbar} we have then that the
bifurcation points where $y_1=y_2$ for $x={\rm constant}$
are given by 
Eq.~\eqref{more=0x>1}.
We are interested in finding $y_1=y_2$ when $x\to1$ from above.
Instead of working with Eq.~\eqref{more=0x>1}
we work with 
Eq.~\eqref{defSbar} which is more direct.
We find
Then, from Eq.~\eqref{y0} one has 
$y_{\rm e}=1-2{\bar\delta}\label{appy0approx1recold}$,
plus higher order.
For that, let $x=1+{\bar\delta}$ as in Eq.~\eqref{appdeltadefx>1}. 
Then, from Eq.~\eqref{y0} one has 
$y_{\rm e}=1-2{\bar\delta}\label{appy0approx1rec}$,
plus higher order terms.
Now we have to work out the vicinity of $y_{\rm e}$.
Then put
$y=y_{\rm e}-{\epsilon}
\label{appvicinityofy0rec}$
where again $y_{\rm e}$ is the  root of the
equation
$y^2+y+1-\frac{3}{x^2}=0$, with solution
given by Eq.~\eqref{y0}, i.e.,
$y_{\rm e}=-\frac{1}{2}+\sqrt{3}\sqrt{\frac{1}{x^2}-
\frac{1}{4}}$, and note
that $\epsilon$ and $\bar\varepsilon$ are different quantities.
Then, these two equations 
yield
$y=1-2{\bar\delta} -{\epsilon} 
\label{appy0approx2rec}$.
From $x=1+{\bar\delta}$ of Eq.~\eqref{appdeltadefx>1}
we find that Eq.~\eqref{defSbar}
is
${S}=(y-1)^3(y^2+3y+4)+4{\bar\delta}
(y-1)^2(y+1)(1+y+y^2)+(2y^2+y^5)4{\bar\delta}^2$.
From  $y=1-2{\bar\delta} -{\epsilon}$ we have just found,
we find that Eq.~\eqref{defSbar} is now
${S}=12{\bar\delta}^2
+24{\bar\delta}{\epsilon}^2-8{\epsilon}^3
\label{appnewdefS1rec}$.
The coincident
root is given
when $\frac{\partial w}{\partial y}=0$, i.e.,
${S}=0$. Then, 
$
12{\bar\delta}^2
+24{\bar\delta}{\epsilon}^2-8{\epsilon}^3$, i.e.,
$
{\bar\delta}^2
+2{\bar\delta}{\epsilon}^2-\frac23{\epsilon}^3=0$.
The solution is $
{\bar\delta}=-{\epsilon}^2
+\sqrt{
{\epsilon}^4+\frac23{\epsilon}^3
}
$.
It turns out that equation $\frac{\partial w}{\partial y}=0$ gives a
self-consistent solution if ${\bar\delta}\ll{\bar \epsilon}$, see
below.
Since ${\epsilon}\ll1$ the dominant term gives
${\bar\delta}=(\frac23{\epsilon}^3)^\frac12$.  We defined
${\bar\varepsilon}$ as $y=1-{\bar\varepsilon}$ and also had
$y=1-2{\bar\delta} -{\epsilon}$, so ${\bar\varepsilon}=2{\bar\delta}
+{\epsilon}$. But ${\bar\delta}$ goes with ${\epsilon}^\frac32$, so in
this order ${\bar\varepsilon}={\epsilon}$, so that
\begin{equation}
{\bar\delta}=\left(\frac23{\bar\varepsilon}^3\right)^\frac12\,.
\label{appsolutionvarepbardeltax>1}
\end{equation}
Thus, $\bar\delta$ goes with $\bar\varepsilon^\frac32$ or, if one
prefers, $\bar\varepsilon$ goes with $\bar\delta^\frac23$, indeed
$\bar\varepsilon=\left(\frac32\bar\delta^2\right)^\frac13$.  Recall
that $x=1+\bar\delta$, and so $x^2=1-2\bar\delta$ in this
approximation, and that $y=1-\bar\varepsilon$.  Then, substituting
$\bar\delta$ for $x^2$ and $\bar\varepsilon$ for $y$ in
Eq.~\eqref{appsolutionvarepbardeltax>1}, one has
\begin{equation}
x^2=1+\left(\frac83(1-y)^3\right)^\frac12\,.
\label{appsolutionapp0invertedx>1}
\end{equation}
We see that Eq.~\eqref{appsolutionapp0invertedx>1} is
Eq.~\eqref{more=0x>1} in the limit $x\to1$ from above and $y\to1$.
Inverting Eq.~\eqref{appsolutionapp0invertedx>1} one has that the
coincident solution takes the form
\begin{equation}
y_1=y_2= 1-\left(\frac38 \left(x^2-1\right)^2\right)^\frac13\,,
\label{appsolutionapp1x>1}
\end{equation}
valid for $x^2-1\ll1$,
which is Eq.~\eqref{solutionapp11}, the equation we wanted to deduce.

Finally, we calculate the implications of Eqs.~\eqref{change1}
and \eqref{change2}.
We have defined $u$ as 
\begin{equation}
u\equiv \frac{r_{\rm c}}{R}\,,
\label{appudef}
\end{equation}
as the cosmological radius in units of $R$, which
from Eq.~\eqref{rcofr+Lambda} means
that $u$ obeys 
\begin{equation}
u=-\frac{y}{2}+\sqrt{\frac{3}{x^2}-\frac{3}{4}y^2}\,.
\label{appuequation}
\end{equation}
Calculating $\frac{du}{dy}$
at the neighborhood of the
point $x=1$, $y=1$, $u=1$, and  recalling 
that we are interested
in the large black hole, i.e., the one
with subscript 2, one finds from
Eq.~\eqref{appuequation}, with the help of
Eqs.~\eqref{appdydxmu} and \eqref{appdydxmu2} , that
\begin{equation}
\left(\frac{du_2}{dx}\right)_{x=1_-}=
\left(\frac{dy_2}{dx}\right)_{x=1_+}
\label{appchange1}
\end{equation}
and 
\begin{equation}
\left(\frac{du_2}{dx}\right)_{x=1_+}=
\left(\frac{dy_2}{dx}\right)_{x=1_-}
\label{appchange2}
\end{equation}
where $1_-$ and $1_+$ means that one is taking the
limit $x\to1$ from below, i.e., $x<1$,
and from above, 
i.e., $x>1$, respectively.
We
can now deduce some further specific properties of the cosmological
horizon.
In Eq.~\eqref{appdydxnewnew} we found
$\left(\frac{dy_2}{dx}\right)_{x=1_+} =-\frac{w}{\sqrt{w^2-4}}-1$
so that from Eq.~\eqref{appchange1} we have
$\left(\frac{du_2}{dx}\right)_{x=1_-}=-\frac{w}{\sqrt{w^2-4}}-1$.
Thus, 
$\left(\frac{du_2}{dx}\right)_{x=1_-}
\to-\infty$ for $w\to 2$,
and $\left(\frac{du_2}{dx}\right)_{x=1_+}\to -2$
for $w\to \infty$.
In Eq.~\eqref{appdydxnew} we found
$\left(\frac{dy_2}{dx}\right)_{x=1_-} =\frac{w}{\sqrt{w^2-4}}-1$
so that from Eq.~\eqref{appchange2} we have
$\left(\frac{du_2}{dx}\right)_{x=1_+}=\frac{w}{\sqrt{w^2-4}}-1$.
Thus, 
$\left(\frac{du_2}{dx}\right)_{x=1_+}
\to\infty$ for $w\to 2$,
and $\left(\frac{du_2}{dx}\right)_{x=1_+}\to 0$
for $w\to \infty$. This was stated in the main text.


\begin{thebibliography}{99}

\bibitem{bek}
J. D. Bekenstein, ``Black holes and entropy'', Phys. Rev. D {\bf 7},
2333 (1973).

\bibitem{h}
S. W. Hawking, ``Particle creation by black holes'', Commun.
Math. Phys. {\bf 43}, 199 (1975).

\bibitem{ghpath}
G. W. Gibbons and S. W. Hawking, ``Action integrals and partition
functions in quantum gravity'', Phys. Rev. D {\bf 15}, 2752 (1977).

\bibitem{gh}
G. W. Gibbons and S. W. Hawking, ``Cosmological event horizons,
thermodynamics, and particle creation'', Phys. Rev. D {\bf 15}, 2738
(1977).

\bibitem{gp}
P. Ginsparg and M. J. Perry, ``Semiclassical perdurance of the de
Sitter space'', Nuclear Physics B {\bf 222}, 215 (1983).

\bibitem{y86}
J. W. York, ``Black hole thermodynamics and the Euclidean Einstein
action'', Phys. Rev. D {\bf 33}, 2092 (1986).

\bibitem{wy}
B. F. Whiting and J. W. York, ``Action principle and partition
function for the gravitational field in black hole topologies'',
Phys. Rev. Lett. {\bf 61}, 1336 (1988).

\bibitem{hwds}
G. Hayward, ``Euclidean action and the thermodynamics of manifolds
without boundary'', Phys. Rev. D {\bf 41}, 3248 (1990).

\bibitem{bbwy}
H. W. Braden, J. D. Brown, B.  F. Whiting, and J. W. York, ``Charged
black hole in a grand canonical ensemble'', Phys. Rev. D {\bf 42},
3376 (1990).

\bibitem{can}
O. B. Zaslavskii, ``Canonical ensemble for arbitrary configurations of
self-gravitating systems'', Phys. Lett. A {\bf 152}, 463 (1991).

\bibitem{jpsl}
J. P. S. Lemos, ``Thermodynamics of the two-dimensional black hole in
the Teitelboim-Jackiw theory'', Phys. Rev. D {\bf 54}, 6206 (1996);
arXiv:gr-qc/9608016.

\bibitem{prl}
O. B. Zaslavskii, ``Extreme state of a charged black hole in a grand
canonical ensemble'', Phys. Rev. Lett. {\bf 76}, 2211 (1996).

\bibitem{prd97}
O. B. Zaslavskii, ``Geometry of nonextreme black holes near the
extreme state'', Phys. Rev. D {\bf 56}, 2188 (1997);
arXiv:gr-qc/9707015.

\bibitem{pecalemos}
C. S. Pe\c ca and J. P. S. Lemos, ``Thermodynamics of
Reissner-Nordstr\"om-anti-de Sitter black holes in the grand canonical
ensemble'', Phys. Rev. D {\bf 59}, 124007 (1999); arXiv:gr-qc/9805004.

\bibitem{Andre:2020}
R.~Andr\'{e} and J.~P.~S.~Lemos, ``Thermodynamics of five-dimensional
Schwarzschild black holes in the canonical ensemble'', Phys. Rev. D
{\bf 102}, 024006 (2020); arXiv:2006.10050 [hep-th].

\bibitem{Andre:2021}
R.~Andr\'{e} and J.~P.~S.~Lemos, ``Thermodynamics of $d$-dimensional
Schwarzschild black holes in the canonical ensemble'', Phys. Rev. D
{\bf 103}, 064069 (2021); arXiv:2101.11010 [hep-th].

\bibitem{fernandeslemos}
T. V. Fernandes and J. P. S. Lemos, ``Grand canonical ensemble of a
$d$-dimensional Reissner-Nordstr\"om black hole in a cavity'',
Phys. Rev. D {\bf 108}, 084053 (2023); arXiv:2309.12388 [hep-th].

\bibitem{lz2023}
J. P. S. Lemos and O. B. Zaslavskii, ``Black holes and hot shells in
the Euclidean path integral approach to quantum gravity'', Classical
Quantum Gravity {\bf 40}, 235012 (29023); arXiv:2304.06740 [hep-th].

\bibitem{wanghuang1}
B.-B. Wang and C.-G. Huang, ``Thermodynamics of de Sitter spacetime in
York's formalism'', Mod. Phys. Lett. A {\bf 16}, 1487 (2001); 
arXiv:2304.06740 [hep-th].

\bibitem{wanghuang2}
B.-B. Wang and C.-G. Huang, ``Thermodynamics of
Reissner–Nordstr\"om-de Sitter black hole in York's formalism'',
Classical Quantum Gravity {\bf 19}, 2491 (2002).

\bibitem{ghezelbashmann}
A. M. Ghezelbash and R. B. Mann, ``Action, mass and entropy of
Schwarzschild-de Sitter black holes and the de Sitter/CFT
correspondence'', J. High Energy Phys. JHEP 01 (2002) 005;
arXiv:hep-th/0111217 [hep-th].

\bibitem{saida}
H. Saida, ``De Sitter thermodynamics in the canonical ensemble'',
Prog. Theor. Phys. {\bf 122}, 1239 (2009); arXiv:0908.3041 [gr-qc].

\bibitem{draperfarkas}
P. Draper and S. Farkas, ``De Sitter black holes as constrained states
in the Euclidean path integral'', Phys. Rev. D {\bf 105}, 126022
(2022); arXiv:2203.02426 [hep-th].

\bibitem{banihashem}
B. Banihashem and T. Jacobson, ``Thermodynamic ensembles with
cosmological horizons'', J. High Energy Phys.
JHEP 07 (2022) 042; arXiv:2204.05324
[hep-th].

\bibitem{bjsv}
B. Banihashemi, T. Jacobson, A. Svesko, and M. R. Visser,
``The minus sign in the first law of de Sitter horizons'',
J. High Energy Phys. JHEP 01 (2023) 054; arXiv:2208.11706 [hep-th].

\bibitem{jacviss2}
T. Jacobson and M. R. Visser, ``Partition function for a volume of
space'', Phys. Rev. Lett. 130, 221501 (2023); arXiv:2212.10607
[hep-th].

\bibitem{jacviss}
T. Jacobson and M. R. Visser, ``Entropy of causal diamond ensembles'',
SciPost Phys. 15 (2023) 023; arXiv:2212.10608 [hep-th].

\bibitem{morvan}
E. K. Morvan, J. P. van der Schaa, and M. R. Visser, ``On the
Euclidean action of de Sitter black holes and constrained
instantons'', SciPost Phys. 14, 022 (2023); arXiv:2203.06155 [hep-th].

\bibitem{pcwdavies}
P. C. W. Davies, ``Thermodynamic phase transitions of Kerr-Newman
black holes in de Sitter space'', Classical Quantum Gravity {\bf 6},
1909 (1989).

\bibitem{romans}
L. J. Romans, ``Supersymmetric, cold and lukewarm black holes in
cosmological Einstein-Maxwell theory'', Nuc. Phys. B {\bf 383}, 395
(1992).

\bibitem{boussohawking2}
R. Bousso and S. W. Hawking, ``(Anti-)evaporation of Schwarzschild–de
Sitter black holes'', Phys. Rev. D {\bf 57}, 2436 (1998);
arXiv:hep-th/9709224.

\bibitem{maeda}
K. Maeda, T. Koike, M. Narita, and A. Ishibashi, ``Upper bound for
entropy in asymptotically de Sitter space-time'', Phys. Rev. D {\bf
57}, 3503 (1998); arXiv:gr-qc/9712029.

\bibitem{wu}
Z. C.  Wu, ``Entropy of a black hole with distinct surface
gravities'', General Relativ. Gravit. {\bf 32}, 1823 (2000);
arXiv:gr-qc/9911078.

\bibitem{yueqinlichunren}
W. Yueqin, Z. Lichun, and Z. Ren, ``Black hole and cosmic entropy for
Schwarzschild–de Sitter spacetime'', Int. J. Theor. Phys. {\bf 40},
1001 (2001).

\bibitem{bousso}
R. Bousso, ``Positive vacuum energy and the $N$-bound'',
J. High Energy Phys. JHEP 0011
(2000) 038; arXiv:hep-th/0010252.

\bibitem{cai1}
R.-G. Cai, ``Cardy-Verlinde formula and thermodynamics of black holes
in de Sitter spaces'', Nucl. Phys. B {\bf 628}, 375 (2002);
arXiv:hep-th/0112253.

\bibitem{shanka}
S. Shankaranarayanan, ``Temperature and entropy of Schwarzschild–de
Sitter spacetime'', Phys. Rev. D {\bf 67}, 084026 (2003);
arXiv:gr-qc/0301090.

\bibitem{teitel1}
C. Teitelboim, ``Gravitational thermodynamics of Schwarzschild-de
Sitter space'', in {\it Strings and Gravity: Tying the Forces
Together}, Proceedings of the Fifth Francqui Colloquium in Brussels in
2001 (de Boek Universit\'e, Bruxelles, 2003), p. 291; hep-th/0203258.

\bibitem{teitel2}
A. Gomberoff and C. Teitelboim, ``De Sitter black holes with either of
the two horizons as a boundary'', Phys. Rev. D {\bf 67}, 104024 (2003);
hep-th/0302204.

\bibitem{diaslemospaircreationdesitter}
O. J. C. Dias and J. P. S. Lemos, ``Pair creation of de Sitter black
holes on a cosmic string background'', Phys. Rev. D {\bf 69}, 084006
(2004); arXiv:hep-th/0310068.

\bibitem{diaslemoscardosonariaiinhigherd}
V. Cardoso, O. J. C. Dias, and J. P. S. Lemos, ``Nariai,
Bertotti-Robinson and anti-Nariai solutions in higher dimensions'',
Phys. Rev. D {\bf 70}, 024002 (2004): arXiv:hep-th/0401192.

\bibitem{sekiwa}
Y. Sekiwa, ``Thermodynamics of de Sitter black holes: Thermal
cosmological constant'', Phys. Rev. D {\bf 73}, 084009 (2006);
arXiv:hep-th/0602269.

\bibitem{roychoudhurypadmanabhan}
T. R. Choudhury and T. Padmanabhan, ``Concept of temperature in multi
horizon spacetimes: Analysis of Schwarzschild–de Sitter metric'',
Gen. Relativ. Gravit. {\bf 39}, 1789 (2007); arXiv:gr-qc/0404091.

\bibitem{myung}
Y. S. Myung, ``Thermodynamics of Schwarzschild-de Sitter black hole:
Thermal stability of Nariai black hole'', Phys. Rev. D {\bf 77}, 104007
(2008); arXiv:0712.3315 [gr-qc].

\bibitem{pappas}
T. Pappas and P. Kanti, ``Schwarzschild–de Sitter spacetime: The role
of temperature in the emission of Hawking radiation'', Phys. Lett. B
{\bf 775}, 140 (2017); arXiv:1707.04900 [hep-th].

\bibitem{fil}
F. Simovic and R. B. Mann, ``Critical phenomena of Born-Infeld-de Sitter
black holes in cavities'', J. High Energy Phys. JHEP 05 (2019) 136;
arXiv:1904.04871 [gr-qc].

\bibitem{qiutraschen}
Y. Qiu and J. Traschen, ``Black hole and cosmological particle
production in Schwarzschild-de Sitter'', Classical Quantum Gravity
{\bf 37}, 135012 (2020); arXiv:1908.02737 [hep-th].

\bibitem{singha}
C. Singha, ``Thermodynamics of multi horizon spacetimes'',
Gen. Relativ. Gravit. {\bf 54}, 38 (2022); arXiv:2108.11704
[gr-qc].

\bibitem{volovik2}
G. E. Volovik, ``Double Hawking temperature: From black hole to de
Sitter'', Universe {\bf 8}, 639 (2022), arXiv:2205.06585 [gr-qc].

\bibitem{akhmedov}
E. T. Akhmedov and K. V. Bazarov, ``On the backreaction issue for the
black hole in de Sitter spacetime'', Phys. Rev. D {\bf 107}, 105012
(2023); arXiv:2212.06433.

\bibitem{andreassonbohmer}
H. Andr\'easson and C. G. B\"ohmer, ``Bounds on $M/R$ for static
objects with a positive cosmological constant'', Classical Quantum
Gravity {\bf 26}, 195007 (2009); arXiv:0904.2497 [gr-qc].



\end{thebibliography}
\end{document}